\documentclass[journal=jacsat,manuscript=article]{achemso}

\usepackage{a4wide}
\usepackage{graphicx,amssymb}
\usepackage{amsmath}
\usepackage{subfig}
\usepackage{float}
\usepackage[dutch,english]{babel}
\usepackage{color}
\usepackage[utf8]{inputenc}
\usepackage{graphicx}

\usepackage{amsmath}
\usepackage{color}
\usepackage{url}
\usepackage[version=4]{mhchem}
\makeatletter
\usepackage{float}
\usepackage[section]{placeins}
\usepackage[symbol]{footmisc}

\DeclareCaptionLabelFormat{adja-page}{\hrulefill\\#1 #2 \emph{(previous page)}}

\usepackage{xr}

\usepackage{mathtools}

\graphicspath{{figures/}}

\author{James L. Martin Robinson*\footnote[0]{*J.L.M.R. and N.M. contributed equally to this work.}\footnote{Present address: Department of Orthopedics, University Medical Centre Utrecht, 3584 CT Utrecht, The Netherlands}} 

\affiliation[Utrecht University]
{Van’t Hoff Laboratory for Physical and Colloid Chemistry, Debye Institute for Nanomaterials Science, Utrecht University, 3584 CH Utrecht, The Netherlands}

\author{Neshat Moslehi*\footnote{Present address: Laboratory of Self-Organizing SoftMatter, Department of Chemical Engineering and Chemistry, Eindhoven University of Technology, 5600 MB Eindhoven, The Netherlands}}
\affiliation[Utrecht University]
{Van’t Hoff Laboratory for Physical and Colloid Chemistry, Debye Institute for Nanomaterials Science, Utrecht University, 3584 CH Utrecht, The Netherlands}

\author{Nikolaos Dramountanis \footnote{Present address: Allnex Netherlands BV, 4600 AB, Bergen op Zoom, The Netherlands}}
\affiliation[Utrecht University]
{Van’t Hoff Laboratory for Physical and Colloid Chemistry, Debye Institute for Nanomaterials Science, Utrecht University, 3584 CH Utrecht, The Netherlands}

\author{Lennart van den Hoven\footnote{Present address: Department of Pharmaceutics, Utrecht Institute for Pharmaceutical Sciences (UIPS), Utrecht University, 3584, CG, Utrecht, the Netherlands}}
\affiliation[Utrecht University]
{Van’t Hoff Laboratory for Physical and Colloid Chemistry, Debye Institute for Nanomaterials Science, Utrecht University, 3584 CH Utrecht, The Netherlands}

\author{Alexander M. van Silfhout\footnote{Present address: TNO Environmental Modelling, Sensing and Analysis, Princetonlaan 6, 3584 CB Utrecht, The Netherlands}}
\affiliation[Utrecht University]
{Van’t Hoff Laboratory for Physical and Colloid Chemistry, Debye Institute for Nanomaterials Science, Utrecht University, 3584 CH Utrecht, The Netherlands}

\author{Kanvaly S. Lacina}
\affiliation[Utrecht University]
{Van’t Hoff Laboratory for Physical and Colloid Chemistry, Debye Institute for Nanomaterials Science, Utrecht University, 3584 CH Utrecht, The Netherlands}

\author{Mies van Steenbergen}
\affiliation[PharmacyUU]{Department of Pharmaceutics, Utrecht Institute for Pharmaceutical Sciences (UIPS), Utrecht University, 3584 CG, Utrecht, the Netherlands}

\author{Wessel Custers}
\affiliation[Utrecht University]
{Van’t Hoff Laboratory for Physical and Colloid Chemistry, Debye Institute for Nanomaterials Science, Utrecht University, 3584 CH Utrecht, The Netherlands}

\author{Bas G. P. van Ravensteijn}
\affiliation[PharmacyUU]{Department of Pharmaceutics, Utrecht Institute for Pharmaceutical Sciences (UIPS), Utrecht University, 3584 CG, Utrecht, the Netherlands}

\author{Willem K. Kegel}
\email{w.k.kegel@uu.nl}
\affiliation[Utrecht University]
{Van’t Hoff Laboratory for Physical and Colloid Chemistry, Debye Institute for Nanomaterials Science, Utrecht University, 3584 CH Utrecht, The Netherlands}

\title{\textbf{Cooperative Ligand-Mediated Transitions in Simple Macromolecules} \begin{footnotesize} \\ \textcolor{red}{\textbf{Please use the following link to access the published, citable, and \underline{corrected} version of this manuscript: }} \url{https://pubs.acs.org/doi/10.1021/acs.jpcb.5c05386}\end{footnotesize}}

\begin{document}

\newpage
\begingroup
\let\center\flushleft
\let\endcenter\endflushleft
\maketitle 
\endgroup

\clearpage

\section{Abstract}

In biology, ligand mediated transitions (LMT), where the binding of a molecular ligand onto the binding site of a receptor molecule leads to a well-defined change in the conformation of the receptor, are often referred to as 'the second secret of life'. Sharp, cooperative transitions arise in many biological cases, while examples of synthetic cooperative systems are rare. This is because well-defined conformational states are hard to ‘program’ into a molecular design. Here, we impose an external constraint in the form of two immiscible liquids that effectively define and limit the available conformational states of two different synthetic and relatively simple macromolecules. We show that the mechanism of the observed cooperative transitions with ligand concentration is the coupling of ligand binding and conformation, similar to more complex biological systems. The systems studied are:\\ 
(1) Hydrophobic polyelectrolytes (HPE), which are (bio) polymers that consist of hydrophobic as well as ionizable (proton and hydroxyl ligand-binding) functional groups. \\ 
(2) Oligomeric metal chelators (OMC), which are oligomers composed of metal ion chelating repeating groups that are able to bind metal ions (considered as the ‘ligands’), resulting in gel-like networks of oligomers crosslinked by coordinated metal ions.\\
We find that in HPE, interactions between ligands and individual macromolecules explain the observed cooperative transitions. For OMC, coordinated bonds significantly enhance the degree of cooperativity, compared to HPE.

\section{Keywords}

Allosteric transitions, Hydrophobic polyelectrolytes, Monod-Wyman-Changeux (MWC) theory, Terpyridine-functionalized polymers. 
     
\vfill
\pagebreak

\section{Introduction}

Many of the sharp, switch-like transitions seen in biological macromolecules can be attributed to the ubiquitous Monod-Wyman-Changeaux (MWC) mechanism \cite{Monod1965,Marzen2013}. The central concept of MWC is the competition between conformational states of the protein and the strength of the ligand binding interaction. In general, the ground-state conformation has a relatively low affinity for ligands, while a conformationally unfavorable state has a high(er) affinity. Upon increasing the concentration of the ligands, a sharp transition can be observed from the ground state to the unfavorable conformation, which is stabilized by ligand binding. This relies on highly ordered macromolecular protein structures, which constrain the number of thermally accessible conformational states. In that way, transitions only occur between these well-defined conformations.  

\begin{figure}[H]
    \centering
    \includegraphics[width=1\linewidth]{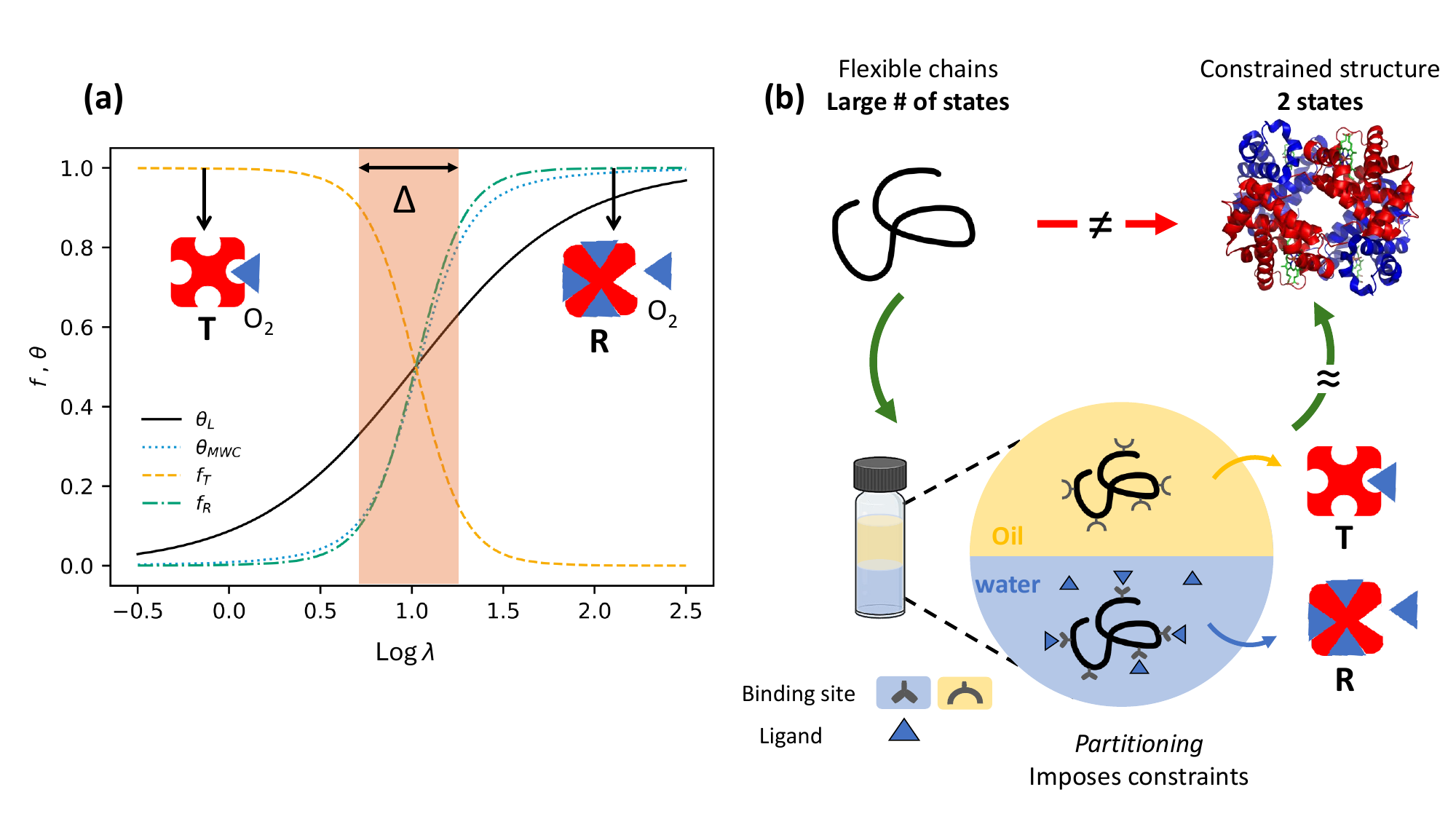}
    \caption{\textbf{Sharp, cooperative, ligand mediated transitions are predicted for simple oligomeric species. (a)} Plot of the fraction of hemoglobin in the Relaxed ($f_R$) and Tense ($f_T$) states, and the fraction of bound oxygen onto hemoglobin, $\theta_{MWC}$ as a function of $\lambda$. $\lambda$ is the oxygen fugacity and proportional to the partial oxygen pressure. For details see our previous work \cite{martin2023cooperative}. The Langmuir isotherm $\theta_L$ is Eq.~(\ref{eq:theta_Langmuir}), with $\beta g = -6$. $\Delta$ is a measure for the width, or sharpness, of the oxygen-ligand mediated transition (LMT). \textbf{(b)} Schematic of the analogy between a conformationally constrained oligomer and hemoglobin. The oligomer in isolation presents many thermally accessible conformations compared to the two main conformations for hemoglobin. However, when placed in a two-state oil and water system the oligomer is funneled into a reduced set of accessible conformations, which coupled to the preferential ion binding in the aqueous phase leads to sharp, cooperative transitions.}
    \label{fig:Intro}
\end{figure}

The archetypal cooperative ligand-mediated transition is the binding of oxygen to hemoglobin. This protein, found in most vertebrates, is composed of four individual globin protein units. Each of these contain a heme group, which act as oxygen binding sites. The explanation for the cooperativity hinges on the notion of a competition between the conformational state of the protein and the binding energy of oxygen. The T (Tense) state is the conformational ground state but presents weaker binding, conversely, the R (Relaxed) state is an unfavorable conformation for the protein but presents a higher affinity for oxygen. The situation is schematically summarized in Fig.~\ref{fig:Intro}(a). More details can be found in a previous work\cite{martin2023cooperative}. At this point we emphasize that the only requirements for MWC-like transitions are: (1) at least two conformational states, where each state has a characteristic binding affinity to the ligands, and (2) the conformationally unfavorable state has a stronger affinity for the ligands \cite{Marzen2013, martin2023cooperative}. 

Several classes of cooperative transitions can be distinguished, for example, phase transitions, superconductivity, magnetization, and several more in more complex molecular systems \cite{Rest2015, Dubacheva2019, Marzen2013}. Here we focus on ligand-mediated transitions. Observations of significant cooperativity in ligand-mediated transitions in non-biological systems are rare. Exceptions we are aware of are: (1) the binding of oxygen and other small molecules onto solid iron-porphyrin derivatives \cite{Collman1978} and onto helical poly-L-lysine heme complex \cite{Tsuchida1976}, and (2) synthetic, low-molecular-weight ligands onto aggregates of modified cyclodextrins have been observed to bind cooperatively, possibly by an MWC mechanism\cite{Petter1990}. In these examples, the observed degree of cooperativity (defined in the Theory section)  is usually less than the situation in oxygen binding onto hemoglobin.

We have recently shown that  hydrophobic polyelectrolytes (HPE) that are able to solubilize bilayer membranes or undergo micellization  as a function of $pH$ operate by a competition between two or more conformational states and their degree of ionization. These systems fit within the framework of MWC theory, showcasing the requirement of a limited number of conformational states to realize cooperative transitions \cite{martin2023cooperative}. The key idea is that ionization (deprotonation of acid groups or protonation of basic groups) can be seen as the binding of ligands in the form of hydroxyl ions or protons. The previously mentioned requirement for the number of conformational states of the macromolecules to be constrained is, in the case of HPE, provided by external conditions or conditions inherent to the macromolecular architecture. In the micellization systems\cite{Zhou2011,Li2016_2,Li2016,Ma2014}, the polymer conformations are constrained by the hydrophilic blocks attached to them, which drives self-assembly into micelles. The core of these micelles now becomes a hydrophobic reservoir for the HPE and a homogeneous environment allowing for a two-state description for the conformational state of the polymer: freely dissolved and in the micelle core. In the case of membrane solubilization, the hydrophobic reservoir which will constrain the conformations of the HPE is the core of the lipid membranes. For these systems the presence of the lipid vesicles funnels the many potential conformations of the HPE into three dominant ones: freely dissolved, around the edge of a lipid nanodisk and fully immersed in the hydrophobic core of the lipid membranes\cite{thomas1995tuningTirrel,Scheidelaar2016}. 

While coupling between conformational states and ligand binding explains the observations, the analysis in \cite{martin2023cooperative} also points to an important difference with respect to similar transitions in biological macromolecules. In hemoglobin, for example, the conformational states are single molecular states that are available irrespective of the larger environment of hemoglobin. In HPE, on the other hand, hydrophobic and aqueous reservoirs are required, which are realized either by local phase separation of HPE-containing diblocks in the form of micellar cores \cite{Li2016} or by the availability of lipid bilayer vesicles \cite{thomas1995tuningTirrel,Scheidelaar2016}. In this work, we avoid the complexity inherent in micelle formation and solubilization of bilayer vesicles, and pin down the mechanism of LMT in well-defined hydrophobic and aqueous reservoirs. We design a two-phase water/oil experimental set-up in which the partitioning of a weakly acidic hydrophobic polyelectrolyte (HPE) and an oligomeric metal chelator (OMC) between oil and water are studied, see Fig.~\ref{fig:Intro}(b), for a schematic representation of the experimental set-up. The partitioning of the macromolecules between oil and water, as a function of the relevant ligand concentration (to be varied by $pH$ or concentration of the metal ions), is analogous to the ligand-mediated transitions between the conformational states of more complex macromolecules, such as the T and R states in hemoglobin. 

In the following we study experimentally two chemically distinct polymers and ligands, and also address the influence of composition dispersity, specifically the dispersity in the monomer ratio of a binary copolymer. The results will be discussed within a generalization of MWC \cite{Monod1965} and our recent extension for HPE \cite{martin2023cooperative}. 

We show that these two quite different systems present cooperative partitioning transitions. Upon increasing ligand concentration (hydroxide ions via $pH$ in HPE, and iron ions in the OMC) in the water phase, the binding energy of the ligands overcomes the unfavorable interaction between hydrophobic moieties on the polymers with water, with the effect that the polymer migrates from the oil phase (hydrophobic state) to the water phase (aqueous state). Note that we consider the metal ions as ligands. The competition between these two driving forces, analogous to the hemoglobin-oxygen system, leads to a sharp transition from the oil to the water phase over a narrow range of free ligand concentration. We emphasize here that the experimental setup, with well-defined hydrophobic (oil) and aqueous regions, has been chosen as a simplification and generalization of the more complex two-states systems mentioned above \cite{Zhou2011,Ma2014, Li2016,Li2016_2,thomas1995tuningTirrel,Scheidelaar2016},
and effectively select two distinct conformational states: a hydrophobic and an aqueous state for the polymers. Thus, oil-water partitioning reflects conformational change in this context.
With the experiment design mentioned above, we show that our two constructed synthetic polymers show cooperative transitions and that this cooperative behavior fits well within our generalization of the MWC model. 

\section{Theory}
The theoretical framework used in this paper to explain the cooperative binding of ionic ligands onto macromolecular templates is based on the MWC model \cite{Monod1965, Marzen2013} mentioned in the Introduction, and which has been applied recently to HPE \cite{martin2023cooperative}. In this work, the model has been extended to include OMC as well as composition dispersity. Detailed derivations are written in the SI (Section~\ref{theory_SI}). Here we briefly summarize the predictions of the model.

We consider partitioning of oligomers with $M$ effective binding sites for ligands over equal volumes of aqueous ($aq$) and oily (hydrophobic, $H$) liquid. The particular liquid phase the oligomer is dissolved in defines the conformational state of the oligomer. The binding affinity for ligands in the hydrophobic state is assumed to be negligibly small compared to the situation in the aqueous state. The ground state of the oligomers is the hydrophobic state. This assumes oligomers with a fairly hydrophobic composition, which will preferably partition into an oily phase over an aqueous phase when the ligand concentration is low in the aqueous phase. Oligomers pay a hydrophobic penalty, $G$, analogous to a conformational penalty, when dissolved in the aqueous state. At the same time, in the aqueous state, ligand binding is favorable with binding free energy $g$ per ligand. As derived in the SI, the fraction of oligomers in the aqueous state is given by $f_{aq} = \frac{\exp{(-\beta G)}(1 + \lambda \exp{(-\beta g)} )^{M}}{1+\exp{(-\beta G)}(1 + \lambda \exp{(-\beta g)} )^{M}}$. The value of $M$ determines the sharpness of the transition from $f_{aq} = 0$ to $f_{aq} = 1$ as a function of $\lambda$ and is defined as the degree of cooperativity. The fugacity or activity of the ligand, $\lambda = \exp{(\beta \mu)}$, with $\beta = 1 /k_B T$. $k_B$ is Boltzmann's constant and $T$ the absolute temperature. $\mu$ is the chemical potential of the ligand that adsorbs (or binds) onto the macromolecular template. $\mu$ is related to the (free) ligand concentration or partial pressure of the ligand.

Further, we show that the fraction of bound ligands is given by $\theta = \langle N \rangle/M = \frac{\lambda \exp{(-\beta g)}}{1 + \lambda \exp{(-\beta g)}} f_{aq}$, with $\langle N \rangle$ the average number of occupied binding sites. This implies a strong correlation between $f_{aq}$ and $\theta$, as around the transition we have $\lambda \exp{(-\beta g)} \gg 1$ and $f_{aq} \approx \theta$, similar to the situation with $f_R$ and $\theta$ for hemoglobin, see Fig.~\ref{fig:Intro}(a). Under the same condition $\lambda \exp{(-\beta g)} \gg 1$, we find that $\theta \approx f_{aq} \approx \frac{(\lambda \exp{(-\beta (g+ g_H))} )^{M}}{1+(\lambda \exp{(-\beta (g+ g_H)))} )^M}$. Here we defined the conformational penalty per ligand binding site $g_H$ via $G = M g_H$. This result is isomorphic to the Hill equation \cite{Hill1910}

\begin{equation}
\theta_H = \frac{(K[L])^{n_{H}}}{1 + (K[L])^{n_{H}}} ~~~~\text{(Hill)}
\label{eq:Hill}
\end{equation}

with $K$ the ligand binding constant, $[L]$ the free ligand concentration and $n_H$ the Hill exponent. If only a single state is available, the fraction of bound ligands reduces to the Langmuir equation

\begin{equation}
	\theta_L= \frac{\lambda\exp(-\beta g)}{1+\lambda\exp(-\beta g)} \label{eq:theta_Langmuir}
\end{equation}

which has been plotted in Fig.~\ref{fig:Intro}(a).

Similarly to the Hill exponent, we consider a transition with a value of $M=1$, which is isomorphic to a Langmuir isotherm, to correspond to a non-cooperative transition. Values of $M$ higher than unity correspond to cooperative behavior.
Compared to a, in general, strictly empirical value of the Hill exponent, we note that the value of $M$ in the equations described above is directly linked to the number of binding sites on the template as well as to other properties such as the presence of intermediate states. Moreover, the theory allows for the effects of polymer (template) length (via the value of $M$) and length (size) dispersity, as well as chemical dispersity to be predicted.

\section{Methods}
In order to study the cooperative partitioning transitions between oil and water, two case studies were designed; the first uses weakly acidic hydrophobic polyelectrolytes (HPE) and the second use an iron-binding oligomeric metal chelator (OMC).
For each case study, the macromolecules were designed and synthesized, followed by thorough polymer characterization. For the HPE systems, poly(6-(acryloyl)aminohexanoic acid) (PAHA, DP$=18$, \emph{\DJ}$=1.05$) and poly(\textit{n}-butyl acrylate-\textit{r}-acrylic acid) (PBA-AA, DP$=22$, \emph{\DJ}$=1.1$) were synthesized using Reversible Addition-Fragmentation Chain Transfer (RAFT) and Single Electron Transfer Living Radical Polymerization (SET-LRP) techniques, respectively. In the OMC case, a terpyridine-functionalized oligomer (abbreviated as PT, DP$=16$, \emph{\DJ}$=1.04$) was synthesized using the SET-LRP technique. For the details of synthesis and characterization of the HPE and OMC, see Sections~\ref{polymer_synth_supp} and \ref{SynthesisPT16} of the SI.

Next, the transition of the synthesized macromolecules between the hydrophobic (oil; pentanol for HPE, and dichloromethane (DCM) for OMC) and aqueous phase was monitored as a function of the relevant ligand concentration in the aqueous phase ($pH$ for HPE and iron ion concentration for OMC), and compared to that of a monoprotic carboxylic acid (MPA) for the HPE or monomeric terpyridine (T) for the OMC. The partitioning of PAHA, PBA-AA, PT, and T between the oil and water phase was quantified using UV-Vis absorption, as described in detail in the SI for HPE: Sections \ref{buffered_exp} and \ref{SI_fraction_data}, and for OMC: Sections \ref{OMC_partition_T} and \ref{OMC_partition_PT}. For HPE, the ionization fraction was measured via potentiometric titrations (Section \ref{titration_exp} and \ref{SI_titration} of the SI), and for OMC, the free iron concentration was measured via UV-Vis absorption and atomic emission spectroscopy (Section \ref{quant_theta_SI} of the SI). Full description of the experiments, quantification, and data treatment are listed in the SI.

\section{Results and discussion}

\section{Experimental observation of cooperative ligand mediated transitions}
In this section, we show experimentally in two chemically quite different oligomers, HPE and OMC, that the fundamental signatures of LMT that occur in oxygen binding onto hemoglobin, see Fig.~\ref{fig:Intro}, are reproduced in partitioning experiments. These features are: (1) the transition is cooperative, that is, it occurs over a narrow range of ligand concentration and is quantified by a Hill exponent $n_H > 1$. (2) strong correlation between the fraction of the macromolecule in the conformational state with the strongest binding affinity for the ligand and the fraction of bound ligand. The last feature can be verified by comparing the correlated $f_R$ and $\theta$ in Fig.~\ref{fig:Intro} for hemoglobin with $f_{aq}$ and $\theta$, see Theory section, for HPE as well as for OMC. \\ 
Next we present our observation that compositionally disperse HPE, specifically copolymers with disperse comonomer ratios, display significant broadening of the ligand mediated transition in HPE. This effect is quantitatively explained in our extension of the MWC model.

\subsection{Partitioning behavior of a homopolymer hydrophobic polyelectrolyte (HPE)}

A simple, qualitative confirmation that HPE present cooperative partitioning behavior would be a marked increase in the sharpness of such a transition compared to a simple monoprotic acid or base. To investigate this, we synthesized a homopolymeric acidic HPE, poly(6-(acryloyl)aminohexanoic acid) (PAHA), of $18$ units in length (DP$=18$). 6-(acryloyl)aminohexanoic acid was synthesized and then polymerized using a trithiocarbonate RAFT agent to yield an oligomer with a low polydispersity index, \emph{\DJ}$=1.05$ (see SI Section~\ref{PAHA_synth} for details). We then performed titration experiments in a two-phase pentanol-water system, with equal volumes of the phases (1.5 - 2 mL) and over a wide range of $pH$ values (3-12) with a stock concentration of 1 mg/ml for PAHA in pentanol. We measured the fraction of polymer chains in the pentanol phase via UV-Vis absorption and the ionization fraction via potentiometric titration  as fully described in the SI (Section~\ref{HPE_exp_proc_supp}). 

\begin{figure}[H] 
	\centering
	\includegraphics[width=\textwidth]{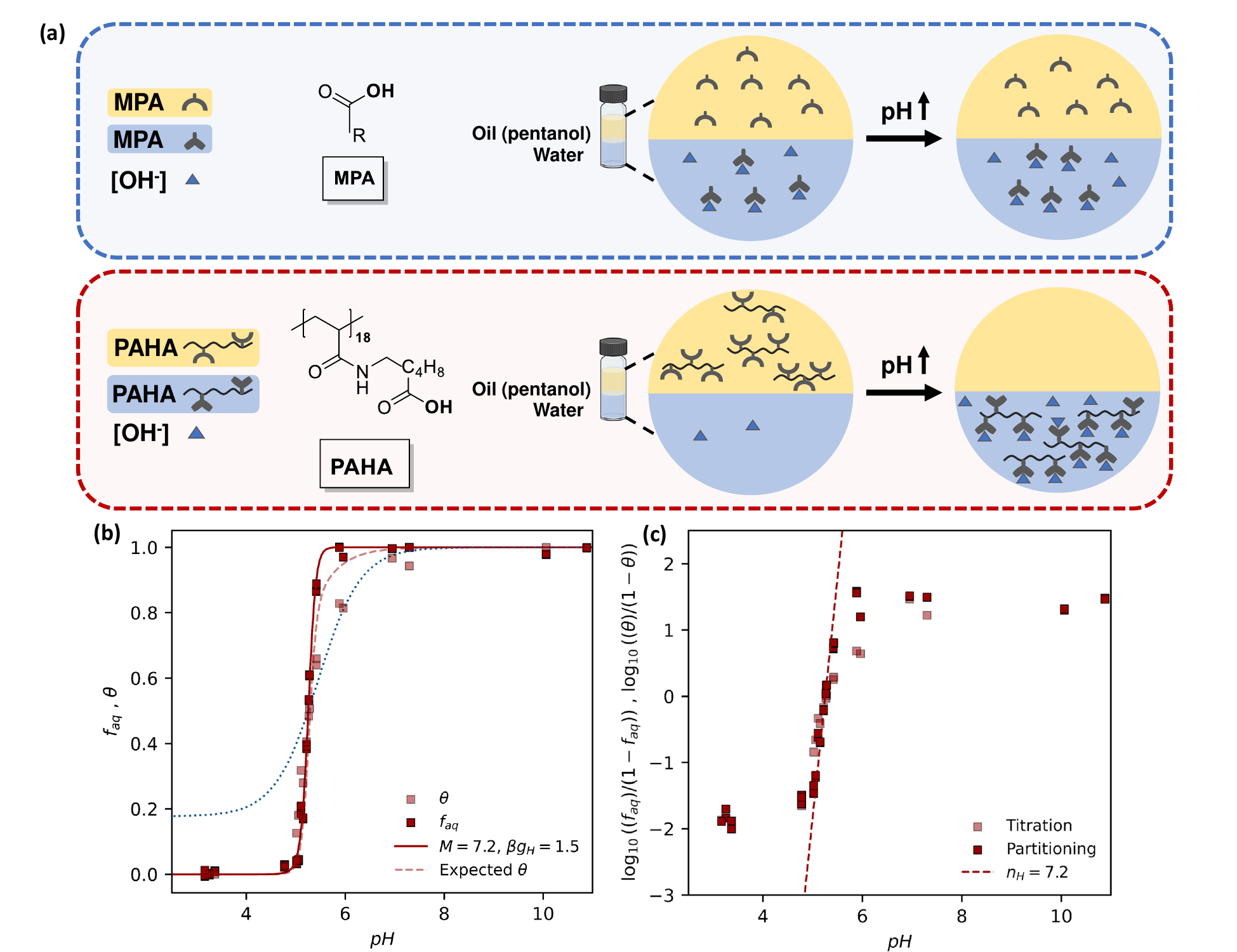}
	\caption{\textbf{Homopolymeric HPE present sharp, cooperative partitioning transitions.}
		\textbf{(a)} Schematic of the comparison of the $pH$-dependent oil and water partitioning between a monoprotic acid (MPA) and the HPE PAHA. During a small change in $pH$ ($\Delta$) there is a complete transfer of HPE between phases, while there is only a small change in partitioning for the MPA.\textbf{(b)} Plot of $f_{aq}$ ($1-f_{H}$) (red squares) and $\theta$ (light red squares) for PAHA ($DP=18$). The data is a combination of two individual runs where all repeat measurements are plotted. The $f_{aq}$ curve is fit to Eq.~(\ref{eq:fractionHPE}) (red line) and using the $pK_a$ value for propanoic acid: $4.69$ \cite{harris_quantitative_2007}. Fit parameters are shown in the legend. As a comparison a (calculated) non-cooperative transition with $M=1$ is plotted as a dotted blue line. Predicted $\theta$ curve (light red dashed line) using the fit parameters for Eq.~(\ref{eq:fractionHPE}) of the $f_{H}$ data into Eq.~(\ref{eq:theta_HPE}). \textbf{(c)} Hill plots for the fraction of ionized sites and the fraction of chains in the oil for PAHA. A straightline with a Hill parameter ($n_H$) of $7.2$ is consistent with the straightline section of this graph. See SI (Section~\ref{data_analysis_HPE}) for details on data treatment and data scaling.}
	\label{fig:HPE_fig}
\end{figure}

The experiment is illustrated in Fig.~\ref{fig:HPE_fig}(a) where a schematic of the partitioning of a monoprotic carboxylic acid (MPA) and PAHA are shown. Note that upon the same increase in the free ligand concentration of the aqueous phase (which is equivalent to increasing the $pH$ of the aqueous phase), PAHA is expected to predominantly shift from the hydrophobic to the aqueous phase, while an MPA will show a broad transition. The partitioning behavior of some simple monoprotic carboxylic acids is presented in SI Fig.~\ref{fig:monomeric_acids}. The adapted literature data shows that the main feature of the $pH$-dependent partitioning of simple acids is a wide transition over around $3$ $pH$ unit.

For an acidic HPE partitioning system our general ligand-mediated cooperative transition equations are applied to a HPE with weak acid groups (SI Section~\ref{HPE_theory_supp}): 

\begin{equation}
	f_{aq}=\frac{[\exp{(-\beta g)} \left(1 + 10^{pH - pK_{a}} \right)]^M}{1+[\exp{(-\beta g)} \left(1 + 10^{pH - pK_{a}} \right)]^M}~~~~\text{(HPE)} 
	\label{eq:fractionHPE}
\end{equation}

The fraction of occupied binding sites is then:
\begin{equation}
     \theta=\frac{10^{pH - pK_{a}}}{1+ 10^{pH - pK_{a}}}f_{aq}~~~~\text{(HPE)}  \label{eq:theta_HPE}
\end{equation}

In Eqs.~(\ref{eq:fractionHPE},\ref{eq:theta_HPE}), $pK_a = -\log_{10}{K_a}$, with $K_a$ the dissociation constant of carboxylic acid groups. Ionization is considered as binding of hydroxyl ion ligands, hence the $pH$ dependent term in Eq.~(\ref{eq:fractionHPE}) \cite{martin2023cooperative}. The correlation between the ionized fraction and fraction of HPE in the aqueous state is strongest when $pH - pK_{a} >> 0$.  The ionization state of the chains was determined from a potentiometric titration of the aqueous phase while the fraction of chains in the oil phase was followed via UV-Vis absorption of the UV-active end group (see SI Section~\ref{PAHA_synth} for details). 

The number of ionizable groups per oligomer serves as an upper bound to the value of $M$. A value of $M$ lower than the number of ionizable groups can be due to Coulomb interactions between ionized groups, partially ionized intermediate HPE states between the aqueous and hydrophobic states \cite{martin2023cooperative}, and composition dispersity, to be addressed shortly. 

Figure~\ref{fig:HPE_fig}(b) presents in red the fraction of the polymer in the aqueous phase ($1-f_H=f_{aq}$) of the two-phase system. The curve was derived from the UV-Vis absorbance ($308$ nm) of the RAFT agent in each pentanol solution compared to the stock solution. Figure~\ref{fig:PAHA-20_fraction_treatment} and SI Section~\ref{PAHA_f_h_data} show the UV-Vis absorption curves and explain the data treatment employed to extract the data. The main feature, we may initially point out, is the sharp transition occurring over around $0.5$ $pH$ units leading to a complete transfer from one phase to the other. As a comparison to a non-cooperative transition, a curve (blue dotted) with a $M=1$ value has been plotted with the same $pK_a$ and hydrophobic penalty, $\beta g_{H}$, as was found for the polymer. 

Fitting the curve for the HPE partitioning with Eq.~(\ref{eq:fractionHPE}), leads to a value of $M$ of $7$. Although clearly higher than for a non-cooperative transition this value is far from the expected $18$ derived from the length of the polymer. As discussed further in the SI (Section \ref{theory_SI}), this effect is most likely due to the presence of intermediate conformational states of the polymer which broaden the transition significantly. For this system in particular, we might expect that if the polymer is not fully ionized when it transitions to the aqueous phase, the Coulombic repulsions between the different charged groups will change as a function of the ionization fraction, leading to changes in the conformations of the polymer. Our experimental procedure does not allow for the elucidation of the particular conformational state of the polymer in either of the phases, however, the titration data derived from this experiment, in parallel to the hydrophobic fraction data, gives us more insight into the nature of this transition.

The light red squares in Fig.~\ref{fig:HPE_fig}(b) shows the curve of the fraction of ionized states, $\theta$, derived from potentiometric titration data and a repeat experiment for the same PAHA polymer. See Fig.~\ref{fig:PAHA-20_ionization_treatment} and SI Section~\ref{PAHA_theta_data} for the titration data analysis. The transition is markedly sharper than would be expected for a simple monoprotic acid. This is most apparent during the initial onset of the transition from low to high $pH$, as there is a clear tail to the curve in the latter half of the transition. 

An important result derived from the model in the theoretical description of this system is that cooperative transitions for HPE present correlated conformational state and ionization transitions. It is evident from the curves that for our two-phase system this is indeed the case. As the $pH$ increases we see close correlation between the fraction of the polymer in the hydrophobic state and the ionization fraction of the polymer. Considering we do not expect significant ionization of the polymer in the pentanol phase, due to its much lower dielectric constant ($\sim$15), a joint onset in both transitions is consistent with expectations. There are, however, clear deviations between the ionization and the hydrophobic fraction of the polymer at the higher $pH$ side of the transition. A certain degree of difference between the curves is to be expected, as is shown in the graph through the predicted $\theta$ curve (light red dashed line) where we have inputted the fit parameters for Eq.~(\ref{eq:fractionHPE}) of the $f_{aq}$ data into Eq.~(\ref{eq:theta_HPE}). The origin of this effect is that the $pK_a$ of the acid groups ($4.69$ is estimated in this case) is fairly close to the transition-$pH$ of the polymer and therefore the $pH$ is too low to lead to full ionization when the polymer transitions. We hypothesize that the deviation between the number of ionizable groups and the measured degree of cooperativity $M$ is mainly due to the presence of intermediate conformational states in the water phase that occur when we have intermediate ionization states. Similar deviations between the number of ionizable groups and $M$ in transitions coupled to HPE ionization have been observed in micelle formation by diblocks containing a HPE block \cite{Li2016, martin2023cooperative}, and in the solubilization of bilayer vesicles \cite{thomas1995tuningTirrel, martin2023cooperative}. We therefore conclude that our simple oil-water partitioning setup reflects the behavior of (much) more complex systems where the HPE conformational (or ionization) state couples to structure and function.

As a comparison to our MWC approach we also present a Hill plot for the partitioning transition (Fig.~\ref{fig:HPE_fig}(c)). We can transform the $f_{aq}$ and $\theta$ data into Hill plot form, which gives rise to clear and aligned straightline sections. It is clear that these transitions present cooperativity, due to the large straightline section and larger than unity gradient (Hill parameter, $n_H$). We have overlayed an $n_H=7.2$ gradient line to show the consistency between the MWC fit and the Hill treatment. We find that the Hill plot, a commonly used "test" for cooperativity \cite{Ercolani2003, liGao2018cooperativity, Petter1990, Collman1978, Honda}, does not afford as much further information as the MWC-like model we present here. Specifically, the merging of the free energy variables for the hydrophobicity and ionization within the Hill equation, does not reflect the importance of the hydrophobic penalty in HPE transitions as a distinct variable to the ionization of the acid groups. Moreover, it does not allow for rationalizations on the difference between the fraction of polymer in the hydrophobic phase and the ionization state of the polymer, that occur when the transition-$pH$ is close to the $pK_a$ of the acid groups.

\subsection{Degree of cooperativity in binding of iron ligand onto terpyridine monomers and terpyridine-functionalized polymers (OMC)}

To investigate the degree of cooperativity in binding of iron ligand onto terpyridine oligomers we again make use of a two-phase (with distinct hydrophobicities) experimental set-up. Here we compare the partitioning of terpyridine (abbreviated as T, which should not be confused with the Tense state) in a two-phase water/oil (dichloromethane, DCM) system, to that of a terpyridine-functionalized oligomer with 16 units in length and a narrow polydispersity (DP = 16 and \emph{\DJ}$\approx1.04$, abbreviated as PT16). {To synthesize PT16, first poly(\textit{tert}-butyl acrylate) (P\textit{t}BA) was synthesized by the SET-LRP method, then deprotected to poly(acrylic acid) (PAA), and finally functionalized by terpyridine groups. The two-phase set-up was designed with equal volumes of the phases, over a wide range of initial concentration of iron ions in the aqueous phase (10 - 200$\mu$M), and with a fixed concentration of terpyridine groups (200 $\mu$M) in DCM (oil phase).
Figure~\ref{fig:OMCfoandhillplots}(a) shows an illustrative summary of the experiments. For the detailed description of synthesis steps and characterization of PT16 as well as the experimental details of the partitioning in the two-phase system, see Sections \ref{SynthesisPT16} and \ref{OMC_experimentalset-up_SI} of the SI, respectively. 

With a reported binding strength $g_2 \approx$~-23 $k_BT$ between iron (II) ions and two terpyridine binding sites \cite{Hannewald2020}~\footnote{1:1 complexes of terpyridine and iron have not been observed \cite{Hannewald2020, Holyer1965Terpyridine}}, and $K = \exp({- \beta g_2})$, for OMC with ligand concentration $[Fe^{2+}] \gtrsim 10^{-9}M$, the general theoretical equations in the SI Eqs.~(\ref{foP}, \ref{thetaP_supp}) become, to an excellent approximation,

\begin{equation}
	f_{aq} = \frac{(K_p K [\text{Fe}^{2+}])^{M}}{1 + (K_p K[\text{Fe}^{2+}])^{M}}~~~~\text{(OMC)}
	\label{eq:fractionOMC}
\end{equation}
and the total fraction of the occupied binding sites 
\begin{equation}\label{thetaP}
	\theta = \frac{K [\text{Fe}^{2+}]}{1 + K [\text{Fe}^{2+}]} f_{aq} \approx f_{aq}~~~~\text{(OMC)}.
\end{equation}

\noindent In these equations, $K_p = \exp{(-2\beta g_H)}$ with $g_H$ the reversible work to transfer a terpyridine moiety from the oil phase to the aqueous phase, analogous to a conformational penalty in the aqueous phase. Unlike for the HPE case, it is nontrivial to link the precise value of $M$ to the number of terpyridine groups per oligomer, as the aqueous state of the OMC is in fact a cross linked (by iron ions) gel of oligomers, see Fig.~\ref{fig:CooperativePolyTerpy_SI} in the SI. Since iron ions are bound by two terpyridine residues, formation of each terpyridine - iron complex leads to a binding free energy $g_2$ accompanied by a hydrophobic penalty $2 g_H$. We note that Eqs.~(\ref{eq:fractionOMC},\ref{thetaP}) are manifestations of the Hill equation, Eq.~(\ref{eq:Hill}), reflecting a limiting case where oligomers are either fully bound or unbound to iron (II) ions. Eqs.~(\ref{eq:fractionOMC},\ref{thetaP}) depend on the combined quantity $K_p K = \exp{(-\beta(2g_H + g_2))}$. We fit our experimental data to Eqs.~(\ref{eq:fractionOMC},\ref{thetaP}) for both terpyridine monomers as well as for oligomers. The value of $g_H$ for terpyridine monomers has been determined by the solubility in water and in DCM, see SI Section \ref{TerpygH_SI}.

The binding behavior of iron ligands onto terpyridine groups was investigated by quantification of free iron and terpyridine concentrations in the water and oil phases using UV-Vis and ICP-AES, to obtain the free ligand (iron) concentration, fraction of occupied binding sites, and fraction of terpyridine groups transitioned from DCM to the water phase. The details of the quantification procedure are described in Sections \ref{quant_theta_SI} and \ref{quant_fH_SI} of the SI, with the associated data shown in Fig.~\ref{fig:RandomTerpymonomer_SI} and ~\ref{fig:CooperativePolyTerpy_SI} for terpyridine monomers and the terpyridine-functionalized oligomer, respectively.

The fraction in the aqueous phase and fraction of bound (to iron (II) ions) terpyridine monomers and oligomers are shown in the form of $f_{aq}$ and $\theta$ as a function of free iron concentration in Fig.~\ref{fig:OMCfoandhillplots}(b). A traditional Hill plot is shown in Fig.~\ref{fig:OMCfoandhillplots}(c) as a comparison. The experimental data were fitted to Eqs.~(\ref{eq:fractionOMC},\ref{thetaP}), shown by solid lines in Fig.~\ref{fig:OMCfoandhillplots}(b). Compared to the monomers, the terpyridine-functionalized oligomers show 1) a significantly sharper transition (over only a narrow range of free iron concentrations) and 2) at higher free ligand (iron) concentration where the transition occurs, see the squares versus circles, respectively, in Fig.~\ref{fig:OMCfoandhillplots}(b). The first observation is consistent with the results for the HPE in the previous section, with a significantly larger degree of cooperativity for the oligomers $M \approx 16$ compared to the monomers with $M \approx 1$. The second observation points to a smaller negative value of $2g_H + g_2$ for the oligomers compared to the monomers: Eqs.~(\ref{eq:fractionOMC},\ref{thetaP}) predict equal values for the transition defined by $f_{aq} = \theta = 1/2$ at equal $2g_H + g_2$, independent of the value of $M$. From the fits we obtain $\beta (2g_H + g_2) \approx -12.8$ for the monomers, and $\beta (2g_H + g_2) \approx -10.8$ for the oligomers. From the solubilities of terpyridine monomers in (acidified) water and in DCM, see SI Section \ref{TerpygH_SI}, we find $\beta g_H \approx 6.8$, which leads to $\beta g_2 \approx -26.4$, for the monomers, being comparable to the value of $\beta g_2 =-23$ reported previously \cite{Hannewald2020} for monomeric terpyridine in (neutral) water. The slightly lower negative value of $\beta (2g_H + g_2) \approx -10.8$ for the oligomers could be caused by an additional hydrophobic contribution to $g_H$ by the oligomeric backbone, or by a decreased (less negative) iron-terpyridine binding strength $g_2$ due to Coulomb repulsion of neighboring bound iron (II) ions, or both. We conclude that the value of $2g_H +g_2$ extracted by comparing the experimental data to Eqs.~(\ref{eq:fractionOMC},\ref{thetaP}) is consistent with independently obtained values of $g_H$ (by solubility measurements) and of $g_2$ (in \cite{Hannewald2020}).

As pointed out in Section \ref{T_monomervsdimer} in the SI, for the binding of two terpyridine monomers onto iron (II) ions, it is expected that $M \approx 1/2$, see Eq.~(\ref{foMono}) in the SI. However, the experimental data are not consistent with that scenario, see Fig.~\ref{fig:T_monomervsdimer} in Section \ref{T_monomervsdimer} of the SI. Eqs.~(\ref{eq:fractionOMC}, \ref{thetaP}) for $M=1$ is analogous to iron (II) ion binding units in the form of terpyridine dimers, which is consistent with the experimental data as well as the value of $2g_H + g_2$ extracted from independent sources. This observation implies that terpyridine forms dimers in DCM, even in the absence of iron ions. If this behavior is general and also occurs in other types of oils in unclear at this point. 

For the oligomers, as pointed out in the introduction of this section, the value of $M$ cannot trivially be linked to the number of terpyridine residues per oligomer. There, crosslinked, gel-like networks form in the aqueous phase where bonds between two terpyridine sites onto different oligomers and iron ions form. Similar to the situation for the terpyridine monomers, we expect small clusters of oligomers due to terpyridine dimerization to form in the oil phase. The value of $M \approx 16$ that we find for PT16 points to clusters of on average two oligomers. Obviously, further network formation mediated by iron (II) ions occurs in the aqueous phase.\\
We observed that the gels in the aqueous phase (see Fig.~\ref{fig:CooperativePolyTerpy_SI}) can easily be ripped apart with tweezers. Upon gentle shaking, the gel fragments rapidly 'heal' to a single blob. In contrast, upon diluting the aqueous phase below the iron (II) concentration where the gels are stable, slow (over the course of weeks) disappearance of the gels is observed. These findings indicate that the preferred conformational state of the oligomers as a function of ligand concentration can translate into functional behavior: the aqueous gel state can sharply be switched between 'self healing' and 'self - destructing', as a function of the ligand concentration.

\begin{figure}[H]
    \centering
    \includegraphics[width=\textwidth]{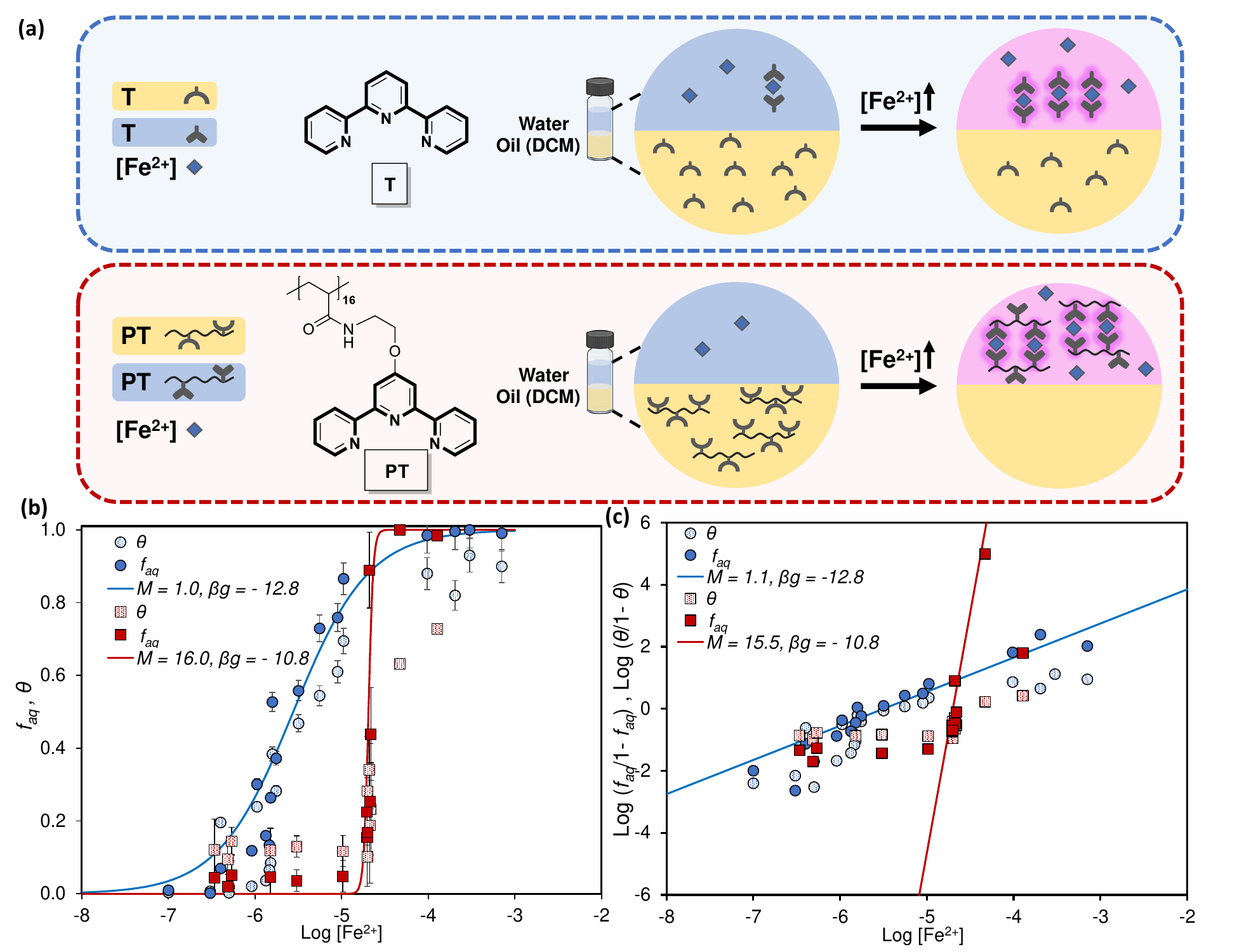}
    \caption{\textbf{Terpyridine oligomers shown much sharper, cooperative transitions than the corresponding monomer (a)} Schematic of the comparison of the partition experiments of monomeric (T) and oligomeric (PT) terpyridine in oil (DCM) and water, with iron (II) ions in the aqueous phase as 'ligands'. \textbf{(b)} Fraction of terpyridine monomers (filled blue circles) and oligomeric terpyridine (filled red squares) in the aqueous state ($f_{aq}$) as a function of the free iron concentration in the water phase. The blue solid line represents the best fit to terpyridine to Eqs.~(\ref{eq:fractionOMC},\ref{foGCdim}) with $\beta g = \beta (2g_H + g_2) = - 12.8$ and $M = 1$. The red line is the same equation with $\beta g = \beta (2g_H + g_2) = - 10.8$ and $M = 16$, which describes well the experimental data on poly(terpyridine). The independently measured fraction of occupied terpyridine sited (by iron (II) ions) in the water phase, $\theta$, are indicated by light blue circles for monomeric terpyridine and light red squares for PT16. \textbf{(c)} The corresponding Hill plots for terpyridine monomer and poly(terpyridine) where the blue and red lines represent Hill coefficients $M$ equal to $1$ and $16$, respectively.}
    \label{fig:OMCfoandhillplots}
\end{figure}
 
\subsection{Partitioning behavior of a compositionally disperse copolymer}

We now present the partitioning results for a copolymer HPE, poly(\textit{n}-butyl acrylate-\textit{r}-acrylic acid) (PBA-AA). As detailed in the SI Section~\ref{PBA-AA_synth}, this polymer is synthesized from a 1:1 ratio \textit{n}-butyl acrylate and \textit{t}-butyl acrylate. These two monomers present very similar reactivity in a (controlled) radical polymerization therefore we can assume that their distribution within a polymer chain is random \cite{reactivity}. The selective removal of the \textit{t}-butyl group leads to a random distribution in the ratio of hydrophobic (\textit{n}-butyl acrylate) and ionizable (acrylic acid) groups per chain throughout the HPE sample sample. This will lead to an ensemble of polymers with spread values of ionizable groups ($M$), which will vary the total ionization energy per chain and, likewise, a spread of hydrophobic penalty values ($g_{H}$).

In the SI (Section \ref{composition_dispersity_theory_supp}) we derive the analogues for the fraction of HPE in the aqueous state ($f_{aq}$, Eq.~(\ref{eq:fractionHPE})) as well as the fraction of ionized carboxyl groups $\theta$, Eq.~(\ref{eq:theta_HPE}), for HPE with a binomial distribution of ionizable groups and hydrophobic groups. The analysis includes composition dispersity as well as length dispersity, in the form of a lognormal distribution. In short, we consider the transitions for each set of values of ionizable groups and total length, and then apply the weights from the corresponding distributions to them. The inclusion of composition dispersity is an important test-case for the theory that we developed. Moreover, controlled composition dispersity in combination with fractionation opens up a new methodology to tune the $pH$ at which the hydrophobic - aqueous transition occurs as well at the width of the transition. We will follow up on this in the next section.

\begin{figure}[H]
	\centering
	\includegraphics[width=\textwidth]{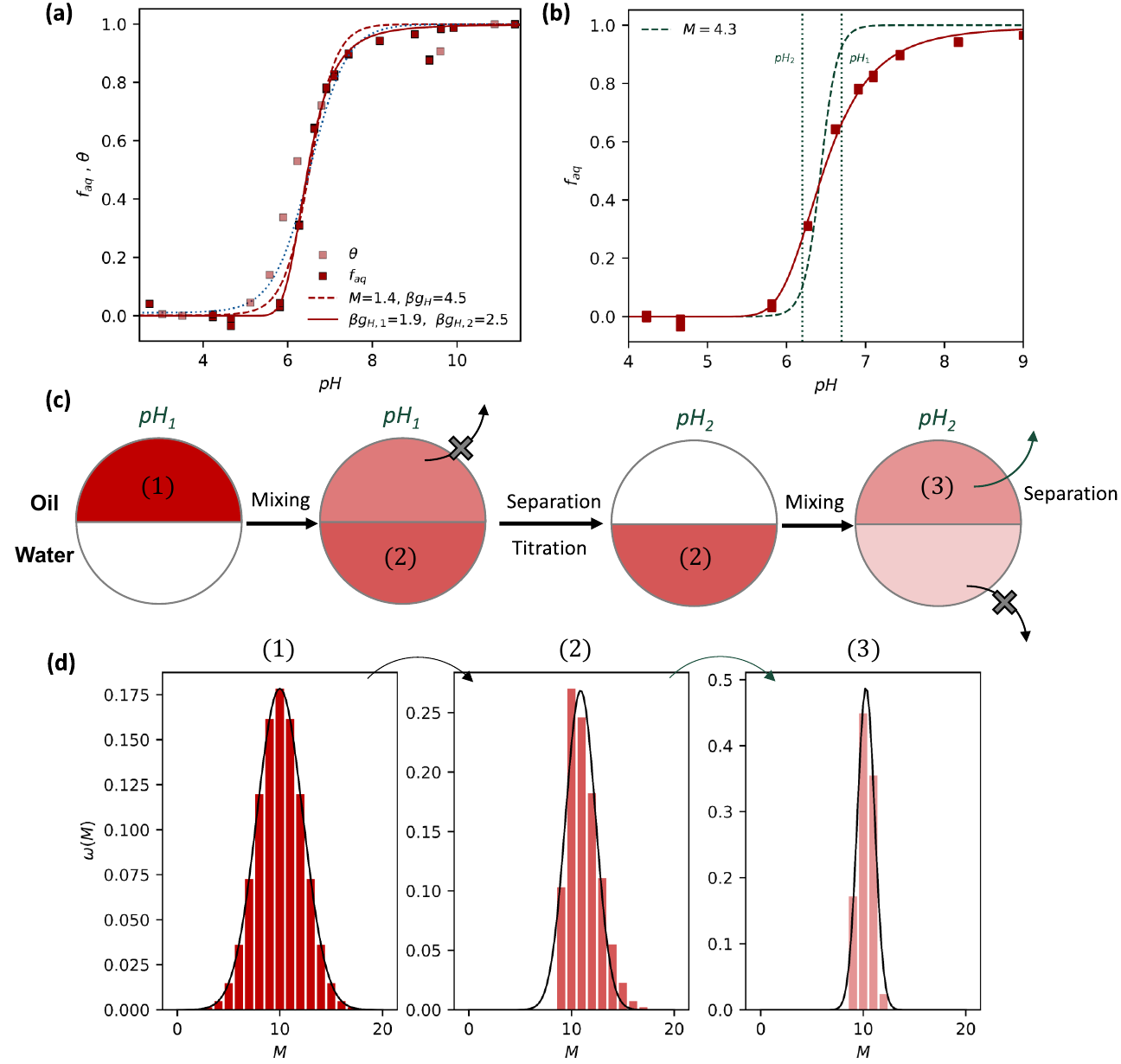}
	\caption{\textbf{Chemically disperse copolymeric HPE presents broad, non-cooperative partitioning transitions. Fractionation can return sharp transitions. (a)} Plot of $f_{aq}$ (red squares) and $\theta$ (light red squares) for PBA-AA. The data for each variable was independently collected from separate experimental systems. Repeat measurements of $f_{aq}$ are individually plotted to illustrate the experimental spread of the data. The curves are a fit using Eq.~(\ref{eq:fractionHPE}) (dashed red) and SI Eq.~(\ref{eq:M_dispersity_fraction}) (solid red), using the $pK_a$ values for acetic acid: $4.56$ \cite{harris_quantitative_2007}. Fit parameters are shown in the legend. As a comparison a (calculated) non-cooperative transition with $M=1$ is plotted as a dotted blue line. \textbf{(b)} $f_{aq}$ for the PBA-AA system and its corresponding fit (SI Eq.~(\ref{eq:M_dispersity_fraction})) are shown as red squares and a red line. Note the asymmetric shape of the curve around the transition point ($f_{aq}=0.5$). The green dashed line shows a calculated $f_{aq}$ transition from the resulting HPE sample after the fractionation procedure (shown in (c)).}
	\label{fig:HPE_fig_2}
\end{figure}

\begin{figure}
    \captionsetup{labelformat=adja-page}
    \ContinuedFloat
    \caption{\textbf{(c)} Schematic for a potential fractionation procedure using a two-phase oil and water system. After equilibration at each fractionation $pH$ point one of the phases of discarded (marked as a gray cross beside an arrow). Shades of red within a phase reflect polymer concentration. $pH_1>pH_2$ \textbf{(d)} Distribution of the number of ionizable sites ($M$) on the polymer chains at different stages of the fractionation procedure. Only the distribution for the chains with 20 total monomers ($M_t$) is shown. Gaussian curves are plotted to guide the eye. See SI Section~\ref{data_analysis_HPE} and \ref{sec:fract_appen} for details on data treatment and scaling, and fractionation calculations, respectively.}
\end{figure}

The data was acquired through a buffered partitioning experiment ($pH$ $3-11$) using the coumarin-modified poly(\textit{n}-butyl acrylate-\textit{r}-acrylic acid) copolymer PBA-AA$^c$, of around $20$ units in length (DP$=22$, \emph{\DJ}$=1.1$), in a two-phase pentanol-water system as described in the SI (Section \ref{buffered_exp}). The coumarin dye is added to the chain-end of these polymers to allow for UV-Vis spectrometry to be used to track their concentration, as was done earlier for the PAHA system. The $f_{aq}$ data was derived from the UV-Vis absorbance ($320$ nm) of the coumarin dye in each pentanol solution, for two individual runs, compared to the stock solution ($0.5$ mg/ml). }Figure~\ref{fig:HPE_fig_2}(a) presents the fraction of the polymer in the aqueous phase, $f_{aq}$, of the two-phase system. Figure \ref{fig:appendix-pba-aa} and SI Section \ref{PBAA_f_h_data} shows the UV-Vis absorption curves and explains the data treatment employed to extract the curve. In stark contrast to the results in Fig.~\ref{fig:HPE_fig} for the PAHA HPE, the transition is spread out over a wide $pH$ range, around $4$ units. We may refer to this, from a purely phenomenological perspective, as non-cooperative behavior. However, we expect each individual polymer chain to behave cooperatively but the chemical dispersity between the different chains leads to a spread in transition-$pH$ values across the polymer sample and therefore leads to an overall broad transition.

Using Eq.~(\ref{eq:fractionHPE}) we can fit the transition with our non-disperse model as a comparison and extract a value for the effective value of $M$, which yields $1.40$. Therefore, at least phenomenologically the ensemble of chains behaves fairly non-cooperatively. This symmetric fitted curve (around $f_H=0.5$) does not match the asymmetric experimental transition well. The transition flattens out at the higher $f_{aq}$, confirming the expected asymmetry found in the numerical predictions in the SI (Section \ref{composition_dispersity_theory_supp}).

Using SI Eq.~(\ref{eq:M_dispersity_fraction}) we can fit the transition taking into account the known chemically disperse nature of the polymer. We assume a binomially distributed pair of monomers, use the expected value of the dispersity ($1.04$)\footnote{See SI Section \ref{polymer_synth_supp} for a discussion on the dispersity value used.} and calculated total length of $20$. We then fit the distribution with two free parameters, namely the hydrophobic penalty for each of the monomers. With values of $g_{H,1}=1.94$, $g_{H,2}=2.48$ for the acrylic acid ($g_{H,1}$) and the  \textit{n}-butyl acrylate ($g_{H,2}$) groups, a good match between the expected transition predicted by the numerical model and the experimental system.  We would indeed expect the \textit{n}-butyl acrylate groups to have a larger hydrophobic penalty than the acrylic acid groups. 

We, again, extract the fraction of ionized groups on the polymer from a potentiometric titration experiment of the two-phase system and the similar polymer PBA-AA$^0$ (DP$=21$, \emph{\DJ}$=1.1$) (see SI Section \ref{Experimental_HPE_supp} for synthesis details). This polymer sample was not modified with coumarin. The resulting data (see SI Section \ref{PAHA_theta_data} and Fig.~\ref{fig:appendix-pba-aa}) is plotted in light red in Fig.~\ref{fig:HPE_fig_2}(a). Compared to what was seen for the homopolymer HPE, the degree of correlation between these two transitions is not as high. A slight offset between them is predicted and explained in the theoretical section, but the experimental results do not match this behavior. It is possible that the addition of the coumarin dye for the partitioning data ($f_{aq}$) increases the hydrophobicity of the polymer, shifting the transition to a higher $pH$ value. Moreover, the ionization fraction and hydrophobic fraction data were acquired from independent unbuffered and buffered experiments, respectively. It is possible there is an offset in the $pH$ calibration between the experiments. There is also a lack of data points in the titration data at higher $pH$ values which hinders any detailed discussion about the difference between the curves.

Overall, there is good agreement between the predictions for a binomially distributed binary copolymer and the experimental data presented here. The broadening effect is substantial and therefore the monomer composition dispersity of copolymers is an important design aspect to be considered when designing polymers to present a particular $pH$ response.

Due to the direct relationship between the comonomer dispersity in the polymer and the theoretical framework, we can predict the transitions for a variety of different copolymers with different monomer distributions. This finding is potentially relevant to applications involving (random) copolymers of Styrene and Maleic acid (SMA), broadly used in isolating membrane proteins by bilayer disk formation \cite{Esmaili2020, Dorr2016, BadaJuarez2019, Xue2018}. There, a broad transition may be beneficial, with applications over a relatively wide (several units) pH range. Moreover we may be able to leverage the partitioning of the different chains to create polymer samples which transition at a particular $pH$ value and sharpness.    

\subsubsection{Fractionation of compositionally disperse copolymers}
As detailed above, polymers presenting chemical dispersity between the chains in a single polymer sample present broader transitions than homopolymers. However, the transitions of each individual chain are still expected to be as sharp as a similar homopolymer. If we were able to selectively fractionate only a specific subset of the chains in a composition disperse polymer sample, this subset may present sharp transitions.

Such an approach may also provide a solution for the difficulty of targeting a specific transition-$pH$ using homopolymers. If restricted to only commercially available monomers, it can be challenging to synthesize a HPE with specific hydrophobic penalty value. The fractionation approach would start with a polymer sample with a broad comonomer distribution, leading to a broad transition, and from it extract a fraction of chains with the desired transition-$pH$. This would be much less atom efficient, but may give access to both tailored transition-$pH$ values and transition broadness that may be harder to access from direct polymer synthesis. In this section we will focus on the fractionation of a HPE with a disperse ratio of ionizable groups ($M$) to total monomers ($M_t$). An example of this class of polymer is the poly(\textit{n}-butyl acrylate-\textit{r}-acrylic acid) (PBA-AA) HPE examined above, whose broad oil-water partitioning transition is the basis of the following discussion.

A schematic illustrating a simplified experimental procedure to potentially achieve this is shown in Fig.~\ref{fig:HPE_fig_2}(c). We postulate the use of a two-phase oil and water system to carry out this fractionation. By carrying out two separate partitioning steps of the polymer sample, at different $pH$ values, we can isolate a fraction of the polymer that transitions between the two chosen $pH$ values. Fig.~\ref{fig:HPE_fig_2}(d) graphically describes the changes to the width of the chemical dispersity distribution during the fractionation procedure. After the first separation step one of the edges of the distribution is removed, in this case the most hydrophobic chains remain in the oil phase which is then discarded. The second step at a lower $pH$ value removes the most hydrophilic chains from the other edge of the distribution. The final sample remains in the oil phase of the second step and presents a much narrower chemical dispersity. 

The specific transition that the fractionated sample will now exhibit depends on the specific shape of the length and comonomer ratio distribution of the initial sample. If the average value of $M$ for the fractionated chains is large ($M>10$), then the bulk of the transition for the fractionated polymer will occur within the two fractionation $pH$ values chosen.

Assuming the polymerization of a particular polymer sample is well understood and its length dispersity is measured, a good estimation of the relative weights for the different chain configurations (number of repeating units of each type) can be made. A measurement of the transition of the whole polymer sample between the phases of the oil-water system (shown in Fig.~\ref{fig:HPE_fig_2}(a)), allows us to fit the hydrophobic penalty parameter for the different ($M$,$M_t$) configurations. This then allows for a prediction of exactly what chains transition at which $pH$.

As shown in Fig.~\ref{fig:HPE_fig_2}(b) it is possible to extract a copolymer sample from an initial disperse copolymer sample that transitions over $0.5$ $pH$ units. In the case of the fractionation procedure leading to the blue line in this figure, the predicted ratio of chains that remain from the initial sample is $39\%$. Details on the numerical procedure to predict the transition curve of the fractionated polymer are presented in the SI Section~\ref{sec:fract_appen}.

\section{Conclusion}

In the present work we pinned down the conditions for cooperative ligand-mediated transitions (LMT) and their underlying mechanism. We showed that in two very different and relatively simple macromolecules, coupling between conformational states and ligand binding leads to strongly cooperative transitions in oligomers with $16$-$20$ ligand binding sites. As a comparison, no cooperativity has been observed in monomeric analogues under the same conditions. The key is to select two well-defined conformational states: a ground state with low ligand affinity, and a conformationally unfavorable state, yet with relatively high ligand affinity. These two states have been realized by demixed oil-water systems, where in both studies, the state with oligomers dissolved in oil represents the ground state. Because of the hydrophobic nature of the oligomers, dissolution in water constitutes an unfavorable state quantified by a conformational penalty. However, the aqueous state can be stabilized by ligand binding. If the binding of several ligands is required to overcome the unfavorable interactions with water, the transition is cooperative and presents a sharp response to the ligand concentration in the system. 

We have proved this scenario for hydrophobic polyelectrolytes (HPE) and oligomeric metal chelators (OMC). Several earlier observations of cooperative transitions exist where HPE are involved. In these studies, the HPE is either one of the blocks in diblock copolymers \cite{Li2016, liGao2018cooperativity}, is part of a crosslinked network \cite{Siegel1993, Philippova1997}, or the transition is coupled to lipid bilayer solubilization \cite{thomasT1994membraneTirrel, Scheidelaar2016}. In these systems, the conformational states of the HPE are well-defined, and couple to structure and function: the formation or dissolution of pH-dependent micelles in HPE containing diblocks \cite{liGao2018cooperativity}, pH-dependent solubilization of lipid bilayer membranes \cite{thomasT1994membraneTirrel}, and swollen / collapsed states of crosslinked HPE \cite{Siegel1993, Philippova1997}. In our work we reduced complexity to the essentials: single-chain HPE with conformational states defined by a hydrophobic and an aqueous reservoir. In that way we have been able to prove directly the coupling between conformational states and ligand binding, an essential feature predicted by MWC theory \cite{Monod1965, martin2023cooperative}, see Fig.~\ref{fig:Intro}. Moreover, we showed that monomer compositional dispersity dramatically influences the width of the conformational transition for binary copolymers, which is potentially relevant in their applications \cite{Scheidelaar2016}. Fractionation of compositional disperse HPE in principle provides a way to select desired $pH$ ranges where the transition occurs 'to order'.

As far as we are aware, we provide the first observation of a cooperative transition in OMC. The essential difference with HPE is that here, iron (II) ions form bonds with two terpyridine residues on the oligomers. We find a strongly cooperative transition as a function of free iron (II) ion concentration where the OMC moves from a dispersed state in oil to a gel in the aqueous phase. Even in this simple setup, coupling of the conformational state to structure and function is apparent: gels form and self-heal (after being ripped apart) rapidly beyond a critical iron (II) ion concentration, and 'self-destroy' below that concentration. The gels will only very slowly disappear upon diluting the water phase below the critical iron ion concentration. These findings open up potential for 'self healing' and slow release systems triggered by very small variations in ligand concentration.

The general framework of ligand- mediated transitions can in principle be applied to a wide range of ligand-polymer systems, as long as the systems present a constrained number of conformations for the polymer chains. Through the use of a simple two - phase oil and water system we have shown that MWC-like cooperative transitions can occur in relatively simple 'non-biological' macromolecules with a variety of ligands.

\section{Supporting Information Available}
General theoretical background (SI-1); Hydrophobic polyelectrolytes (HPE, SI-2); Oligomeric metal chelator (OMC, SI-3)

\section{Authors Contributions}
J.L.M.R and N.M contributed equally to this work.
J.L.M.R and N.M: Conceptualization, Methodology, Investigation, Visualization, Writing - original draft. N.D, L.v.d.H, A.M.v.S, K.S.L, M.v.S, and W.C: Methodology, Investigation. B.G.P.v.R and W.K.K: Conceptualization, Methodology, Supervision, Writing - review, and editing. All authors read and approved the final manuscript.

\section{Funding}
We acknowledge the Dutch Research Council (NWO) for funding (grant no 712.018.003).

\section{Note}
Authors declare no competing interests regarding this research.

\section{Acknowledgments}
We thank Dominique Thies-Weesie for technical assistance during the initiation of this project, and Jan Groenewold for many discussions. We thank Isabelle Meijer for her preliminary work on hydrophobic polyelectrolytes.

\bibliography{references}

\end{document}


\pagebreak
\vfill

\tableofcontents

\pagebreak
\vfill
\section{General theoretical background}
\label{theory_SI}
Here we summarize the model used in the main paper and in the next section we generalize it to include multivalent ligands and composition dispersity.

We use a grand canonical ensemble treatment to describe our systems, that is, we take the system as open to the exchange of ligands. The grand canonical statistical weights of a polymeric template with $M$ binding sites in a hydrophobic (or 'oily') (H) and an aqueous (aq) environment, respectively, are:

\begin{align}
	\Xi_{H}=& \left(1 + \lambda \exp{(-\beta g')} \right)^M \approx 1.  \label{eq:gen_state_oil}
    \\
    \Xi_{aq}=& \exp{(-\beta G)} \left(1 + \lambda \exp{(-\beta g)} \right)^M.  \label{eq:gen_state_aq}
\end{align}

Where the $\Xi_i$ are the coarse-grand partition functions of the $i = H, aq$ state of a template, $\lambda$ is the fugacity of the ligand, $\lambda = \exp{(\beta \mu)}$, with $\beta = 1 /k_B T$. $k_B$ is Boltzmann's constant and $T$ the absolute temperature. $\mu$ is the chemical potential of the ligand that adsorbs (or binds) onto the macromolecular template. $\mu$ is related to the ligand concentration or partial pressure of the ligand. $g$ is the binding free energy of the ligand to a binding site on a chain within the aqueous phase, and $g'$ the binding free energy of a ligand onto the template in the hydrophobic state. $G$ represents the reversible work required to transfer a chain from its hydrophobic state in the oil phase to the aqueous phase, and takes into account the (unfavorable) interactions between hydrophobic moieties on the macromolecules and water. This term is analogous to the self-energy term of the hemoglobin molecule when changing conformations upon adsorption of oxygen ligand. We have set $\Xi_H \approx 1$ as binding of the ligands to the templates in the hydrophobic state is unfavorable and therefore $\lambda \exp{(-\beta g')} << 1$. In the case of hemoglobin, with $M = 4$, the analogues of the hydrophobic and aqueous states are the T and the R states, respectively, and both states have finite affinity for the oxygen ligands.

The expression for the grand partition functions can be derived in the usual way by writing $\Xi = \sum_{N=0}^{M}Z(N,M,T)~\lambda^N$. Here, subscripts have been dropped. The (coarse-grained) canonical partition function is given by $Z(N,M,T)=\exp{(-\beta G)}~\Omega(N,M)~\exp{(-\beta N g)}$ with $G=0$ for a template in the hydrophobic state. Under the assumption that binding sites are uncorrelated, the number of ways $N$ ligands can be distributed over $M$ binding sites is $\Omega = \binom{M}{N}$, which has been used to derive Eqs.~(\ref{eq:gen_state_oil}, \ref{eq:gen_state_aq}) by making use of the binomial theorem $\sum_{N=0}^{M} \binom{M}{N}x^N = (1 + x)^M$. For the terpyridine oligomers with iron ions as ligands, the situation is more complicated as will be discussed in the section on terpyridine-functionalized templates below.

From Eqs.~(\ref{eq:gen_state_oil}, \ref{eq:gen_state_aq}) the fraction of occupied binding sites (by the ligand), $\theta$, and the fraction of polymer chains in each of the phases can be derived. For the polymer fractions (when considering equal volumes), we have $f_i={\Xi_i} \slash {\Xi}$, where the subscript $i$ can be hydrophobic (H) or aqueous (aq), and $\Xi$ is the total partition function , $\Xi = \Xi_{aq}+\Xi_{H}$. For our two-phase oil and water system we have

\begin{align}
	f_{H}=&\frac{1}{1+\exp{(-\beta G)}(1 + \lambda \exp{(-\beta g)} )^{M}}  \label{eq:fH}
    \\
    f_{aq}=&1-f_{H}.  \label{eq:faq}
\end{align}

\begin{figure}
    \centering
    \includegraphics[width=1\linewidth]{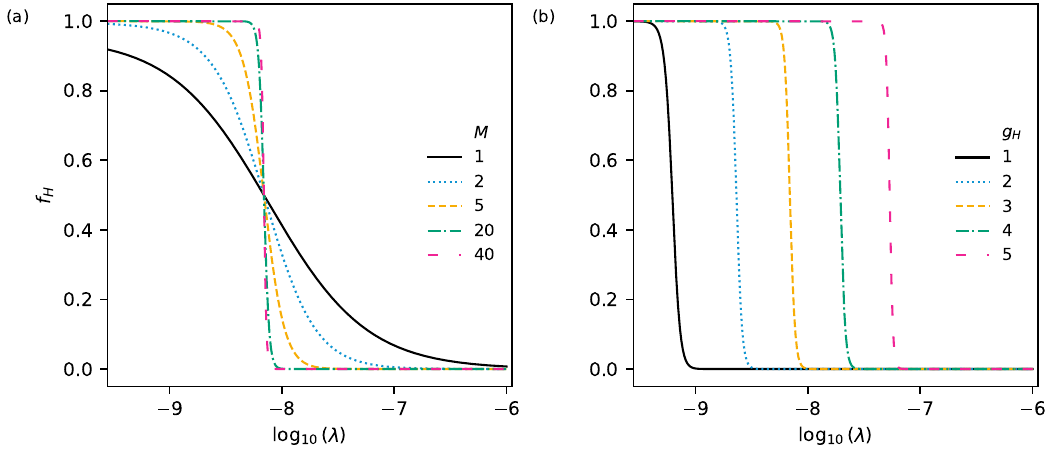}
    \caption{\textbf{General LMT two-state transition behavior. (a)} Increasing the value of $M$ in Eq.~(\ref{eq:fH}) leads to an increase in sharpness of the transition. $M$ can be considered a cooperativity parameter. \textbf{(b)} The parameter $g_H$, defined as $\beta G=Mg_H$ (Eq.~(\ref{eq:fH})) , affects the value of the fugacity at which the transition occurs.}  
    \label{fig:function_behaviour}
\end{figure}

\noindent Figure \ref{fig:function_behaviour}(a) shows the effect of the value of $M$ on the sharpness of the transition, and (b) illustrates the effect of the hydrophobic penalty $G$, for a fixed binding energy $g$. Theoretical framework therefore predicts that oligomeric species are long enough to present sharp, cooperative transitions. As the hydrophobic penalty increases the transition is shifted towards higher fugacity values, as more favorable binding in the aqueous must counteract an increase in the stability of the template in the hydrophobic phase.

\noindent The fraction of occupied binding sites is given by,

\begin{equation}
     \theta=\frac{\langle N \rangle}{M} = \frac{1}{M} \frac{\lambda}{\Xi}\frac{\partial \Xi}{\partial \lambda} = \frac{\lambda\exp(-\beta g)}{1+\lambda\exp(-\beta g)}f_{aq}, \label{eq:theta_general}
\end{equation}

Note the correlation between the fraction of occupied binding sites $\theta$ and the fraction of chains in the aqueous phase, $f_{aq}$ being analogous to the situation for hemoglobin in Fig.~\ref{fig:Intro}(a) where $\theta$ is strongly correlated with $f_R$. This general model can then be applied to a particular ligand-polymer system by finding a suitable expression for the fugacity ($\lambda$). 

In the situation that we have only a single conformational state, so that one of the partition functions in Eqs.~(\ref{eq:gen_state_oil}, \ref{eq:gen_state_aq}) equals zero, we immediately recover the Langmuir adsorption equation. For example if we take $\Xi_H = 0$, we find by using the left-hand side of Eq~(\ref{eq:theta_general})

\begin{equation}
	\theta_L= \frac{\lambda\exp(-\beta g)}{1+\lambda\exp(-\beta g)}, \label{eq:theta_Langmuir_supp}
\end{equation}
\noindent which has been plotted in Fig.~\ref{fig:Intro}.

It is important to note that the equations above represent an idealized case of a cooperative binding-conformation transition for a template with multiple ($M$) binding sites for ligands. The model assumes that the template or receptor only has two stable conformational states throughout the transition. While this is plausible for a highly constrained biological molecule such as hemoglobin, it may be unrealistic for relatively simple macromolecules. The presence of intermediate states between the two dominant conformational states will lead to a broadening of the curve for the fraction of occupied sites and the transition overall. Due to the predicted correlation between the value of $M$ and the sharpness of the transition this can be considered as a reduction in the effective value of $M$. Therefore, $M$, although directly correlated to the number of binding sites, should be seen as a cooperativity parameter. The closer the $M$ value derived from an experimental transition is to the physical number of binding sites on a template, the closer the system abides to a purely two-state transition.

Additionally, in this analysis, we have assumed that the binding energy for every binding site is equal and that there are no interactions between the binding sites. Considering that the binding of charged ligands will lead to the ionization of the site and electrostatic repulsion with neighboring ionized binding sites, this cannot be fully neglected. It is not within the scope of this work to detail the effects of intra-template interactions, but we note that these interactions will lead to a spread in the binding site energy and therefore a broadening of the transition, in fact negative cooperativity. This effect is reflected in a smaller value of $M$ as expected from the number of ligand binding sites. 

There may also be interactions that lead to a larger value of $M$ than expected from the architecture of the polymer template, as will be seen in the situation for the terpyridine OMC system. There, a large value of $M$ can a priori be attributed to the high configurational entropy of the ligands and the templates as reflected in the value of the multiplicity $\Omega(N,M)$ as discussed above. Overall the value of $M$ gives us information about the behavior of the system compared to an ideal two-state, uncorrelated reference situation. 

In ending this section, we point out the condition where the so-called Hill equation \cite{Hill1910_supp, kerr2024beyondhill_supp,gesztelyi2012hill_supp} is recovered. 

Taking $\lambda \exp{(-\beta g)} >> 1$ and writing $G = M g_H$, where $g_H$ should not be confused with $g'$(binding free energy of a ligand onto the template in the hydrophobic state) in Eq.~(\ref{eq:gen_state_oil}), we obtain

\begin{equation}
	\theta= \frac{(\lambda\exp(-\beta (g + g_H)))^M}{1+(\lambda\exp(-\beta (g+g_H)))^M}= \frac{(K_a[L])^{n_H}}{1+(K_a[L])^{n_H}} ~~ (= f_{aq}). \label{eq:theta_Hill}
\end{equation}

In these conditions the variable $M$ is the 'Hill exponent' ($n_H$). In general, $\lambda \exp{(-\beta (g + g_H))} = [L]K_a$ with $[L]$ the unbound ligand concentration and $K_a$ the (effective) association constant. This is commonly used to extract the value of the Hill exponent (in this case $M$) by plotting $\log{\frac{\theta}{1 - \theta}}$ versus $\log{[L]}$. 

Compared to a strictly empirical value such as the Hill constant, we note that the value of $M$ is directly linked to the number of ionizable groups of the template and therefore the theory allows for the effects of polymer length and length (size) dispersity, as well as chemical dispersity to be predicted. The number of ionizable sites serves as an upper limit to the sharpness of a transition, and the deviation from this upper limit of $M$ also holds information about the system.

\subsection{Hydrophobic polyelectrolytes (HPE)}\label{HPE_theory_supp}

In this section we extend the general theoretical description to HPEs. As shown in a previous publication \cite{martin2023cooperative_supp}, for HPE with $M$ weakly acidic ionizable groups, we have $\lambda \exp{(-\beta g)} = \text{10}^{pH - pKa}$ so that Eq.~(\ref{eq:gen_state_aq}) can be written as

\begin{equation}
	\Xi_{aq} = \exp{(-\beta G)} \left(1 + 10^{pH - pK_{a}} \right)^M.  \label{GPF_X}
\end{equation}

Where the $pK_a$ is the negative logarithm of the acid dissociation constant of the ionizable groups on the HPE and the $pH$ is the negative logarithm of the proton concentration in the aqueous phase of the system. Here we have assumed all structural ionizable sites ($M$), that is the physical number of ionizable sites on the chain, have the same ionization constant. Similarly we may express $f_{H}$ and $f_{aq}$ as

\begin{align}
	f_{H}=&\frac{1}{1+(\exp{(-\beta G)} \left(1 + 10^{pH - pK_{a}} \right)^M)}  \label{eq:HPE_state_oil}
    \\
    f_{aq}=&1-f_{H}.  \label{eq:HPE_state_aq}
\end{align}

The fraction of occupied binding sites is then:

\begin{equation}
     \theta=\frac{10^{pH - pK_{a}}}{1+ 10^{pH - pK_{a}}}f_{aq}. \label{eq:theta_HPE_supp}
\end{equation}

The correlation between the ionized fraction and fraction of HPE in the aqueous state is strongest when the exponent $pH - pK_{a}$ is large compared to unity. This corresponds to a situation where $pH_{trans}>pK_a$ by around a couple of $pH$ units. The transition-$pH$, $pH_{trans}$, is defined here as the midpoint of the transition, where $\Xi_H=\Xi_{aq}=1$.

\subsection{Terpyridine monomers and terpyridine-functionalized polymers (OMC)} 

Terpyridine is an extensively-used chelating group, well-known for its high affinity towards complexation with many transition metal ions including iron. While terpyridine can form both mono and bis complexes with iron(II), (FeT)\textsuperscript{2+} and (FeT\textsubscript{2})\textsuperscript{2+}, respectively, the formation of bis complexes is thermodynamically favorable \cite{Holyer1965Terpyridine_supp}. Therefore, the mono complexes are neglected and we assume that iron ions always bind two terpyridine binding sites, where the binding sites can either be two terpyridine monomers located on different oligomeric templates (inter-template bond) or onto the same oligomeric template (intra-template bond).  \\

\subsubsection{Terpyridine monomer} \label{terpyridine monomer} 
Here we first apply chemical equilibrium conditions to the partitioning of terpyridine over an oil and aqueous phase, where iron ions are present in the aqueous phase. We investigate two scenarios. In the first and initially anticipated scenario, terpyridine is overwhelmingly molecularly dissolved in the oil phase in the absence of iron ions in the aqueous phase. In the second scenario, we assume that terpyridine forms dimers in oil. These scenarios lead to fundamentally different partitioning behavior and will be compared to experiments. To make the connection to HPE and OMC, it is shown that the chemical equilibrium approach is consistent with the statistical thermodynamic treatment. \\ 
\textbf{Chemical equilibrium.} Based on our experimental set-up, we consider demixed oil and water that are in contact via an interface. Initially, terpyridine molecules dissolve in the oil phase. However, upon addition of iron (II) ions in the water phase (beyond a certain concentration of iron ions) in the water phase, terpyridine molecules partition in the aqueous (aq) phase:
\begin{align}
	\ce{T_{(oil)} <=> T_{(aq)} }&& \nonumber
\end{align}

\noindent in which the terpyridine molecule (monomer) is abbreviated as T. Since the concentration of terpyridine in the oil phase is more preferred as a variable due to its low solubility in water, we define the partition coefficient of terpyridine between water and oil (in the absence of iron or any other metal ions) as:
\begin{align}
	K_p=\frac{[\text{T}_{(aq)}]}{[\text{T}_{(oil)}]} = \exp(-\beta g_H). && \label{Kp1}
\end{align}

\noindent Here, $g_H$ is the free energy difference between terpyridine in water and terpyridine in oil (more specifically: the difference in standard chemical potential of terpyridine in oil and in water based on 1 M reference concentrations). The subscript $H$ stands for "hydrophobic". Equation \ref{Kp1} is expected to be correct up to the water-solubility of terpyridine which is reported to be approximately 6 mM \cite{WinNT_supp,bretti2008solubility_supp}. 
Additionally, bis (terpyridine) iron (II) complexes, (FeT\textsubscript{2})\textsuperscript{2+}, are formed in the aqueous phase via the following reaction:
\begin{align}
	\ce{2T_{(aq)} + Fe^{2+} <=> (FeT_2)^{2+} \label{FeT}} && \nonumber
\end{align} 
It should be noted that the formation of mono (terpyridine) iron(II) complex, (FeT)\textsuperscript{2+}, is neglected based on the equilibrium constants reported previously \cite{Holyer1965Terpyridine_supp}. The equilibrium constant for formation of a bis (terpyridine) iron(II) complex in the aqueous phase is given by:
\begin{align}
	K=\frac{[(\text{FeT}_2)^{2+}]}{[\text{T}_{(aq)}]^2 [\text{Fe}^{2+}]} = \exp(-\beta g_{2}).&&
\end{align}

\noindent Here $g_{2}$ is the reversible work of formation of a bis (terpyridine) iron(II) complex, (FeT\textsubscript{2})\textsuperscript{2+}, in water. 
Thus, we write the equilibrium concentration of (FeT\textsubscript{2})\textsuperscript{2+} as
\begin{align}
	[(\text{FeT}_2)^{2+}] = K K_p^2 [\text{T}_{(oil)}]^2 [\text{Fe}^{2+}]. &&\label{FeT2}
\end{align}

\noindent Equation \ref{FeT2} shows that in order to form (FeT\textsubscript{2})\textsuperscript{2+} in water, there is an unfavorable step of transferring terpyridine from oil to water ($K_p<<1$), and a favorable step of binding iron to terpyridine ($K>>1$). We find the fraction of terpyridine in oil as a function of (free) iron concentration in water through the mass balance $[\text{T}_{(tot)}] = [\text{T}_{(oil)}](1+ K_p) + 2[(\text{FeT}_2)^{2+}]$ (based on equal volumes of the oil and aqueous phases). Substitution of Equation \ref{FeT2} and solving for $[\text{T}_{(oil)}]$ leads to: 
\begin{align}
	f_H = \frac{[\text{T}_{(oil)}]}{[\text{T}_{(tot)}]} = \frac{-1 + \sqrt{1 + 8 K K_p^2 [\text{T}_{(tot)}] [\text{Fe}^{2+}]}}{4 K K_p^2 [\text{T}_{(tot)}] [\text{Fe}^{2+}]}, &&\label{foMono}
\end{align}
where we assumed that $K_p <<1$. 

Anticipating the experimental results (which point to a significantly steeper dependence of the terpyridine fraction in the oil phase on the free iron (II) ions concentration), we consider the situation that terpyridine is in the form of dimers in the oil phase.
In this case the equilibrium follows:

\begin{align}
	\ce{T_{2(oil)} <=> T_{2(aq)}} && \nonumber
\end{align}
with the partition coefficient defined as $K'_P=[\text{T}_{2(aq)}]/[\text{T}_{2(oil)}]= \exp(-2 \beta g_H) = K^2_P$. The bis (terpyridine) iron (II) complex will then form based on the reaction:

\begin{align}
	\ce{T_{2(aq)} + Fe^{2+} <=> ( FeT_2)^{2+}}. && \nonumber
\end{align}
Thus, the equilibrium concentration of $(\text{FeT}_2)^{2+}$ is given by
\begin{align}
	[(\text{FeT}_2)^{2+}] = K K_p^2 [\text{T}_{2(oil)}] [\text{Fe}^{2+}].&& \label{FeT2dim}
\end{align}
Now the fraction of terpyridine in oil, or in its hydrophobic conformation, becomes
\begin{align}
	f_H = \frac{[\text{T}_{(oil)}]}{[\text{T}_{(tot)}]} = \frac{1}{1 + K K_p^2 [\text{Fe}^{2+}]}. &&\label{fodim}
\end{align}

\noindent Here we have used the mass balance $[\text{T}_{2(tot)}] = [\text{T}_{2(oil)}] + [(\text{FeT}_2)^{2+}]$ where $[\text{T}_{2(tot)}] = [\text{T}_{(tot)}]/2$. 

It will become clear in the results and discussion section that Equation \ref{fodim} describes the experimental data much better than Equation \ref{foMono}. Furthermore,  Equation \ref{fodim} is analogous to the Langmuir adsorption equation as can be seen by writing the fraction of occupied sites ($\theta$) of terpyridine dimers ($\text{T}_2$) by iron ions:
\begin{align}
	\theta = \frac{[(\text{FeT}_2)^{2+}]}{[\text{T}_{2(tot)}]} = \frac{K K_p^2 [\text{Fe}^{2+}]}{1 + K K_p^2 [\text{Fe}^{2+}]} = 1 - f_H. &&\label{thetadim}
\end{align}
\textbf{Statistical thermodynamics}
Here we apply the grand canonical ensemble to obtain the same results as discussed in the chemical equilibrium approach, as it should. The terpyridine molecule (monomer) can be in two states; dissolved in oil (with subscript oil) which is an energetically favorable state for the terpyridine molecule, and an aqueous state (with subscript aq) which is an unfavorable state for the terpyridine molecule (due to its hydrophobicity) in the absence of iron ions. However, the aqueous state can be stabilized due to the higher affinity of the terpyridine molecule in this state to iron ions. It is assumed that terpyridine is in thermodynamic equilibrium between the oil and aqueous phases (states) and that iron ions can be exchanged between terpyridine and the reservoir. Therefore, by using the grand canonical ensemble, the grand partition function of the terpyridine dimers is the summation over all states and occupancy numbers (occupancy with iron). Therefore, the grand partition function reads
\begin{align}
	\Xi= \sum_{state~i}\Xi_i = \Xi_{aq} + \Xi_{H}. && \label{Xitot}
\end{align}
Terpyridine in the aqueous state can be either occupied or unoccupied by iron ions. Thus,
\begin{align}\label{Xiaqdim}
	\Xi_{aq}= \exp(-\beta G_H)\sum_{N=0}^{1} \lambda_F^N z(T,N) = \exp(-2 \beta g_H) \sum_{N=0}^{1} \lambda_F^N \exp(-N\beta g_{2}) \\ \nonumber
	= \exp (-2 \beta g_H)(1 + [Fe^{2+}] \exp (-\beta g_{2}) ), &&
\end{align}
in which $G_H (= 2 g_H)$ is the free energy difference between terpyridine dimers in water and in oil phases (states). Here the fugacity of iron (with subscript $F$) is $\lambda_F = \exp (\beta \mu_F)$ and $\mu_F$ is the chemical potential of iron ions. $z(T,N) = \exp (-N \beta g_{2})$ is the relevant part of the molecular partition function of terpyridine with $N$ bound iron ions (N = [0,1]). In the second step in Equation \ref{Xiaqdim}, $\lambda_F$ is written as the iron concentration, and at the same time the appropriate standard states for terpyridine and iron ions are applied, consistent with the chemical equilibrium approach. We assume that terpyridine in the oil state does not bind iron and thus $\Xi_{H} = 1$. From that the fraction of terpyridine in oil, $f_H$, is given by:
\begin{align}
	f_H = \frac{\Xi_{H}}{\Xi} = \frac{1}{1+ [Fe^{2+}]\exp (-2\beta g_H)\exp (-\beta g_{2})} &&\label{foGCdim}
\end{align} 
and the average fraction of terpyridine occupied by iron (in the aqueous state) reads
\begin{align}
	\theta = \langle N \rangle = \frac{\lambda_F}{\Xi} \frac{\partial \Xi}{\partial \lambda_F} = \frac{[Fe^{2+}]\exp (-2\beta g_H)\exp (-\beta g_{2})}{1+ [Fe^{2+}]\exp (-2\beta g_H)\exp (-\beta g_{2})}.  &&\label{thetaGCdim}
\end{align}
Under the assumption of $\exp(-\beta g_H) = K_p << 1$, it is easily verified that Equations \ref{foGCdim} and \ref{thetaGCdim} are equal to Equations \ref{fodim} and \ref{thetadim}, respectively. 

\subsubsection{Terpyridine-functionalized oligomers (OMC).} \label{pure} 
In the case of oligomeric terpyridine templates, because of the formation of bis (terpyridine) iron (II) complexes by binding of iron ions onto two terpyridine groups, gel-like networks are expected to form in the aqueous phase where the oligomers are effectively 'bridged' (equivalently crosslinked) by iron ions. Furthermore, it should be noted that  we assume that the binding affinity for all terpyridine repeating units (regardless of their local chemical surroundings) are the same. The calculation of the multiplicity $\Omega(N,M)$ in this system is a nontrivial problem. In the unrestricted case, that is, if each terpyridine residue is allowed to bind (via an iron ion) to any other terpyridine residue, that is, residing on the same oligomer or onto any other of the $n_p$ oligomers, there are on the order of $n_p M^2$ possible bonds per oligomer.

However, in a cross linked, gel-like network, the overwhelming majority of these bonds will be geometrically inaccessible. Therefore we approximate $\Omega(N,M) \approx \binom{M}{N}$, just as in the situation for HPE. In this case, the value of $M$ can be expected to be on the order of the number of terpyridine groups per oligomer. Numeric calculations of the oligomer fractions and occupied terpyridine residues via an extreme (and presumably unphysical) scenario $\sum_{N=0}^{M} \binom{nM^2}{N} x^N$, with $x=[Fe^{2+}K]$ as a function of iron ion concentration, reveal significant shifts of the hydrophobic to aqueous transition to lower free iron concentration, but essentially the same value for the cooperativity parameter $M$ compared to the situation with $\Omega(N,M) = \binom{M}{N}$. In the following, we will take the latter approximation for the multiplicity and compare the (effective) values of $g_H$ and $g_2$ with the situation for terpyridine monomers. Large deviations between these values may point to a significant error in the expression we use for $\Omega(M,N)$. However, the results summarized in Fig.\ref{fig:OMCfoandhillplots} in the main paper point to a value of  $2g_H + g_2$ for the oligomers that are consistent with independent measurements and reports, as well as with the values obtained for the monomers. The value of $M \approx \text{16}$ is consistent with the existence of on average dimeric clusters of oligomers, in DCM, where the 16 terpyridine dimers act as binding sites with multiplicity $\Omega(N,M) = \binom{16}{N}$. \\
Getting the analogues of Eqs.~(\ref{eq:fH}, \ref{eq:theta_general}), with the definitions of $K_p$ and $K$ from the previous section, the fraction of poly (terpyridine) in the oil phase $f_H = \Xi_{H}/\Xi$ is given by
\begin{equation}\label{foP}
	f_H = \frac{1}{1 + K_p^M (1 + [\text{Fe}^{2+}] K)^{M}}, 
\end{equation}

and for the total fraction of the occupied binding sites we find 
\begin{equation}\label{thetaP_supp}
	\theta = \frac{\langle N \rangle}{M} = \frac{K [\text{Fe}^{2+}]}{1 + K [\text{Fe}^{2+}]} (1 - f_H).
\end{equation}

So, as long as $[\text{Fe}^{2+}] >> K^{-1}$, (with concentrations in M) we have $\theta = 1 - f_H$, even at very low iron concentrations. 

\subsection{Composition dispersity in HPE copolymers}\label{composition_dispersity_theory_supp}


Within HPE we may identify two distinct classes. Homopolymers, composed of a single repeat unit, where the ionizable and hydrophobic group are on the same unit, and copolymers, where there are multiple types of repeating units with potentially different ionization and hydrophobic penalty energies. If the copolymerization of the units is statistical or random there will be a dispersity in the ratio of the repeating units on each chain over the whole polymer sample. In the following we will examine a (compositionally monodisperse) homopolymer, poly(6-(acryloyl) aminohexanoic acid) (PAHA), and compare its pH-dependent partitioning to that of a (compositionally polydisperse) random copolymer, poly(\textit{n}-butyl acrylate-\textit{stat}-acrylic acid) (PBA-AA). The ability to predict the effect of this dispersity on the transitions of HPE is an important test case of our theoretical framework, which includes specific polymer composition information such as the number of repeating groups and the total hydrophobicity of a chain. 

Within copolymers, we may identify two groups of relevance for our analysis: those whose repeating units always contain ionizable groups and those which contain some monomers without ionizable groups. This is an important distinction as the broadening in the latter group is expected to be much more severe and it is our focus here. This is the case for the PBA-AA used in our experiments and whose $pH$-dependent partitioning is shown in Fig.\ref{fig:HPE_fig_2}.

The origin of the expected broadening lies within the large difference in both ionization and hydrophobic energy between the two monomers. An acrylic acid group is weakly acidic and weakly hydrophobic in its protonated form, while the butyl acrylate group is not ionizable and fairly hydrophobic. The varying ratio of each of the groups within the compositionally disperse chains, leads to a large difference in the balance of ionization and hydrophobic energy. Subsequently the transition-$pH$ of each of the chains will differ, leading to an overall spread-out transition.

To apply our theoretical framework to this class of HPE it is necessary to define the total number of repeat units on the polymer, $M_t$. For a binary copolymer we then have $M_t=N'_1+N'_2$, where $N'_i$ is the number of units for the $i^\text{th}$ monomer, and if we denote the monomer type 1 as the one with ionizable groups then $M=N'_1$.

\begin{figure}[h]
    \centering
    \includegraphics[width=1\linewidth]{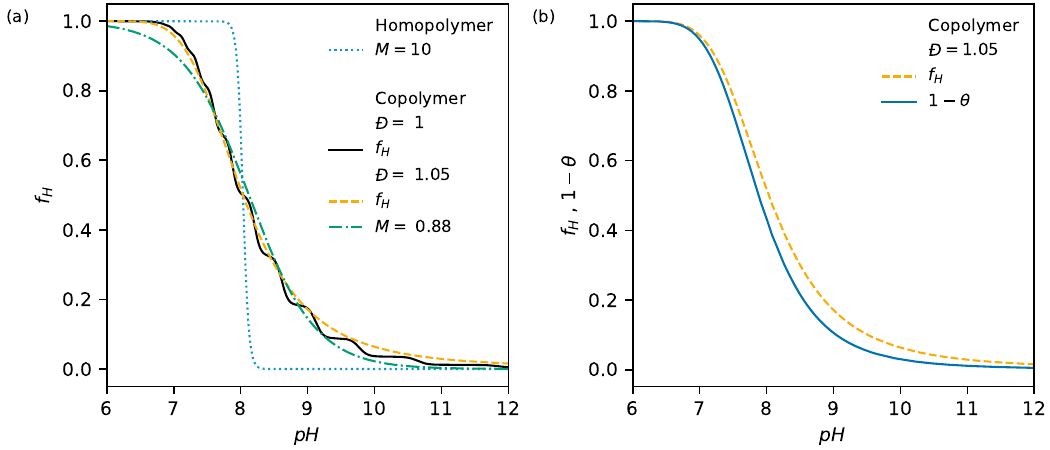}
    \caption{\textbf{Compositional dispersity leads to a spread in the ionization transitions of hydrophobic polyelectrolytes.} \textbf{(a)} Calculated plots (Eq.~(\ref{eq:M_dispersity_fraction})) of the fraction of HPE in the hydrophobic state for a homopolymeric (blue dotted line) and compositionally disperse polymers with (dashed yellow line) and without length dispersity (solid black line). Equation (\ref{eq:HPE_state_oil}) was used in Eq.~(\ref{eq:M_dispersity_fraction}) for the $f_H'$ curve of an individual chain. The hydrophobic penalty is set as $\beta G=g_HM=g_{H,1}N'_1+g_{H,2}N'_2$. For the above curves the global variables used are: $\ev{M_{t}}=20$, $g_{1}=g_{2}=4.0$, $\ev{M}=10$. The dashed green line is a fit of the calculated curve for the length and compositionally disperse system with the general Eq.~(\ref{eq:HPE_state_oil}). \textbf{(b)} $f_H$ (Eq.~(\ref{eq:M_dispersity_fraction})) and $1-\theta$ (Eq.~(\ref{eq:M_dispersity_theta})) curves for a compositionally and length disperse copolymer. Note the lack of full overlap between the curves. The same global variables are used as in (a).} 
    \label{fig:copolymer_dispersity}
\end{figure}

We will also include length dispersity in this model and therefore we have two distributions for the make-up of the polymer, a length and composition distribution. The inclusion of a small amount of length dispersity, which will always be present in conventionally synthesized polymers, increases the smoothness of the curves and improves the match with the experimental data. We illustrate this effect in Fig.~\ref{fig:copolymer_dispersity}(a).

The usual experimental measure of the length dispersity of a polymer extracted from, for example, size exclusion chromatography (SEC), is a length dispersity value, \emph{\DJ}. It is possible to relate \emph{\DJ}~to the mean and variance of a distribution through the following expression \cite{harrisson_downside_2018_supp}:

\begin{equation}
    \emph{\DJ}=1+\frac{\sigma^2}{\mu^2}.\label{eq:PDI_express}
\end{equation} 

Where the mean, $\mu$, and variance, $\sigma^2$, are sample variables in this case. 

Throughout this work we will be considering short ($<100$ repeat units) polymers and therefore we have opted for a log-normal distribution of the polymer lengths, when treating the length dispersity in these systems. This distribution has been shown to be suitable for polymers synthesized using controlled radical polymerization techniques \cite{MONTEIRO2015197_supp}. The weight of each chain length, $M_t$, can be expressed as: 

\begin{equation}
\omega(M_t) = \frac{1}{M_t\sigma'\sqrt{2\pi}} 
  \exp\left( -\frac{1}{2}\left(\frac{\ln{M_t}-\mu'}{\sigma'}\right)^2\right).
    \label{eq:log_normal}
\end{equation}

The variables $\mu$' and $\sigma'$ are related to the actual expectation value and variance of $M$ through the following equations: $\mu=\exp(\mu'+\frac{\sigma'^2}{2})$ and $\sigma^2=(\exp(\sigma'^2)-1)\exp(2\mu'+\sigma^2)$.
$\omega(M)$ then becomes a function of the mean, $\mu$, and variance , $\sigma^2$, of this log-normal distribution and consequently of the \emph{\DJ}~of the polymer.

There are a variety of different ways to take into account the compositional dispersity of copolymers. For our analysis here we will take the simplest approach which is a binomial distribution, leading to an expression for the weight as follows:
\begin{equation}
   \omega'(M_t,N'_1,p)= {\binom{M_t}{N'_1}} p^{N'_1}(1-p)^{M_t-N'_1}. \label{binomial_distribution_2_monomers}
\end{equation} 
Where p is the probability of adding a monomer of type 1 to the end of a chain. A binomial distribution does not take into account the fact that the reactivity for a particular monomer, with respect to the chain end, usually depends on the last monomer which was attached to the chain. This can be taken into account by using a terminal model (or even a penultimate model \cite{klumperman_solvent_1994_supp}), which leads to the well known reactivity ratios \cite{brandrup_polymer_1999_supp}. Knowing these ratios, it would be possible to find the weight  $\omega'(M_t,N'_1)$ for a particular copolymer. The polymers we deal with in this work, however, present mostly very similar chemical reactivities with respect to either of the monomers being at the chain end and therefore a binomial distribution should suffice. The probability, $p$, is therefore simply a ratio of the concentration of each of the monomers in the initial reaction vessel.

The expression for the fraction of chains in the hydrophobic (or aqueous) state follows from Eq.~(\ref{eq:HPE_state_oil}) where we then apply weights for each of the chain lengths and compositions:

\begin{equation}
f'_{H}=\sum^\infty_{M_t=1}\omega(M_t)\sum^{M_t}_{N^{'}_1=0}\omega'(M_t,N'_1)f_H(M_t,N'_1).\label{eq:M_dispersity_fraction}
\end{equation}

The hydrophobic penalty is set as a linear combination of each of the groups ($\beta G=g_{H,1}N'_1+g_{H,2}N'_2$) in Eq.~(\ref{eq:HPE_state_oil}). As before the fraction of polymer chains in the aqueous state is $f_{aq}'=1-f'_{H}$.

As in the previous section, an expression for the total fraction of ionized sites for the polymer sample can be found using the following equation:
\begin{equation}
    \theta=\frac{1}{\ev{M}}\sum^\infty_{M_t=1}\omega(M_t)\sum^{M_t}_{N^{'}_1=0}N_{1}\omega'(M_t,N'_1)\theta'(M_t,N'_1)\label{eq:M_dispersity_theta}
\end{equation} 

Where $\ev{M}$ is the mean number of ionizable groups per chain over the whole polymer sample ($\ev{M}=\sum^\infty_{M_t=1}\omega(M_t)\sum^{M_t}_{N^{'}_1=0}N'_{1}\omega'(M_t,N'_1)$).

Figure \ref{fig:copolymer_dispersity}(a) shows an example of the broadening expected for a length and compositionally disperse binary copolymer, where in this case only one type of copolymer has ionizable groups. Equation (\ref{eq:HPE_state_oil}) was used in Eq.~(\ref{eq:M_dispersity_fraction}) for the $f_H$ curve of an individual chain. The extensive broadening is clearly apparent compared to the homopolymer situation.  It is worth noting that in this particular figure the hydrophobic penalty per group is identical for both monomers and it would be expected that the broadening is even more significant if we allowed the values for the penalties to differ, which will usually be likely due to the difference in the chemical nature of these groups. 

Finally, an interesting feature of these transitions is the asymmetry around the $f_H=0.5$ point. This occurs due to the simultaneous change in transition-$pH$ and sharpness for the different chains in these disperse polymers. Chains with a higher $M/M_t$ fraction will transition at lower $pH$ values and will present sharper transitions. Conversely, chains with a lower $M/M_t$ ratio will transition later and less sharply. This leads to the "tail" seen for the latter half of the transition as the $pH$ increases.

As might be expected from a polymer sample where the number of ionizable groups varies drastically between polymers with similar lengths, the polymers described in this section present deviations between the ionization fraction and hydrophobic fraction curves. Figure \ref{fig:copolymer_dispersity}(b), plots the two curves for an example polymer sample. The ionization fraction curves has an earlier transition-$pH$, again due to chains with a higher proportion of ionizable groups transitioning earlier and these chains having a larger weight in the ionization fraction calculation.

\subsection{Fractionation calculation} \label{sec:fract_appen}
The numerical procedure to calculate the predicted result of an experimental fractionation of a chemically disperse copolymer HPE follows from the equations shown in the previous section. They detail the set of weights of each of the different chains in a chemically disperse polymer, which in the case of the calculation shown in Fig.~\ref{fig:HPE_fig_2} (main text), is a polymer with a disperse ionizable group number to total length ratio.

We start from the set of weights $\omega(M_t,M)$ for the initial disperse polymer, where $M$ and $M_t$ are the number of ionizable groups and the total number of groups on the chain, respectively. This information could be found from an understanding of the polymerization mechanics (e.g. random polymerization) and a measurement of the length dispersity using size exclusion chromatography. In the case of the calculation for PBA-AA in the main text, $M$ corresponds to the number of acrylic acid (AA) groups on a chain. For this system, where we consider the polymerization of the precursor to PBA-AA, PnBA-tBA, between n-butyl acrylate (nBA) and t-butyl acrylate as a random process, $\omega(M,M_t)$ is the combination of Eq.~(\ref{binomial_distribution_2_monomers}) and Eq.~(\ref{eq:log_normal}).

To carry out the numerical fraction procedure we create a set a set of weights for a wide range of potential $(M_t,M)$ values.
We then calculate the fraction of the chains in the aqueous phase ($f_{aq}$) for each $(M_t,M)$ couple at the first fractionation $pH$ in Fig.~\ref{fig:HPE_fig_2}. A general equation for this is as follows:

\begin{equation}
    f_{aq}(M_t,M,pH_1)=\frac{\exp(-\beta G(M_t,M)))(1+10^{pH_1-pK_a})^M}{1+\exp(-\beta G(M,M_t))) (1+10^{pH_1-pK_a})^M} \label{eq:fractionation_appendix_faq}
\end{equation}

The expression for the conformational penalty is $\beta G=g_{H,1}N'_1+g_{H,2}N'_2$, where $N'_1$ is the number of hydrophobic groups defined as $N'_1=M_t-M$. The values of the hydrophobic penalties per group, $g_{H,1}$ and $g_{H,2}$, can be found, for a particular polymer sample, from a fit of the initial partitioning transition of the polymer in the two-phase oil and water system. See Fig.~\ref{fig:HPE_fig_2}(a) for fits of the PBA-AA transition the fractionation calculation is based on.

The product of $f_{aq}(M_t,M,pH_1)$ and $\omega(M_t,M)$ now gives the weight of each chain configuration in the aqueous phase. We denote this set of weights $\omega'(M_t,M)$. In the fractionation scheme in Fig.~\ref{fig:HPE_fig_2}(c), this aqueous phase is then exposed to a new, empty oil phase and its $pH$ is modified. We now carry out an analogous calculation to calculate the weights of the chains that partition into the oil phase. We use the expression for the fraction of chains in the hydrophobic state this time, given by:

\begin{equation}
    f_{H}(M_t,M,pH_2)=\frac{1}{1+\exp(-\beta G_H(M_t,M))(1+10^{pH_2-pK_a})^M} \label{eq:fractionation_appendix_fh}
\end{equation}

The product of $\omega'(M,M_t)$ and $f_{H}(M_t,M,pH_2)$ now gives the final set of weights in the oil phase $\omega''(M_t,M)$. The full operation to find the set of weights for the fractionated polymer is therefore:

\begin{equation}
    \omega''(M_t,M)=\sum_{M_t=1}^{\infty}\sum_{M=0}^{M_t}\omega(M_t,M)f_{aq}(M_t,M,pH_1)f_{H}(M_t,M,pH_2)
\end{equation}

The polymer sample at this step in the fractionation procedure is now extracted from the two-phase system and isolated, affording us a sample with a distribution of chains described by $\omega''(M_t,M)$.

The calculation for the chain efficiency (fraction of chains in the final sample with respect to the initial sample) can be found from the sum of all the values in the final weight matrix before normalization is carried out.

If this fractionated sample is now introduced into a separate two-phase system, as shown in Fig.~\ref{fig:HPE_fig_2}(c), its partitioning behavior can be described by introducing this matrix of weights $\omega''(M_t,M)$ (after normalization) into Eq.~(\ref{eq:M_dispersity_fraction}). This is identical to calculating $f_H$ from Eq.~(\ref{eq:fractionation_appendix_fh}) for every $(M_t,M)$ couple and doing a summation of the product of this value and $\omega''(M_t,M)$, for every $pH$ value.

\section{Hydrophobic polyelectrolytes (HPE)}\label{Experimental_HPE_supp}

\subsection{Monoprotic acids}

\begin{figure}[h]
    \centering
    \includegraphics[width=\textwidth]{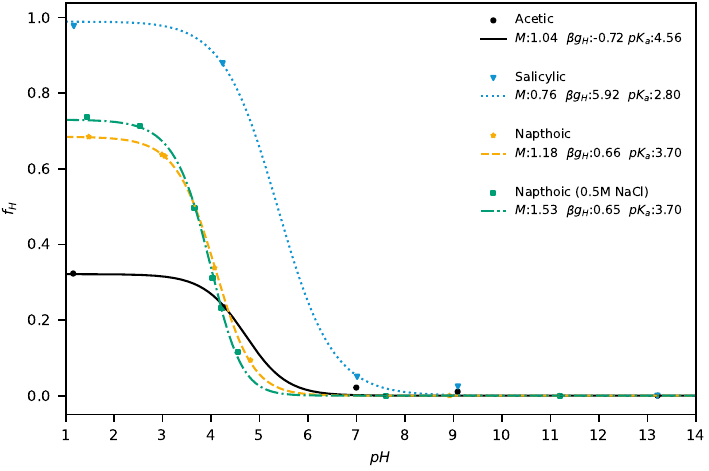}
    \caption{\textbf{Monoprotic acids show broad partitioning behavior in oil-water systems.} Fraction of monoprotic acids in the oil-phase ($f_H$) of two-phase oil and water systems. Acetic and salicylic acid octanol-water partitioning data is taken from reference \cite{disdier_effect_2022_supp}. The partitioning data at the highest $pH$ value  was scaled to reach $0$ for these acids. Napthoic acid isooctane-water partitioning data is taken from reference \cite{standal_partition_1999_supp} for two different salt concentrations. The data series have been fitted (plotted curves) with Eq.~(\ref{eq:HPE_state_oil}) and using $pK_a$ values from \cite{disdier_effect_2022_supp} and \cite{harris_quantitative_2007_supp} for the two sets of data, respectively. Fit parameters are shown in the legend.}
    \label{fig:monomeric_acids}
\end{figure}

\subsection{Materials and instrumentation}\label{Materials_methods}
\subsubsection{Materials}
1-pentanol ($\ge99\%$, ACS reagent), \emph{N,N}-dimethylformamide (DMF, $\ge99.8\%$, ACS reagent), potassium chloride (\ce{KCl}, Analysis grade), potassium carbonate (\ce{K2CO3}, anhydrous, $99\%$, ACS reagent), sodium azide (\ce{NaN3}, $99.5\%$), boric acid ($99.5\%$, ACS reagent), phosphoric acid ($99\%$), copper(II) bromide (\ce{CuBr2}, $99\%$), citric acid ($99.5\%$), \emph{n}-butyl acrylate (nBuA, stabilized, synthesis grade), silica gel (Davisil grade $633$, $60$\AA ~pore size, 200-425 mesh particle size), ethyl $\alpha$-bromoisobutyrate (EBIB, $98\%$), propargyl bromide ($80$ wt.$\%$ in toluene, stabilized), \emph{N,N,N',N'',N''}-penta\-methyl\-di\-ethylene\-triamine (PMDETA, $99\%$), tris[2-(dimethylamino)ethyl]amine (Me\textsubscript{6}TREN, $98\%$), acryloyl chloride ($97\%$, stabilized) and 2,2'-azobis(2-methylpropionitrile) ($98\%$) were purchased from Sigma-Aldrich.

Potassium hydroxide (\ce{KOH}, Analysis grade), trisodium citrate dihydrate (analysis grade, ACS reagent), ethanol (absolute for analysis, ACS reagent), acetic acid (glacial $100\%$, ACS reagent), potassium iodide (\ce{KI}, analysis grade), trifluoroethanol (TFE, $99\%$) and sodium hydroxide (\ce{NaOH}, Analysis grade) were purchased from Merck.

Copper wire ($0.25$mm diameter, $99.98\%$), \emph{t}-butyl acrylate($99\%$,stabilized), copper(I) bromide (\ce{CuBr},$98.1\%$) and dichloromethane (DCM, anhydrous, $99.7\%$,stabilized) were purchased from Alfa Aesar.

7-hydroxy-4-methylcoumarin (4MU, $97\%$), basic alumina (Brockmann 1, $60$\AA pore size, 40-300 mesh particle size) and fuming hydrochloric acid ($37\%$ solution in water)  were purchased from Acros Organics.

Methanol (HPLC grade), n-hexane (HPLC grade), chloroform (HPLC grade, stabilized), diethyl ether (HPLC grade, stabilized) and ethyl acetate (analytical reagent grade) were purchased from Biosolve B.V.

6-aminohexanoic acid ($99\%$), magnesium sulfate (\ce{MgSO4},anhydrous, $\ge99.5\%$) and tetrahydrofuran (THF, anhydrous, $99.9\%$, stabilized) were purchased from Thermo Fischer Scientific.

Acetone ($\geq 99.5\%$) was purchased from VWR chemicals.

Trifluoracetic acid (TFA, $99\%$) was purchased from Honeywell.

Deuterated dimethylsulfoxide-d6 (DMSO-d6, $99.8$ atom$\%$D) and deuterated chloroform-d1 (\ce{CDCl3}, $99.8$ atom$\%$D, stabilized) were purchased from Carl Roth.

4-cyano-4-[(ethylsulfanylthiocarbonyl)sulfanyl]pentanoic acid (CECTP, $\ge98\%$) was purchased from Polymer Source inc.

Milli-Q water (deionized, $18.2 \text{M}\Omega\text{cm}$, $21$ \textdegree C) was used throughout synthetic procedures when referring to (deionized) water.

\subsubsection{Instrumentation} \label{sec:instrumentaion_4}

UV-Vis absorbance spectroscopy was carried out on the CLARIOstar microplate reader (BMG LABTECH) using UV-STAR 96-well plates (Greiner). 

$pH$ measurements were carried out on a SevenExcellence $pH$-meter (METTLER TOLEDO) using an InLab Micro probe (METTLER TOLEDO). Standard calibration solutions (Hanna Instruments), $pH$ $4.01$ and $10.01$, were used to calibrate the instrument before the measurements runs. Measurements of the solutions after the experimental run were used to correct for instrumental drift.

During the process of polymer precipitations an Allegra X-12R (Beckman Coulter) centrifuge was used at $3273$g in $50$ mL plastic centrifuge tubes. When centrifuging two-phase pentanol and water solutions $10$ mL glass centrifuge tubes were centrifuged at $931$ $g$.

NMR measurements were carried out on a $400$ MHz spectrometer (Agilent) and integrated using MestReNova (Mestelab Research) software.

SEC measurements of the polymer samples were carried out on an Alliance HPLC e2695 (Waters) connected to a Waters 2414 refractive index detector and a Waters 2489 UV-Vis detector. A $300$ mM ammonium acetate buffer ($pH$ $9$) was made in Milli-Q water and then filtered using a Whatman RC55 $0.45$ $\mu$m pore filter. Samples were dissolved at a concentration of $1$ mg/ml in the eluent and then filtered through a Phenex RC $0.2$ $\mu$m syringe filter. The separation column employed was an Agilent PL-aquagel 30 and a flow rate of $1$ ml/min was used. PEG standards were chosen for the calibration of the instrument.

\subsection{Experimental procedure}\label{HPE_exp_proc_supp}

Two types of experiments were performed to analyze the partitioning transitions of HPE:

The first uses a buffered aqueous solution in contact with the oil phase. This allows for a series of samples at different $pH$ values to be easily made. The effect of possible $pH$ drift over time is reduced and long-term observation of the samples is possible. From this experiment it is possible to extract the fraction of the polymer chains in either of the phases but ionization data is inaccessible. The second experiment gives access to the ionization fraction of the polymer. This requires unbuffered solutions which are carefully titrated to measure the $pH$ response of the aqueous phase of the two-phase system. It is also possible to extract the fraction of the chains in either of the phases using this method.

The choice of method will depend on the desired application. An assay style screenings of a series of polymers may benefit from the simplicity and speed of the buffered method even if ionization data is not measurable. Before describing the particular setup used to yield the experimental results described in this study it is worth stating some general design principles for the phases used and for the polymer being investigated.

The two liquids chosen must be at least partially immiscible as to allow for two phases to form. One must be aqueous, as its $pH$ will be varied and the other must be hydrophobic enough for there to be an important offset in the hydrophobic penalty for the polymer between both phases and for ionization of the acid or basic groups to be suppressed. In the experiments presented here pentanol was chosen as the hydrophobic organic phase. 

The HPE chosen must, by definition, contain weakly ionizable groups which afford it its $pH$ sensitivity. It must contain a certain amount of hydrophobic groups for the hydrophobic penalty to be large enough to shift the transition-$pH$ a few $pH$ units away from the $pK_a$ of the ionizable groups. The HPE investigated in this work are shown in Table~\ref{table:polymer_structures_4}. Being able to attain well-defined polymers, with predictable composition, length and dispersity, is of great importance in our investigation of the properties of hydrophobic polyelectrolytes. We have therefore employed "controlled" radical polymerization techniques to synthesize the polymers used in this study. See, for example \cite{RAFT_review_supp,SET_review_supp,Ribelli_supp}, for an overview of the RAFT and ATRP techniques used and their underlying mechanisms. 

\begin{table}[ht]
     \begin{center}
     \begin{tabular}{c c c}
     \toprule
      Name & Chemical structure & {\makecell{Degree of \\ polymerization (DP)}}\\ 
      \midrule
      \makecell{poly(6-(acryloyl) aminohexanoic acid) \\ (PAHA)} & \includegraphics[scale=0.5, valign=c]{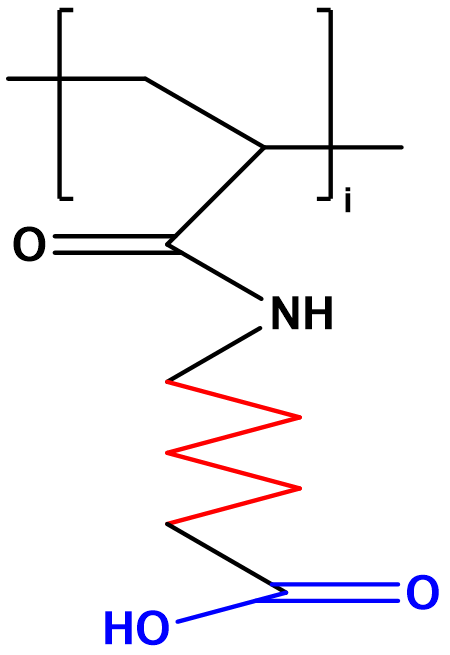} & $18-65$ \\[2cm]
      \makecell{poly(butyl acrylate - acrylic acid)) \\ (PBA-AA)} &  \includegraphics[scale=0.5, valign=c]{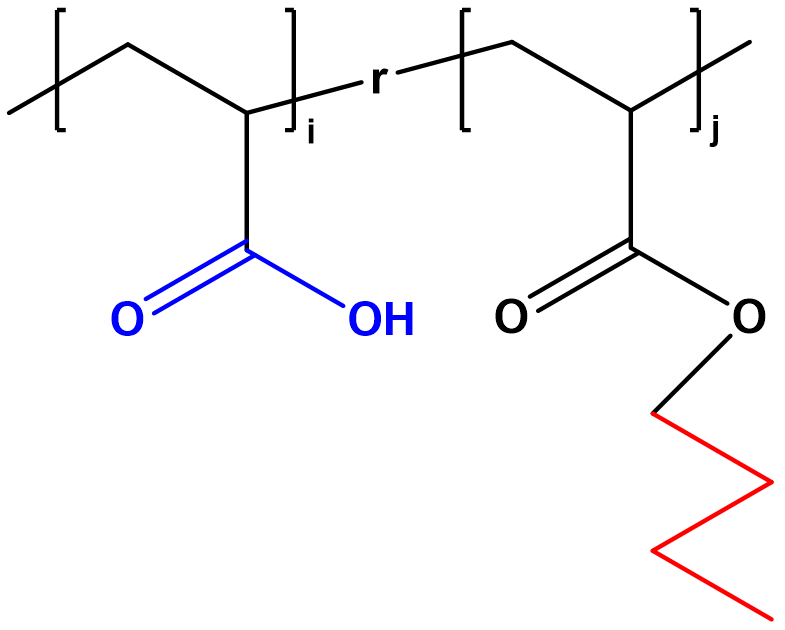} & $20$ \\
       \bottomrule
      \end{tabular}
      \caption{Chemical structures of the hydrophobic polyelectrolytes used in this study. Hydrophobic groups are depicted in red and ionizable groups are depicted in blue.}
      \label{table:polymer_structures_4}
      \end{center}
\end{table}

Finally, a consideration for the measurement methods must be made. UV-Visible (UV-Vis) spectrometry is the method of choice used in this study to measure the fraction of the polymer in the oil phase. The use of a plate reader spectrometer, allowing for the parallel measurement of dozens of samples, simplifies the measurement of large series of samples and therefore this was the method chosen in this study. This requires there to be a chemical moiety on the polymer that has a high absorbance in this spectral range and for the liquid in the phase being measured not to absorb in a similar range. It is worth noting that many dyes are $pH$ sensitive and therefore measurement in the aqueous phase might be unsuitable. It is also possible to use methods that do not rely on the absorption of light. This is the case for refractometry which is a commonly used method to quantitatively measure concentrations of dissolved molecules.

\subsubsection{Buffered system}\label{buffered_exp}
Due to the wide range of $pH$ values ($pH~3-12$) of interest a series of pentanol-saturated Britton-Robinson buffers (equimolar acetic, phosphoric and boric acid, $40$ mM in Milli-Q water) with an added $0.7$ M KCl were made and titrated using $0.7$ M KOH to the desired $pH$ values.. $0.1$ w/v$\%$ sodium azide (NaN3) was added to the buffer series to avoid any bacterial growth followed by an excess amount of pentanol as to saturate them.

The HPE being analyzed was then dissolved in water-saturated pentanol. The concentration of these solutions is usually around $1$ mg/mL. The exact concentration chosen for this stock solution will depend on the absorption value of the specific wavelength we have chosen to analyze. A value of $2.0$ was aimed for. 

Equal volumes ($1.5-2$ mL) of a particular buffer solution and of the pentanol-polymer solution were added together, and they were then thoroughly shaken and/or left on a roller. The solutions were allowed to equilibrate over a couple of days and were periodically shaken throughout or left on a roller.

Solutions made with the chemically disperse polymer, poly(butyl acrylate -  acrylic acid) (PBA-AA), presented persistent emulsions. Therefore when using this polymer, these solutions were centrifuged at $931$ $g$ for $30$ min.

For all polymers the two phases were then separated by pipetting off the top pentanol phase. The fraction of polymer in the oil phase was then measured using UV-Vis spectrometry and the $pH$ of the aqueous phases was measured. Standard calibration solutions (Hannah instruments, $pH$ $4.01$ and $10.01$) measured at the end of the measurement run were used to calibrate the $pH$-meter \footnote{A comment can be made at this point on the interpretation of the $pH$ measurements of a pentanol-saturated solution ($\sim$2.2 $\%$w/w). Following from \cite{pH_measurement_supp,pH_measurement2_supp} we might expect a (small) offset of the measurement with respect to a fully aqueous phase, therefore the magnitude ($\Delta pH$) of the transition region is expected to be unaffected. The same $pH$-meter was used for all experiments so we can consider all experiments directly comparable.}.

\subsubsection{Unbuffered system (titration)}\label{titration_exp}
A $0.7$ M KOH solution and a $0.7$ M KCl solution were made with degassed Milli-Q water. The $0.7$ M KCl solution was then acidified using a degassed $1$ M HCl solution to a $pH$ of around $3$. A stock water-saturated pentanol-polymer solution with a concentration of around $1$ mg/mL was made and then degassed.

Under a flow of \ce{N2} gas, $2$ mL each of the pentanol-polymer and $0.7$ M KCl acidified solution were added to a series of glass centrifuge tubes. A known amount of the $0.7$ M KOH solution was then added to each sample in the series. The amount of each solution added was tracked by measuring the added weight at each step. 

The two-phase samples were then thoroughly shaken over the course of a few hours and allowed to equilibrate overnight. They are then centrifuged at $931$ $g$ for $30$ min.

To avoid possible \ce{CO2} dissolution into the aqueous phase, the $pH$ of the aqueous solutions was measured before separation of the oil and aqueous phases. The $pH$-meter was inserted through the oil phase into the aqueous phase. 

The solutions were then separated and the fraction of polymer in the oil phase is measured using UV-Vis spectrometry.

\subsection{Polymer synthesis and characterization}\label{polymer_synth_supp}
Being able to attain well-defined polymers, with predictable composition, length and dispersity, is of great importance in our investigation of the properties of hydrophobic polyelectrolytes. We have therefore employed "controlled" radical polymerization techniques to synthesize the polymers used in this study. See, for example \cite{RAFT_review_supp,SET_review_supp,Ribelli_supp}, for an overview of the RAFT and ATRP techniques used and the underlying mechanisms. Each polymer's synthesis will now be described in turn including the synthesis of any monomers or dyes used to synthesize these polymers.

\subsubsection{poly(6-(acryloyl)aminohexanoic acid) (PAHA)}\label{PAHA_synth}
The synthesis of the homopolymeric poly(6-(acryloyl)aminohexanoic acid) (PAHA) was carried out using RAFT polymerization.

\subsubsection{Synthesis of 6-(acryloyl)aminohexanoic acid}
The monomer polymerized into the PAHA polymer was synthesized following the description of Hetzer et al.\cite{Hetzer_supp}.

To a $100$ mL round bottom flask NaOH ($4.5$ g, $0.1$ mol, $2.5$ eq.) and 6-aminohexanoic acid ($5$ g, $0.04$ mol, $1$ eq.) were added and then dissolved using $30$ ml of deionized water. After cooling down the mixture in an ice bath, acryloyl chloride ($3.8$ ml, $0.05$ mol, $1.25$ eq.) was added dropwise at a rate of $0.12$ mL/min (using a syringe pump and a teflon tube). The solution was stirred while in the ice bath for 2 h. After removing the solution from the ice bath, it was acidified using an HCl solution ($1:1$ fuming to deionized water) until $pH$ $1$ was reached. A white precipitate formed which subsequently dissolved when $40$ mL of ethyl acetate was added to the flask. The mixture was then transferred to a separatory funnel and after removal of the ethyl acetate phase, the water phase was washed three times with $30$ mL of ethyl acetate. The combined organic phases were then washed with $100$ mL of acidified (pH $1$) water and dried over excess \ce{MgSO4}. After filtration of the mixture it was concentrated \emph{in vacuo} to return a white powder. To further purify the product, it was dissolved in a minimum amount of ethanol ($\sim2$ mL) and precipitate into $40$ mL of cold diethyl ether. This was repeated three times, after which the precipitate was dried under a stream of dry \ce{N2}. Average yields for this procedure were approximately $30\%$.

\begin{figure}
    \centering
    \subfloat[\centering $^1$H NMR ($400$ MHz, DMSO-d6) of the final product of the 6-(acryloyl)aminohexanoic acid synthesis]{{\includegraphics[width=0.8\textwidth]{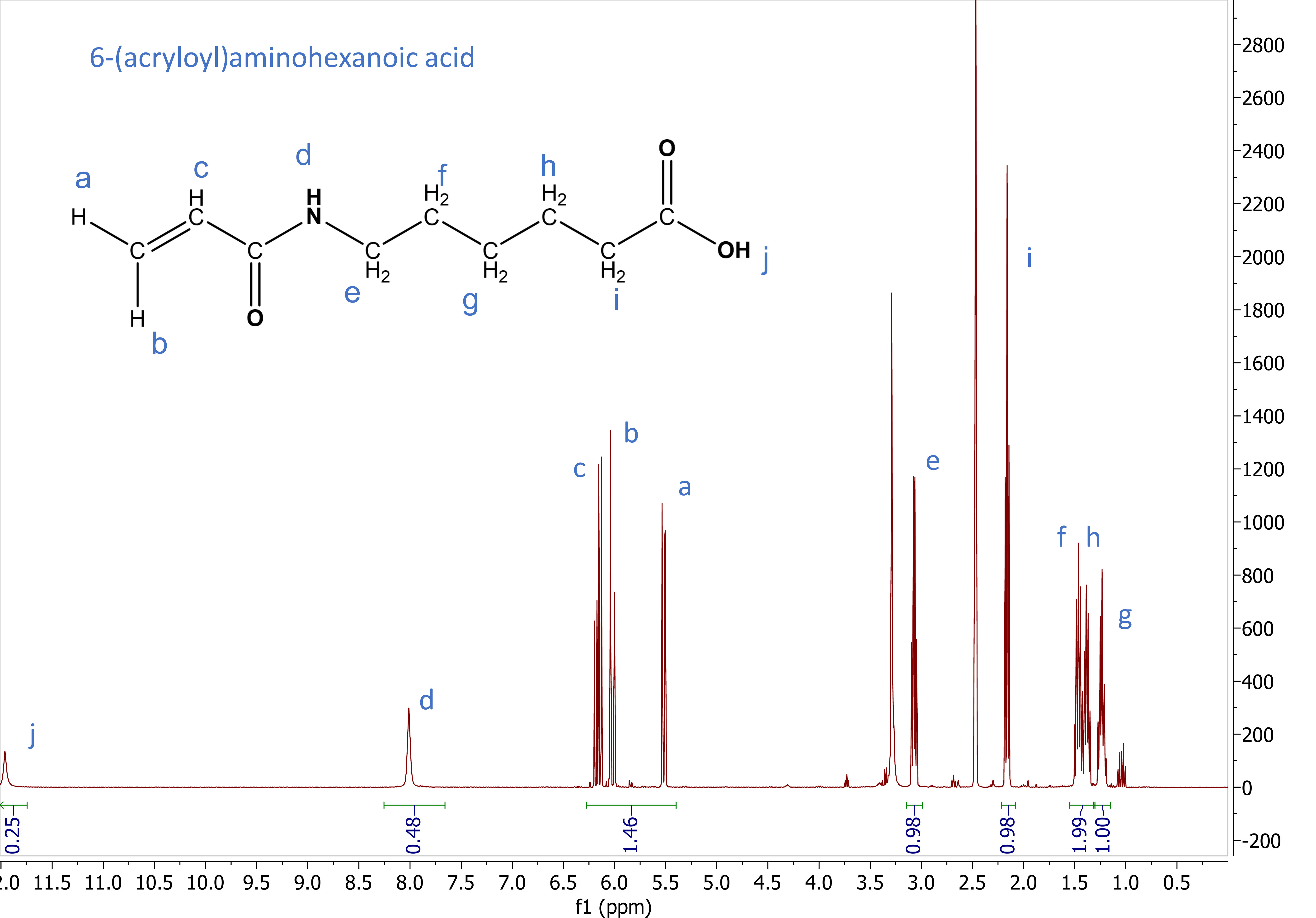}}}%
    \\
    \subfloat[\centering $^1$H NMR ($400$ MHz, DMSO-d6) of the final product of a DP=20 PAHA polymer synthesis.]{{\includegraphics[width=0.8\textwidth]{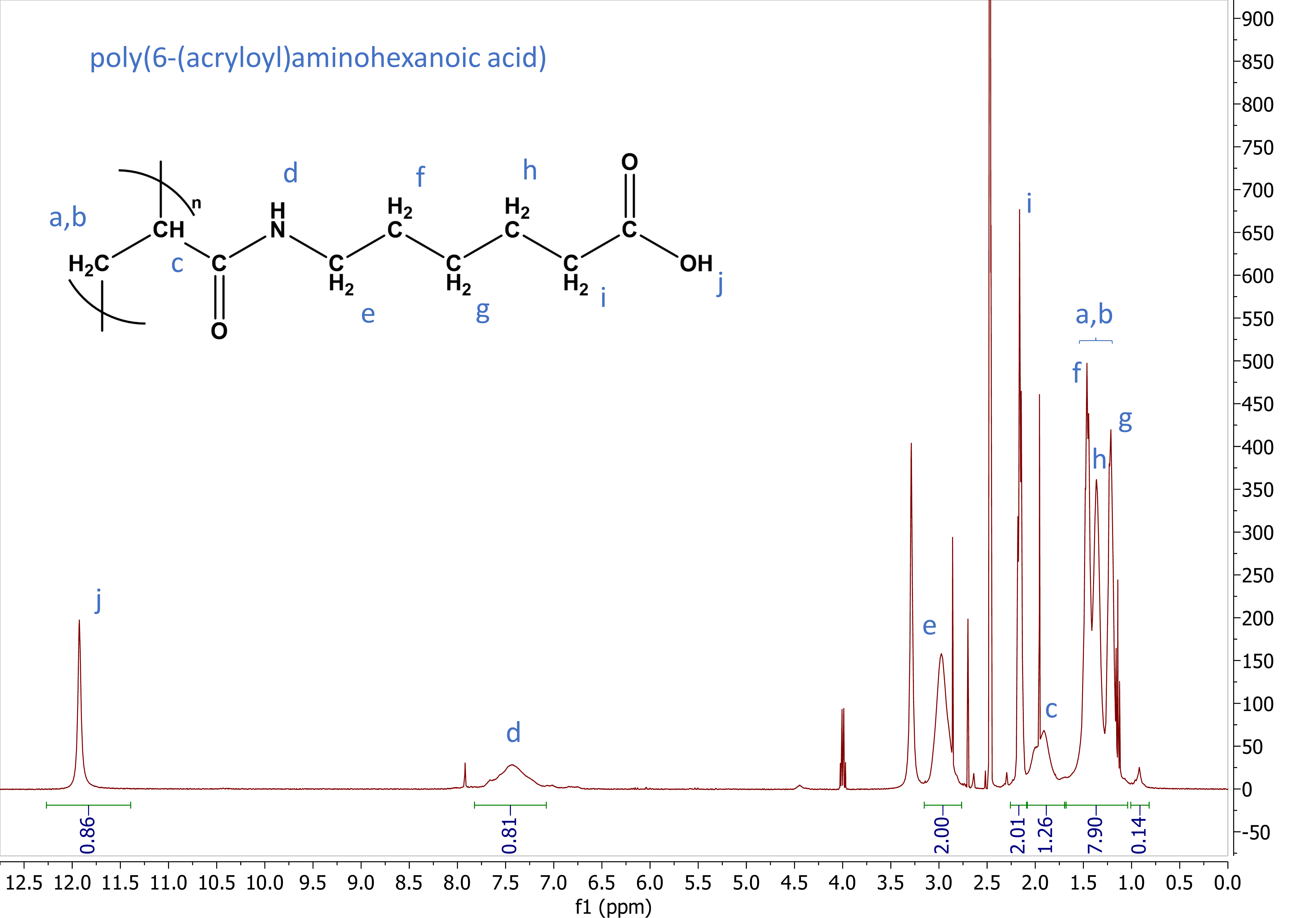}}}%
    \caption{\textbf{$^1$H-NMR spectra} of the PAHA synthesis.}
    \label{fig:monomer_PAHA_NMR}
\end{figure}

$^1$H NMR ($400$ MHz, DMSO-d6), shown in Fig.~\ref{fig:monomer_PAHA_NMR}(a), was carried out to confirm the presence of the desired compound.

\subsubsection{RAFT polymerization of 6-(acryloyl)aminohexanoic acid}
To yield the PAHA polymer, with good control over its length dispersity, RAFT polymerization was employed. The specific procedure employed is based on the polymerization described in Brodszkij et al. \cite{Stadler_supp}.

Three different degrees of polymerization (DP) were targeted, which are achieved by varying the ratio of reagents during the polymerization. Note that only the shortest polymer was used in the partitioning experiments. Data for all three is shown to illustrate the ability to target different DPs. The degree of polymerization of a RAFT reaction can be approximated to the ratio of the molar concentration of the monomer with respect to the CTA, if we assume full conversion \cite{RAFT_review_supp}. Table \ref{tab:RAFT_table} summarizes the different amounts of reactants used to yield the polymers. The following general procedure was used to carry out the polymerization. 

\begin{table}
    \centering
    \begin{tabular}{|c|c|c|c|c|} 
     \hline
     Target DP & monomer (mg/eq.) & AIBN (mg/eq.) & CTA (mg/eq.) & DMF (mL)\\
     \hline
     $20$  &  $255.3$/$20$  &  $2.3$/$0.2$  &  $19.7$/$1$  &  $1.8$\\
     $40$  &  $243.6$/$40$  &  $1.1$/$0.2$  &  $8.8$/$1$  &  $1.8$\\
     $80$  &  $251.1$/$80$  &  $0.6$/$0.2$  &  $5.4$/$1$  &  $1.8$\\
     \hline
     \end{tabular}
    \caption{Amounts of reactants used to yield different degrees of polymerization (DP) of PAHA using RAFT polymerization. Equivalents are normalized with respect to the CTA for each reaction run.}
    \label{tab:RAFT_table}
\end{table}

In a $4$ mL vial the chosen amounts of 6-(acryloyl)aminohexanoic acid monomer, AIBN and the CTA, CECTP, were dissolved in DMF. The solution was then degassed by bubbling dry \ce{N2} through it for $1$ h. The vial was placed in an oil bath set at $75$ \textdegree{C} and left stirring overnight. The whole reaction mixture is then pipetted into $40$ mL of cold diethyl ether. The precipitate formed is reprecipitated a further two times by dissolving it in a minimum amount of ethanol ($\sim3$ mL) and pipetting the solution into cold ethyl acetate. The bulk of the solvent is removed under a stream of dry \ce{N2} and then the polymer is dried \emph{in vacuo} overnight. Average yields for this procedure were approximately $60\%$.

\subsubsection{PAHA polymer characterization}
$^1$H nuclear magnetic resonance (NMR) ($400$ MHz, DMSO-d6), shown in Fig.~\ref{fig:monomer_PAHA_NMR}(b), was carried out to confirm the presence of the desired compound. Note the broadening of the peaks with respect to the monomer and the disappearance of any vinyl peaks between $5.5$ and $6.5$ ppm. The NMR spectrum for all three samples with a different target degree of polymerization were very similar.

Further, size exclusion chromatography (SEC) was used to analyze the hydrodynamic volume distribution, as a proxy for the length distribution, of the synthesized polymers. The SEC was run using an ammonium acetate buffer ($300$ mM, $pH$ $9$) to ensure the full solubilization of the polymer chains in the solvent. Both refractive index (RI) and UV-VIS absorption measurements were employed to detect the polymer eluted from the column. Table \ref{tab:GPC_results} summarizes the results of the SEC analysis of the target DP=$20$, $40$ and $80$ polymers.

\begin{table}
    \centering
    \begin{tabular}{|c|c|c|c|c|} 
     \hline
     Target DP & Mw (kDa) & Mn (Da) & \emph{\DJ} \\
     \hline
     $20$  &  $5.8$  &  $5.5$ &  $1.06$ \\
     $40$  &  $9.0$  &  $7.8$  &  $1.15$ \\
     $80$  &  $13.4$  &  $10.2$  &  $1.31$ \\
     \hline
     \end{tabular}
    \caption{Summary of SEC analysis (derived from PEG standards) of PAHA polymers with different target degrees of polymerization. Only data derived from RI data is shown here.}
    \label{tab:GPC_results}
\end{table}

\begin{figure}
\includegraphics[
width=0.8\textwidth]{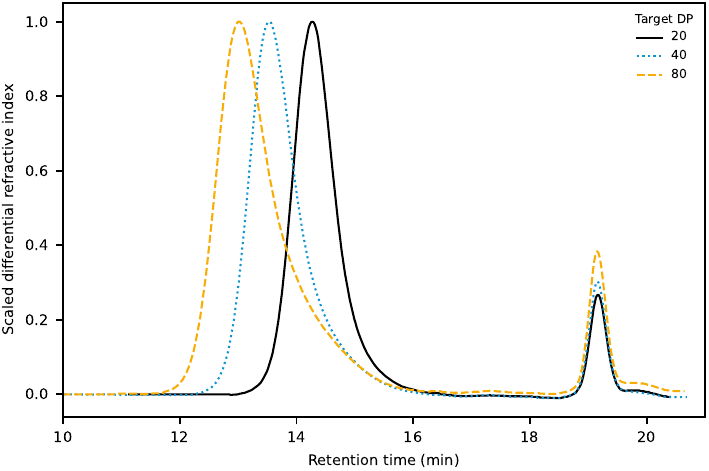}
    \caption{\textbf{RI SEC chromatograms} of the different PAHA polymers.}
    \label{fig:PAHA_GPC_RI}
\end{figure}

As expected the molecular weights of the three different polymers do increase with an increase in the targeted DP. However, due to the large difference in the chemical structure of the PAHA polymer and the PEG standard used, the molecular weights derived from SEC will not be accurate. The length dispersity values (\emph{\DJ}) on the other hand, remain a good estimate of the molecular weight distribution of the sample. There is a marked increase in the dispersity (\emph{\DJ}) of the polymer as we increase the target DP which coincides with a tail appearing in the RI chromatograms for the polymers shown in Fig.~\ref{fig:PAHA_GPC_RI}.

\begin{figure}
    \centering
    \subfloat{{\includegraphics[width=0.8\textwidth]{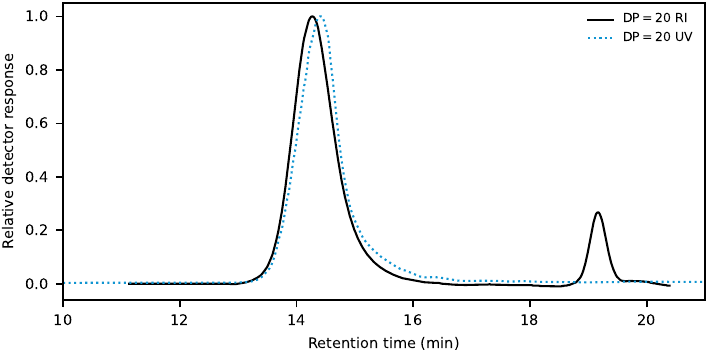}}}%
    \\
    \subfloat{{\includegraphics[width=0.8\textwidth]{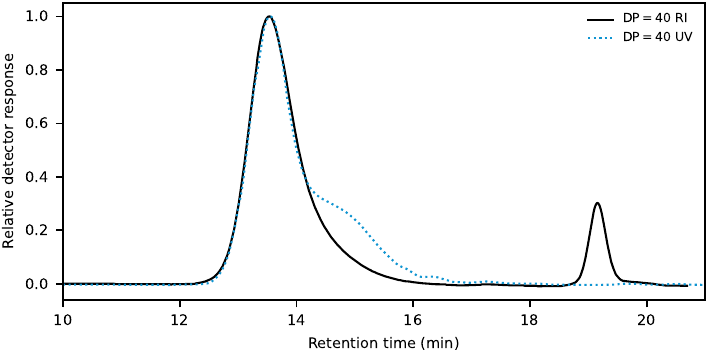}}}%
    \\
    \subfloat{{\includegraphics[width=0.8\textwidth]{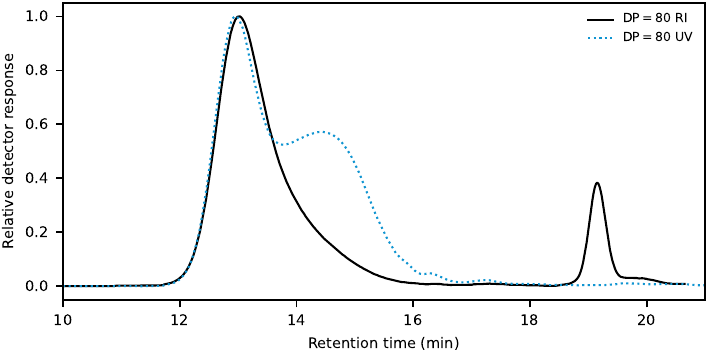}}}%
    \caption{\textbf{Overlay of RI and UV ($308$ nm) SEC chromatograms} of the PAHA polymers with different target degrees of polymerization.}
    \label{fig:PAHA_GPC_UV}
\end{figure}

The RI traces indicate the general mass concentration over time of the polymer eluting from the column, conversely the UV ($308$ nm) traces, shown in Fig.~\ref{fig:PAHA_GPC_UV}, show predominantly the elution of the UV active moiety on the polymer. In this case, the UV active group will be the trithiocarbonyl end group, which originates from the CTA used in the polymerization. It is interesting to note that the shapes of the RI and UV traces do not match well. The UV traces do not present tails in the distribution, but actual bimodal character. This indicates a correlation between the trithiocarbonyl group retention at the end of a polymer and the chain length.

Although, in principle, we can derive quantitative degrees of polymerization from both the NMR and SEC analysis of the final products, the data presented here does not allow for this. The NMR spectra does not have an easily identifiable reference peak from either the initiator or the CTA, and the use of a very structurally distinct reference for the SEC, namely PEG, only allows for qualitative length comparison. To ascertain an, at least, rough measure of the average length of the polymers we can use the monomer conversion calculated from the NMR measurements of the reaction mixture after it is precipitated in cold diethyl ether (Table~\ref{tab:length_results}). 

An estimate for the concentration of total chains that are initiated in the system is given by the initial molar concentration of the CTA, which can then be compared to the estimated number of equivalents of monomer which have reacted.

\begin{table}
    \centering
    \begin{tabularx}{\textwidth}{|X|X|X|X|X|}
     \hline
     Sample target DP & Monomer conversion & Estimated DP (NMR) & Elution time (main peak RI, min) & Estimated DP (SEC)\\
     \hline
     DP=$20$  &  $-$  &  $-$  &  $14.494$ &  $17.9$ \\
     DP=$40$  & $88\%$ &  $35.2$  & $13.778$  &  $-$\\
     DP=$80$  &  $81\%$  & $64.8$  & $13.255$ &  $-$\\
     DP=$20^e$ &  $72\%$  & $14.4$  & $14.72$ &  $-$\\
   
     \hline
     \end{tabularx}
    \caption{Summary of the estimation of the degree of polymerization of PAHA polymers using a combination of monomer conversion data and SEC measurements.}
    \label{tab:length_results}
\end{table}

Monomer conversion data was not available for the target DP=$20$ sample. Instead we can estimate a value for the degree of polymerization of this polymer using the SEC data of the other samples. This analysis includes the target DP=$20^e$, which was synthesized with the same conditions are the other target DP=$20$ sample but with the addition of $1$ equivalent of eosin-Y acrylate (polymer not used in this work). We created a calibration curve using the estimated DP values for the other three samples shown in Table~\ref{tab:length_results}. A $\log (\text{DP})$ vs elution time graph for these samples yields a straight line ($R^2=0.99$) and then a value can be read off for the target DP=$20$ sample. Further analysis could be carried out by performing, for example, MALDI-TOF (matrix-assisted laser desorption/ionization time of flight mass spectrometry) to measure absolute molar masses for the different polymers. 

\subsubsection{poly(n-butyl acrylate-$s$-acrylic acid) (PBA-AA)}\label{PBA-AA_synth}
Unlike the PAHA polymer, whose synthesis is described above, PBA-AA samples were synthesized using a variant of ATRP, namely SET-LRP \cite{SET_review_supp}. The initiator chosen was not UV active, therefore the PBA-AA$^c$ polymer was modified after synthesis using \emph{click} chemistry to add a coumarin dye. The synthesis begins with the synthesis of poly(n-butyl acrylate-$s$-t-butyl acrylate), followed by the azidification of the polymer chain ends and the addition of the coumarin dye to it. Finally the t-butyl acrylate groups are deprotected to yield the desired HPE. A non-tagged variant of the polymer PBA-AA$^0$ was also synthesized in identical fashion omitting the chain-end modification steps. The synthesis of all the components is detailed below.

\subsubsection{Synthesis of a \emph{clickable} $4$MU dye}
An alkyne modified 7-hydroxy-4-methylcoumarin ($4$MU) was synthesized following the procedure described by Chen et al\cite{CHEN20081789_supp}.

To a $250$ mL round bottom flask with two-necks $120$ mL of acetone was added. Then 7-hydroxy-4-methylcoumarin ($5$ g, $25.75$ mmol, $1$ eq.) was added, followed by \ce{KI} ($0.43$ g, $2.57$ mmol, $0.1$ eq.) and \ce{K2CO3} ($7.12$ g, $51.5$ mmol, $2$ eq.) A condenser and an addition funnel were added to the flask and then heated in an oil bath at $60$ \textdegree{C} for $45$ min. The propargyl bromide ($2.9$ mL, $25.7$ mmol, $1$ eq.) was subsequently added dropwise over the period of $1$ hour. The reaction mixture was allowed to stir overnight at $60$ \textdegree{C}. The reaction mixture was transferred into another flask and $50$ mL of DCM was added. This caused a precipitate to form and the supernatant was removed using vacuum filtration. This precipitate was washed two times with a further $50$ mL of DCM. The combined DCM supernatants, a gold colored solution, were then extracted two times with an equal volume of deionized water. The organic phase was dried with an excess of \ce{MgSO4} and then filtered. The resulting solution was concentrated \emph{in vacuo} and then purified via a recrystallization in anhydrous methanol. Finally the crystals were reprecipitated in cold \emph{n}-hexane, after dissolution in a minimal amount of DCM, to give a white powder. The yield of the reaction was $41\%$

\begin{figure}
    \centering
    \subfloat[\centering $^1$H NMR ($400$ MHz, \ce{CDCl3}) of the alkyne modified 7-hydroxy-4-methylcoumarin.]{{\includegraphics[width=0.8\textwidth]{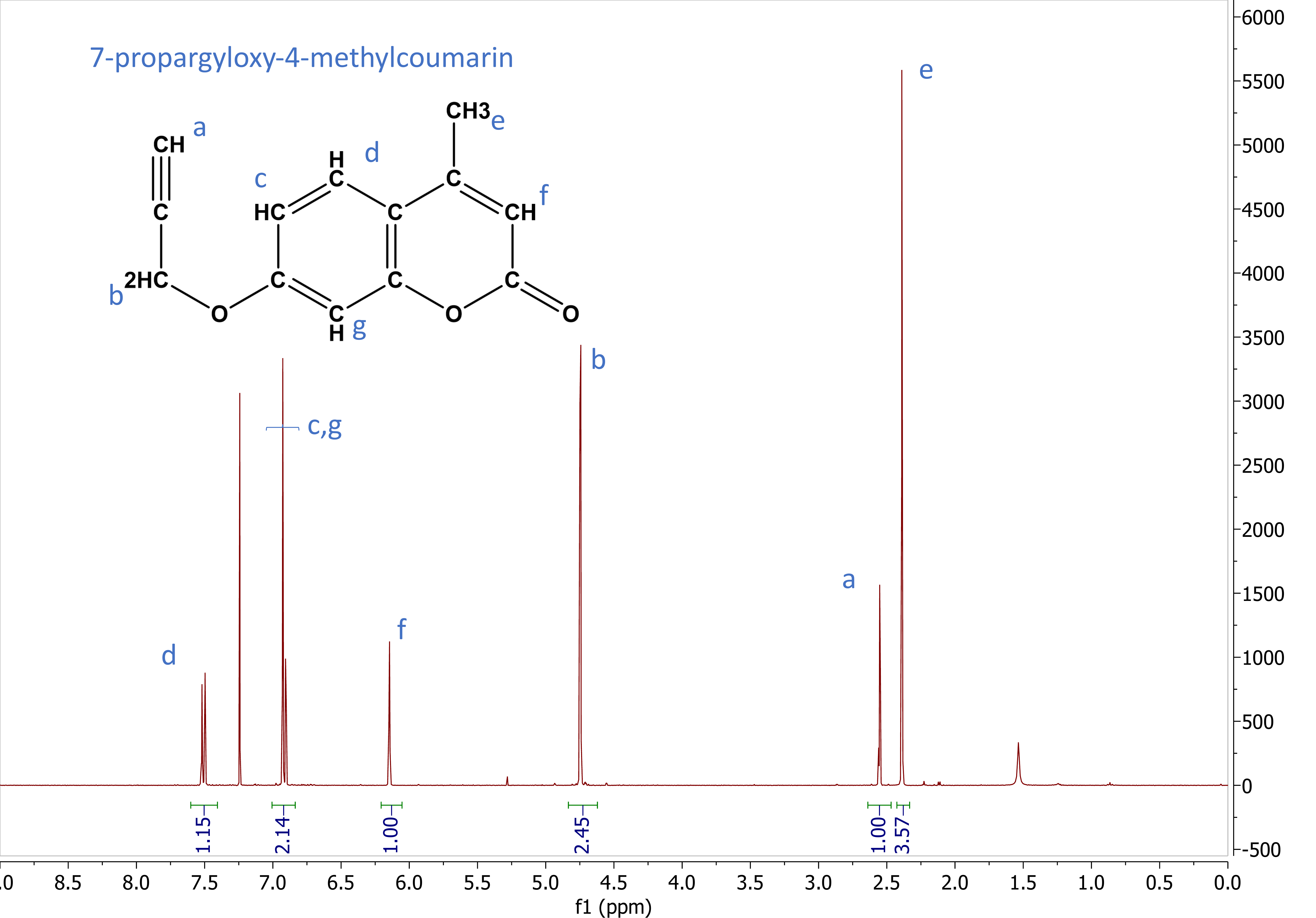}}}%
    \\
    \subfloat[\centering $^1$H NMR ($400$ MHz, \ce{CDCl3}) of the final product of a poly(n-butyl acrylate-$s$-t-butyl acrylate) synthesis (first step of PBA-AA$^c$ synthesis).]{{\includegraphics[width=0.8\textwidth]{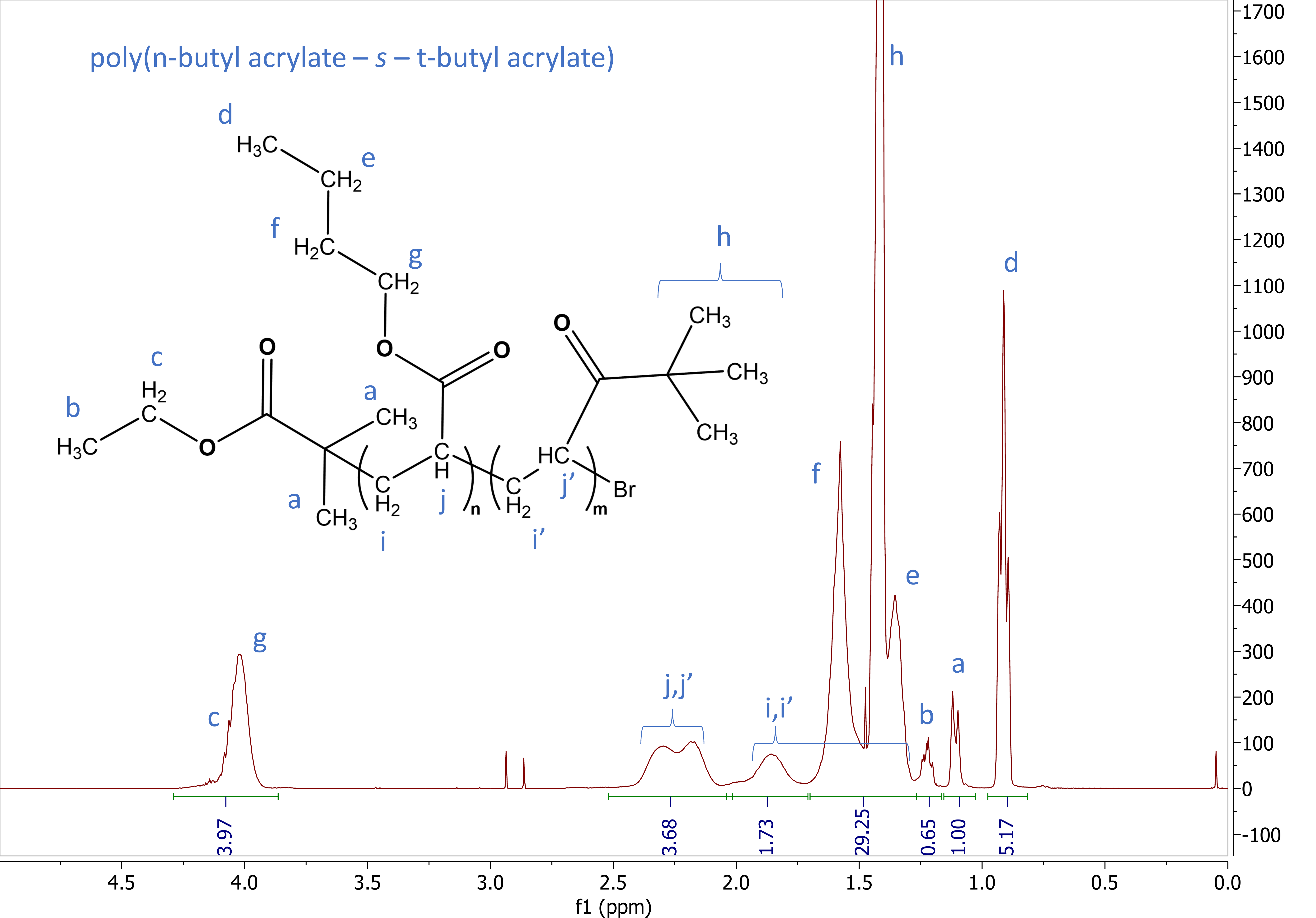}}}%
    \caption{\textbf{$^1$H-NMR spectra} of \textbf{(a)} the clickable coumarin dye and \textbf{(b)} poly(n-butyl acrylate-$s$-t-butyl acrylate).}
    \label{fig:coumarin_pol_NMR}
\end{figure}

$^1$H NMR ($400$ MHz, \ce{CDCl3}), shown in Fig.~\ref{fig:coumarin_pol_NMR}(a), was carried out to confirm the presence of the desired compound.

\subsubsection{SET-LRP synthesis of poly(n-butyl acrylate-$s$-t-butyl acrylate)}
The polymerization synthesis to form our initial protected polymer follows from van Ravensteijn et al.\cite{bas_stars_supp}.

To a $20$ mL vial \ce{CuBr2} ($17.2$ mg, $0.077$ mmol, $0.05$ eq.) and \ce{Me6TREN} ($74$ \textmu L, $0.28$ mmol, $0.18$ eq.) were added and then dissolved, with the aid of sonication, in $4.5$ mL of TFE. Then n-butyl acrylate ($2.79$ mL, $19.2$ mmol, $12.5$ eq.), \emph{t}-butyl acrylate ($2.79$ mL, $19.2$ mmol, $12.5$ eq.) and EBIB ($0.023$ mL, $1.54$ mmol, $1$ eq.) were added to the vial. A $10$cm length of copper wire, which was previously etched in concentrated \ce{HCl}, washed with acetone and dried, was wrapped around a stirrer bar and added to the vial. The solution was sealed with a rubber septum and then purged by bubbling dry \ce{N2} through it for $15$ min. The reaction was allowed to stir at RT for $5$ hours and $40$ min. To quench the reaction the mixture was exposed to air and the copper wire laden stirrer bar was removed. The reaction mixture was diluted with DCM and then run through a short basic alumina column to remove any copper salts. The now colorless solution was concentrated \emph{in vacuo}. The polymer was then dissolved in a minimal amount of diethyl ether (approx. $4$ mL) and precipitated into a $4:1$ methanol to water mixture. This process was repeated a further two times. Finally the polymer was dissolved in DCM, dried with excess \ce{MgSO4} and concentrated \emph{in vacuo}. Average conversions for this polymerization were $80\%$ and average final yields were $60\%$.        

$^1$H NMR ($400$ MHz, CDCl3), shown in Fig.~\ref{fig:coumarin_pol_NMR}(b) for PBA-AA$^c$, was carried out to confirm the presence of the desired compound. Note the broadness of the peaks indicating a polymeric compound.

\subsubsection{Azidification of poly(n-butyl acrylate-$s$-t-butyl acrylate)}
We can take advantage of the good end group retention of SET-LRP to add an azide moeity to the end of the polymer chains \cite{SET_review_supp}. This will allow us to \emph{click} on the previously synthesized alkyne-coumarin dye. A procedure from Anastasaki et al. \cite{end_gruop_mod_supp} was the basis of the following procedure.

poly(n-butyl acrylate-$s$-t-butyl acrylate) ($3.2$ g, $1.25$ mmol, $1$ eq.) was dissolved in $30$ mL of DMF. Then \ce{NaN3} ($0.813$ g, $12.5$ mmol, $10$ eq.) was added to the solution and the whole reaction mixture was stirred at RT overnight. The now cloudy solution was diluted with $60$ mL of \ce{CHCl3} and extracted with $400$ mL of water. The organic phase was separated and the water phase was washed two times with $100$ mL of \ce{CHCl3}. The combined organic phases were then washed two times with $100$ mL of water. Finally the organic phase was dried using \ce{MgSO4} and concentrated \emph{in vacuo}. The average yields for this procedure were approx. $60\%$.

$^1$H NMR ($400$ MHz, \ce{CDCl3}) for PBA-AA$^c$, was carried out to attempt to confirm the presence of the desired compound. Only a small change in the NMR spectra is expected, namely the shift in the proton resonance of the terminal C-H group. Small peaks are found at $3.6$-$3.8$ ppm which were not present before the azidification. We may attribute this to the azidification of the chain-end but the spectra is too noisy to qualitatively look at coupling efficiencies. 

\subsubsection{Coumarin dye - polymer chain end \emph{click} reaction}
The previously synthesized alkyne-coumarin dye can now be coupled to the end of the polymer. The following procedure was adapted from Honda \cite{Honda_supp}.

\ce{CuBr} ($0.252$ g, $1.76$ mmol, $1.5$ eq.) and PMDETA ($0.368$ mL, $1.76$ mmol, $1.5$ eq.) were dissolved in $30$ mL of DMF with the aid of sonication. This resulted in a dark green solution. Then, both of the previously synthesized 7-propargyloxy-4-methylcoumarin ($0.502$ g, $2.34$ mmol, $2$ eq.) and azidified poly(n-butyl acrylate-$s$-t-butyl acrylate) ($3.0$ g, $1.17$ mmol, $1$ eq.) were dissolved in the solution. The solution was sealed with a rubber septum and then purged by bubbling dry \ce{N2} through it for $45$ min. The solution was allowed to stir overnight. The whole reaction mixture was diluted in acetone and then passed through a silica column to remove the copper salts. The solution was then concentrated \emph{in vacuo}. The remaining sample was dissolved in $15$ mL of THF and then precipitated into a $4:1$ methanol to water mixture, with a ratio of $7.5$mL of THF solution to $40$ mL of methanol-water. This precipitation was repeated a further two times dissolving the precipitated polymer in a minimal amount of THF (approx. $2$ mL). The polymer was then dissolved in $20$ mL of DCM and the solution was dried over excess \ce{MgSO4}. Finally the solution was concentrated \emph{in vacuo} to yield a clear polymer.
The average yields for this procedure were around $65\%$ with coupling efficiencies (calculated using $^1$H NMR) of the dye to the end of the polymer of approximately $60\%$\footnote{This efficiency is calculated with respect to the chain end, therefore assuming there is full azide coverage of it.}.

\begin{figure}
    \centering
    \subfloat[\centering $^1$H NMR ($400$ MHz, \ce{CDCl3}) of the coumarin-modified poly(n-butyl acrylate-$s$-t-butyl acrylate) ]{{\includegraphics[width=0.8\textwidth]{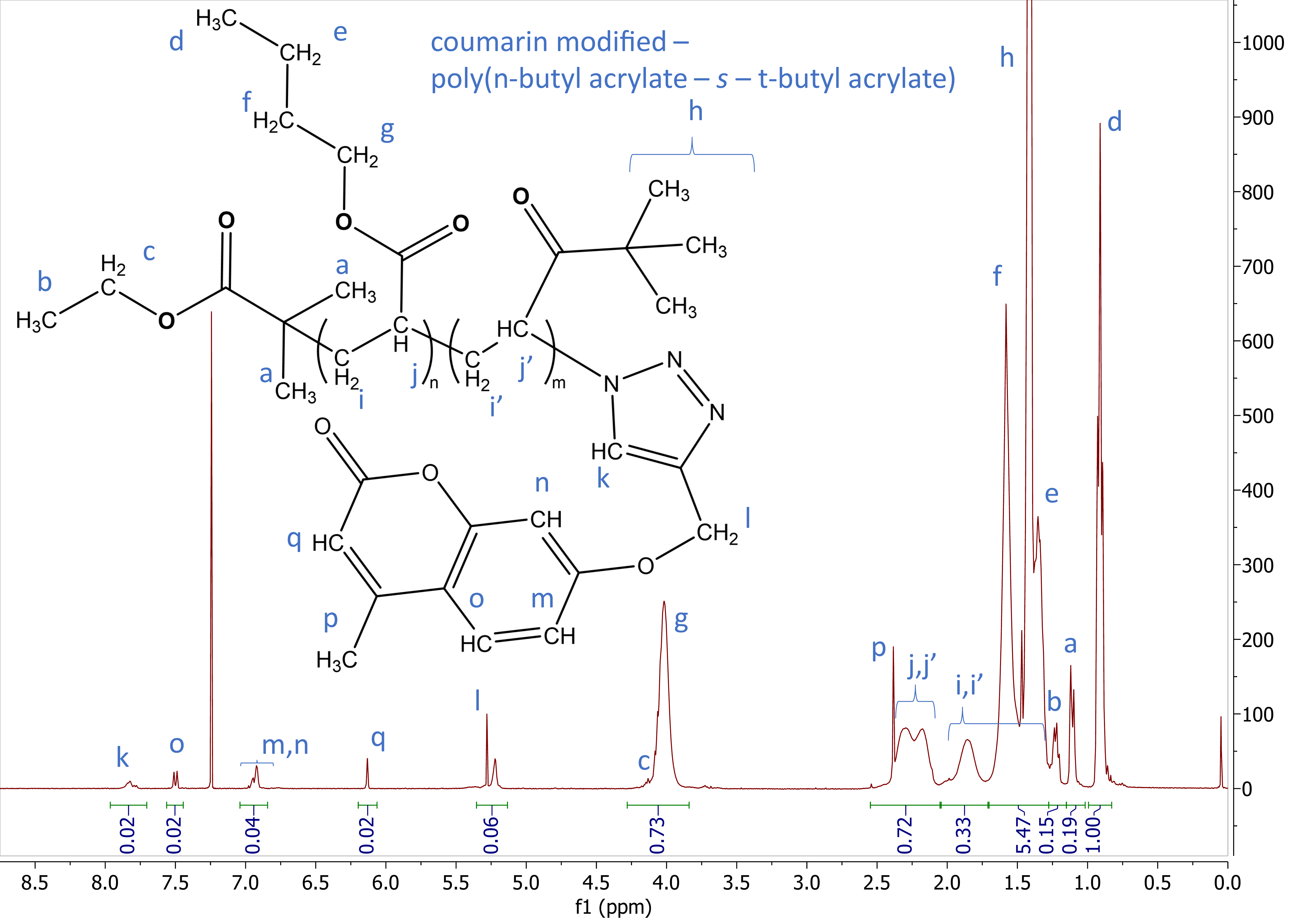}}}%
    \\
    \subfloat[\centering $^1$H NMR ($400$ MHz, DMSO-d6) of the coumarin-modified poly(n-butyl acrylate-$s$-acrylic acid)]{{\includegraphics[width=0.8\textwidth]{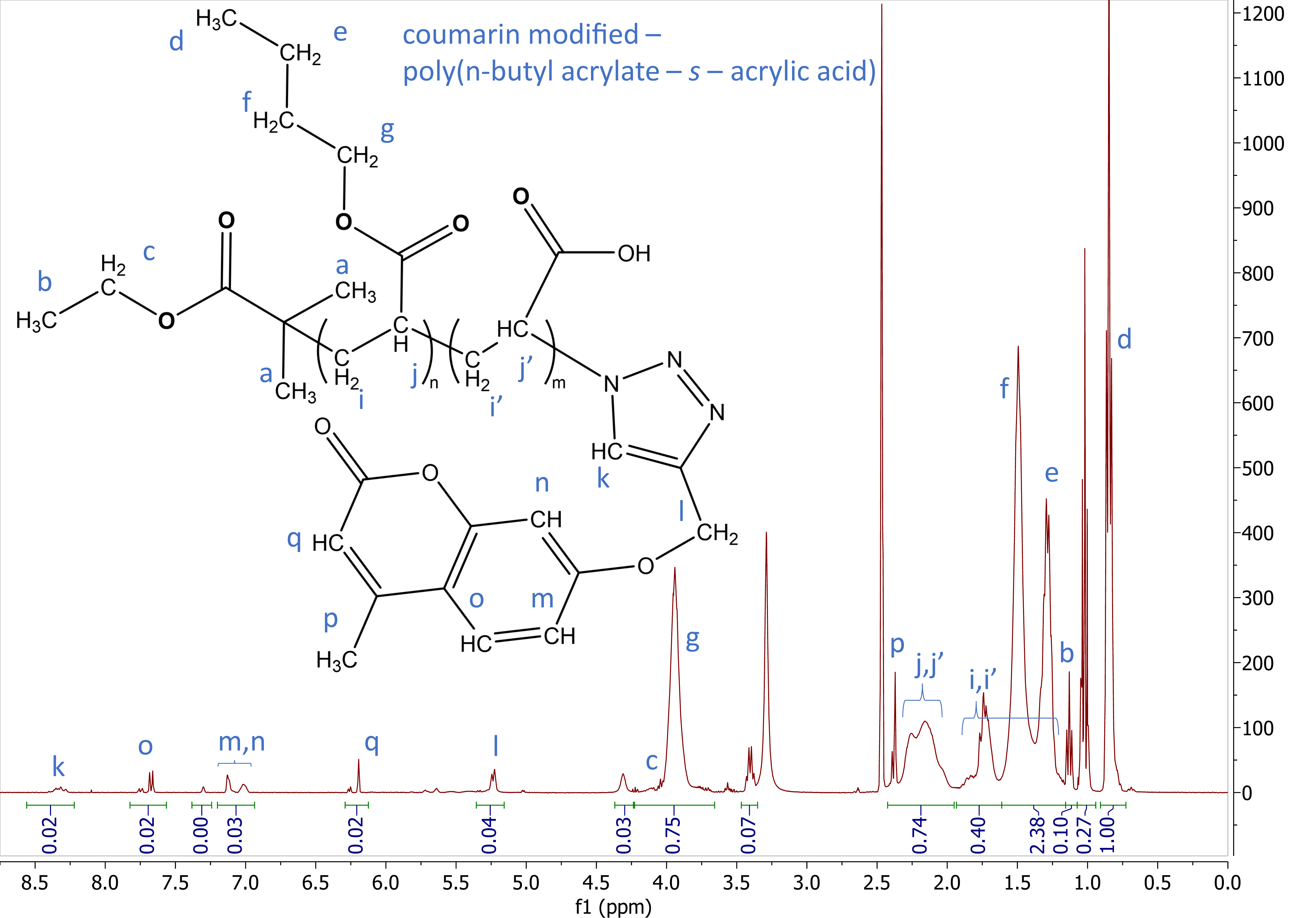}}}%
    \caption{\textbf{$^1$H-NMR spectra} of the protected and deprotected coumarin-modified random copolymer HPE.}
    \label{fig:prot_deprot_NMR}
\end{figure}

$^1$H NMR ($400$ MHz, \ce{CDCl3}) for PBA-AA$^c$, shown in Fig.~\ref{fig:prot_deprot_NMR}(a), was carried out to confirm the presence of the desired compound.

\subsubsection{poly(n-butyl acrylate-$s$-t-butyl acrylate) deprotection}
The final step to yield a hydrophobic polyelectrolyte is to deprotect the t-butyl ester groups on the polymer chain into carboxylic acid groups. The procedure described below was used to synthesize the final PBA-AA$^c$. Synthesis of PBA-AA$^0$ followed the same procedure at a $2/3$ scale.

The, still protected, coumarin-modified polymer ($2.1$ g, $0.82$ mmol, $1$ eq.) was dissolved in $15$mL of DCM in a round-bottom flask. Then trifluoroacetic acid ($6.3$ mL, $82$ mmol, $100$ eq.)\footnote{The amount of trifluoroacetic acid to be added was calculated as 10x the amount of t-butyl acrylate groups in solution. There were approximately $10$ \emph{t}-butyl acrylate groups per polymer chain, therefore leading to an approximately 100x equivalency. This could also simply be stated as excess trifluoroacetic acid.} was added and the solution was allowed to stir overnight. The solution was then concentrated \emph{in vacuo}. The remaining sample was dissolved in $20$ mL of THF and then precipitated into a cold hexane, with a ratio of $5$ mL of THF solution to $40$ mL of hexane. This was repeated a further three times using a minimal amount of THF to dissolve the precipitated polymer each time. Finally, the precipitated polymer was dissolved in excess ethanol and concentrated \emph{in vacuo}. The average yields for this procedure were approx. $65\%$  with deprotection efficiencies (calculated using $^1$H NMR) being close to $100\%$. Removal of any excess trifluoroacetic acid was qualitatively confirmed using $^{19}$F NMR.

\subsubsection{PBA-AA polymer characterization}
$^1$H NMR ($400$ MHz, DMSO-d6), shown in Fig.~\ref{fig:prot_deprot_NMR}(b) for PBA-AA$^c$, was carried out to confirm the presence of the desired compound and SEC was used to investigate the length distribution.

The NMR spectra shows the disappearance of the t-butyl group resonance at around $1.4$ ppm, indicating the full deprotection of the polymer. Overall, the appearance of the spectra is similar to the protected polymer indicating that the rest of the structure of the molecule has remained unchanged.

Unlike with the PAHA polymer discussed earlier in this section it is possible to extract the degree of polymerization of the polymer directly from the NMR spectrum of the initial protected polymer (Fig.~\ref{fig:coumarin_pol_NMR}(b)). Using the reference peak at $1.1$ ppm, which we attribute to the two methyl groups on the initiator moiety, and comparing this integration to the integrated peak of the backbone \ce{C-H_2} protons on the polymer allows us to calculate an estimated degree of polymerization for the whole polymer. Further, using the resonance peak of the terminal \ce{C-H3} group at around $0.9$ ppm on the n-butyl monomer allows us to calculate the ratio of t-butyl to n-butyl in the initial protected polymer. It is expected that this ratio will be retained for any subsequent polymer modifications. The calculated values for the total degree of polymerization, and the individual n- and t-butyl degrees of polymerization for both polymer variants synthesized are summarized in Table \ref{tab:PBAA_results}

\begin{table}
    \centering
    \begin{tabularx}{\textwidth}{|X|X|X|X|X|}
     \hline
     Sample & Total DP (NMR) & n-butyl acrylate DP (NMR) & t-butyl acrylate / acrylic acid DP (NMR) & \emph{\DJ}{} (SEC)\\
     \hline
     PBA-AA$^c$  &  $22.1$  &  $10.3$  &  $11.8$ &  $1.09$ \\
     PBA-AA$^0$  &  $20.7$  &  $9.9$  &  $10.6$ &  $1.11$\\
     \hline
     \end{tabularx}
    \caption{Summary of the estimation of the degree of polymerization and dispersity of the PBA-AA polymer samples.}
    \label{tab:PBAA_results}
\end{table}

Analysis of the length dispersity (\emph{\DJ}) of this polymer is essential for the discussions we carry out in the results section of this study. The SEC of the PBA-AA$^c$ sample carried out in ammonium acetate buffer ($300$ mM, pH $9$) shows a dispersity value of around $1.09$, with a marked tail. It is worth noting however that the chemical dispersity of this polymer, due to the two similarly reactive monomers which compose it, might lead to a spread in the elution times during SEC. This is a result of the fact that the two randomly arranged groups are quite chemically distinct and therefore their interaction with the aqueous buffer will differ, leading to different apparent weights. If we assume that an identical polymerization of the homopolymer poly(n-butyl acrylate) has the same length dispersity, previous experiments not detailed here show a dispersity value of $1.04$ and a SEC trace without a tail.\footnote{It is worth calculating the difference in the standard deviation of the distributions that this, apparently small, difference in \emph{\DJ}~ actually represents. The standard deviations for \emph{\DJ}~ values of $1.04$ and $1.09$ are approx. $3.46$ and $6$ respectively, where have assumed an average length of $20$ units for the polymer. Therefore the change in the implied distribution is not negligible.} 

\begin{figure}
    \includegraphics[width=0.8\textwidth]{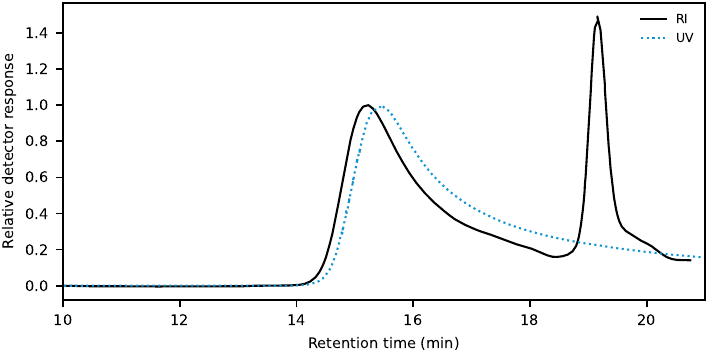}
    \caption{\textbf{SEC chromatograms} of the final product of the PBA-AA$^c$ synthesis.}
    \label{fig:PBAA_GPC}
\end{figure}

\subsection{Measurement of HPE partitioning and ionization behavior}\label{data_analysis_HPE}

The individual data treatment for each type of experiment was kept as consistent as possible to allow for a fair comparison. The individual raw data is now presented for each of the experiments shown in the main text accompanied by a general explanation of the data analysis carried out. 

\subsubsection{Measuring the fraction of chains in the hydrophobic phase}\label{SI_fraction_data}
Data treatment for the extraction of data from UV-Vis absorbance measurements is straightforward. First a blank measurement of the solvent is taken, which in this case is water-saturated pentanol. Then measurements of all the samples are carried out in triplo (or duplo). After subtraction of the blank signal and a further background subtraction is carried out, the relative intensity of the characteristic wavelength of the particular absorbing moiety of each polymer type is normalized with respect to the stock solution of the polymer used for each experiment. This allows for the identification of any anomalous data. The data is then rescaled to set the highest absorption value, which will be at the lowest $pH$ values, to $1$ as we expect all of the chains to reside in the hydrophobic phase at these $pH$ values. The lowest fraction value, if there is a plateau in the data, are set to $0$. A linear relationship between concentration and absorbance (Beer-Lambert law) has been assumed throughout. An example of this procedure is shown for one of the sets of raw data following from this introduction.

\begin{figure}
    \centering
    \includegraphics{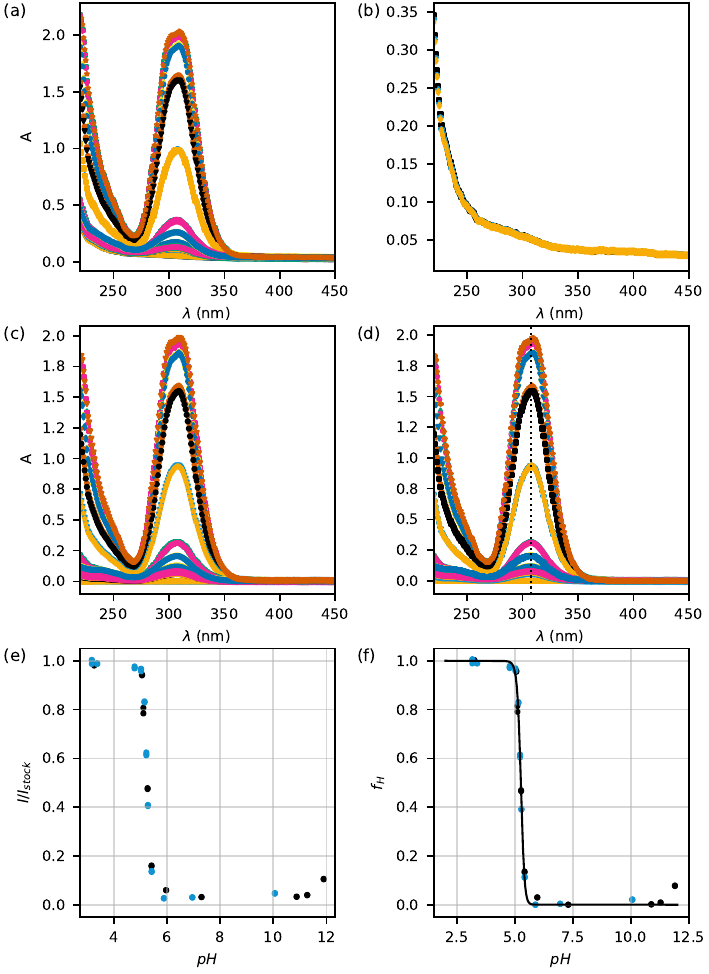}
    \caption{\textbf{Example data treatment for UV-Vis absorption data.} \textbf{(a)} Raw UV-Vis data from the hydrophobic phase from a single experimental run.\textbf{(b)} Spectrum of the solvent used as hydrophobic phase. \textbf{(c)} Spectra of the hydrophobic phases with the solvent signal removed. \textbf{(d)} Spectra after background removal. \textbf{(e)} Normalized intensity of the chosen wavelength with respect to the stock solution of polymer for both experimental runs. \textbf{(f)} Final data after rescaling. }
    \label{fig:PAHA-20_fraction_treatment}
\end{figure}

\subsubsubsection{PAHA (DP=$18$): $f_H$ from titration experiment} \label{PAHA_f_h_data}
Figure~\ref{fig:PAHA-20_fraction_treatment} (a) shows the raw UV-Vis absorbance data for the pentanol phase of a two-phase pentanol and water system which is titrated as described in the last section. The experimental data is shown in \ref{fig:HPE_fig}(b). 

Part (b) of the figure shows the absorbance spectrum for the water saturated pentanol used as the hydrophobic phase in this experiment. Part (c) shows the spectra after the solvent spectrum is removed and part (d) after the background signal found between $375$ and $400$ nm is removed. The vertical dashed line indicates the wavelength ($308$ nm) which is used as the reference wavelength to calculate relative concentrations with respect to the stock solution. These relative concentrations are plotted in part (e). Finally we subtract the minimum absorbance and renormalize to the highest one is part (f).  

\subsubsubsection{PBA-AA$^c$: $f_H$ from buffered experiment}\label{PBAA_f_h_data}
Figure~\ref{fig:appendix-pba-aa}, presents in the top four panels the raw UV-Vis spectra, the blank solvent spectrum, the raw data and the scaled data for the buffered partitioning experiment of PBA-AA$^c$ detailed in Fig.~\ref{fig:HPE_fig_2}. It is worth noting here that the high baseline at high $pH$ values. It is possible that there is residual dye, not coupled to the polymer chain end, that leads to this high baseline. In this case a series of refractive index measurements could determine if the baseline is due to dye impurities or the polymer itself.

\subsubsection{Measuring the HPE ionization fraction from titration experiments}\label{SI_titration}
The procedure to extract the ionization fraction of a HPE from the titration of hydrophobic polyelectrolytes relies on the necessary electroneutrality of the solution. This condition allows for the following equality to be set up:
\begin{align} \label{eq:titration_eq}
\ce{[ionized acidic groups]=[H+] - [OH^-] + [K+]_{added} - [Cl^-]_{initial}}
\end{align}
where the initial concentration of \ce{Cl-} originates from the initial acidification of the aqueous phase of the two-phase system using \ce{HCl} and the added concentration of \ce{K+} is the concentration of potassium ions added when titrating using \ce{KOH}. The concentration of \ce{H+} and \ce{OH-} is found from the $pH$ of the aqueous solution and follows from $[\ce{H+}]=10^{-pH}$ and $pK_w=pH+pOH$.

Similarly to the procedure to find the fraction of chains in the hydrophobic environment, we rely on scaling of the data to eliminate artifacts that arise throughout the experiments. In this case it is assumed that at the starting point of the titration ($pH~3$) all of the polymer resides in the oil phase (following from the UV-Vis absorption data) and therefore no ionization should be present. Although, a priori, it is possible to estimate the number of ionizable groups added to the experiment system, the accuracy of this prediction is not very high. This is due to the potential residual solvent and/or  water swelling the polymer samples used, which skew any potential calculation relying on masses. Therefore the values of the ionization fraction at high $pH$ values are also set to $1$ as to reflect the expect full ionization.

\begin{figure}[H]
    \centering
    \includegraphics{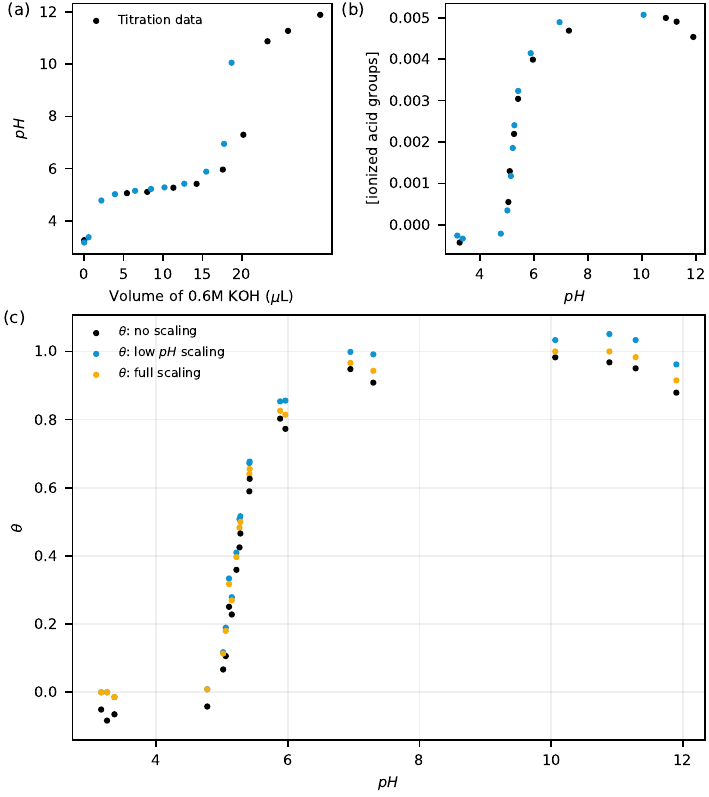}
    \caption{\textbf{Example data treatment for PAHA titration data.} \textbf{a)} Raw titration data for both experimental repeats. \textbf{b)} Concentration of ionized sites calculated from Eq.~(\ref{eq:titration_eq}). \textbf{c)} Ionization fraction data calculated with three different scaling methods: no scaling, scaling to force no ionization at the lowest $pH$ value and scaling that additionally sets the ionization to $1$ for the highest value of the ionization.}
    \label{fig:PAHA-20_ionization_treatment}
\end{figure}

\subsubsubsection{PAHA (DP$=18$): $\theta$ from titration data}\label{PAHA_theta_data}
Figure \ref{fig:PAHA-20_ionization_treatment} illustrates the procedure described above hor the PAHA (DP$=18$) titration presented in Fig.~\ref{fig:HPE_fig}.

\subsubsubsection{PBA-AA$^0$: titration and buffered experiments}\label{PBAA_theta_data}

\begin{figure}[H]
    \centering
    \includegraphics{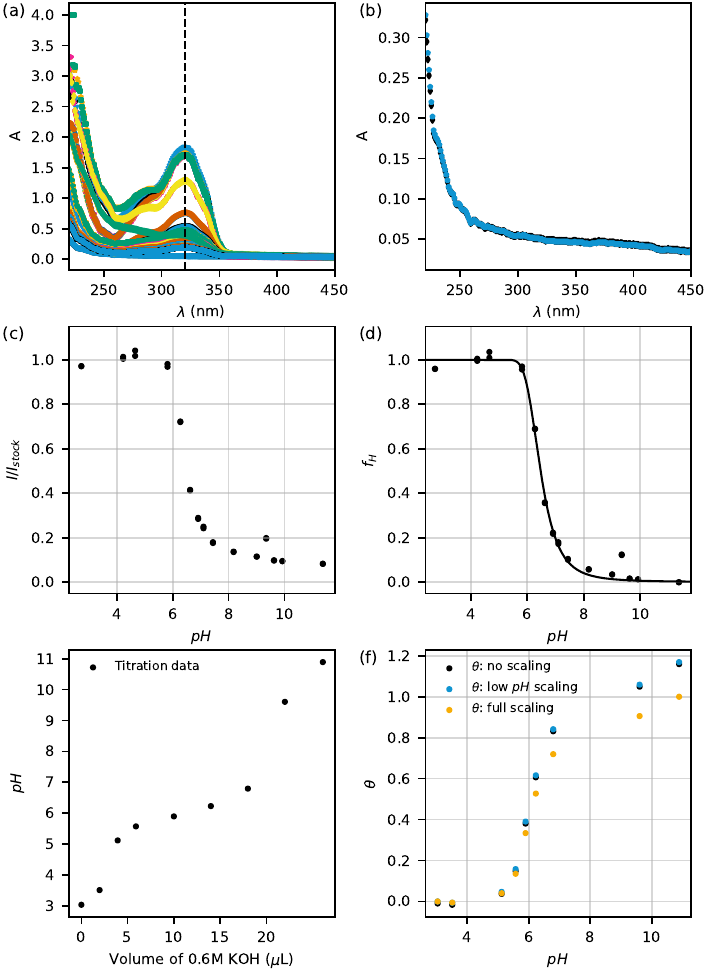}
    \caption{\textbf{Raw data for PBA-AA partitioning.} \textbf{a)} Raw UV-Vis data of PBA-AA$^c$ from the hydrophobic phase of a buffered two-phase oil and water system. \textbf{b)} Spectrum of the solvent used as hydrophobic phase. \textbf{c)} Normalized intensity of the chosen wavelength with respect to the stock solution of polymer. \textbf{d)} Final $f_H$ data after rescaling. \textbf{e)} Raw titration data of PBA-AA$^0$ in a two-phase oil and water system. \textbf{f)} Ionization fraction data calculated with three different scaling methods: no scaling, scaling to force no ionization at the lowest $pH$ value and scaling that additionally sets the ionization to $1$ for the highest value of the ionization.}
    \label{fig:appendix-pba-aa}
\end{figure}

Figure~\ref{fig:appendix-pba-aa} presents, in the bottom two panels, the data treatment for the titration experiments carried out on PBA-AA$^0$ detailed in Fig.~\ref{fig:HPE_fig_2}. Of note is this case is the lack of plateau at high values of ionization, as well as the values over unity for the unscaled ionization. There are no repeats carried out for this set of data, and therefore it is possible there is an erroneous experimental offset for the measurement of the volume of \ce{KOH} added which we are not aware of. The transition-$pH$ value does match the independent buffered experiment so there is a qualitative agreement with the buffered system but repeats of the experimental runs are necessary to confidently describe the system.

\section{Oligomeric metal chelates (OMC)} \label{OMC_SI}

\subsection{Materials}

opper (II) bromide (CuBr\textsubscript{2}, 99\%), tris[2-(dimethylamino)ethyl] amine (Me\textsubscript{6}TREN, 97\%), ethyl $\alpha$-bromoisobutyrate (EBiB, 98\%), silica gel (high-purity grade (Davisil grade 633), pore size 60\AA, 200-425 mesh particle size), 2,2':6',2"-terpyridine (T, 98\%), 4'-chloro-2,2':6',2"-terpyridine (Cl-T, 99\%), methanol-d\textsubscript{4} (99.8\%), dimethyl sulfoxide  (DMSO, anhydrous 99.9\%), iron (II) chloride tetrahydrate (FeCl\textsubscript{2}, 99\%), L-ascorbic acid (L-AA), dimethyl sulfoxide-d\textsubscript{6} (DMSO-d\textsubscript{6}, 99.9\%), ethanolamine (99.5\%), N-(3-dimethylaminopropyl)-N'-ethylcarbodi-imide hydrochloride (EDC-HCl, 98\%), N,N-dimethyl- formamide (DMF, 99.8\%), and N-hydroxysuccinimide (NHS, 98\%) were purchased from Sigma Aldrich (St. Louis, MO, USA). 2,2,2-trifluoroethanol (TFE, 99\%), potassium hydroxide powder (KOH, $\sim$ 85\%), sodium chloride (NaCl) and sodium hydroxide pellets (99\%) were purchased from Merck (Burlington, MA, USA). 

\noindent Dichloromethane (DCM, 99\%), methanol absolute HPLC (MeOH, 99.9\%), tetrahydrofuran (THF, 99.8\%), diethyl ether (Et\textsubscript{2}O, anhydrous 99.5\%), chloroform HPLC (99.9\%) and acetonitrile (ACN, HPLC-R) were purchased from Biosolve B.V. (Valkenswaard, The Netherlands). Copper wire (Cu(0), d = 0.25mm, 99.99\%) was purchased from Alfa Aesar (Haverhill, MA, USA). Hydrochloric acid (HCl, 37\%) was purchased from Acros Organics (Geel, Belgium). Ethanol absolute (100\%), acetone (100\%), n-hexane (99\%) and triethylamine (TEA, high purity grade) were purchased from VWR International (Radnor, PA, USA). \textit{tert}-butyl acrylate (\textit{t}BA) and n-butyl acrylate (nBA) were purified (using a 1:1 Al\textsubscript{2}O\textsubscript{3}:Silica column) and stored at 4 \degree C. Trifluoroacetic acid (TFA, 99\%), aluminum oxide (98\%) and sodium sulfate (Na\textsubscript{2}SO\textsubscript{4}, 99\%) were purchased from Honeywell (Charlotte, NC, USA). Chloroform-d1 (99\%) was purchased from Carl Roth (Karlsruhe, Germany). Dimethyl sulfoxide (DMSO, 99.7\%) was purchased from Thermo Fisher Scientific (Waltham, MA, USA). The Milli-Q (MQ) water used was deionized by a Millipore Synergy water purification system (Merck Millipore, Billerica, MA, USA). 

\subsection{Synthesis of the terpyridine-functionalized polymer (PT)}\label{SynthesisPT16}

\begin{figure}[H]
    \centering
    \includegraphics[width=\linewidth]{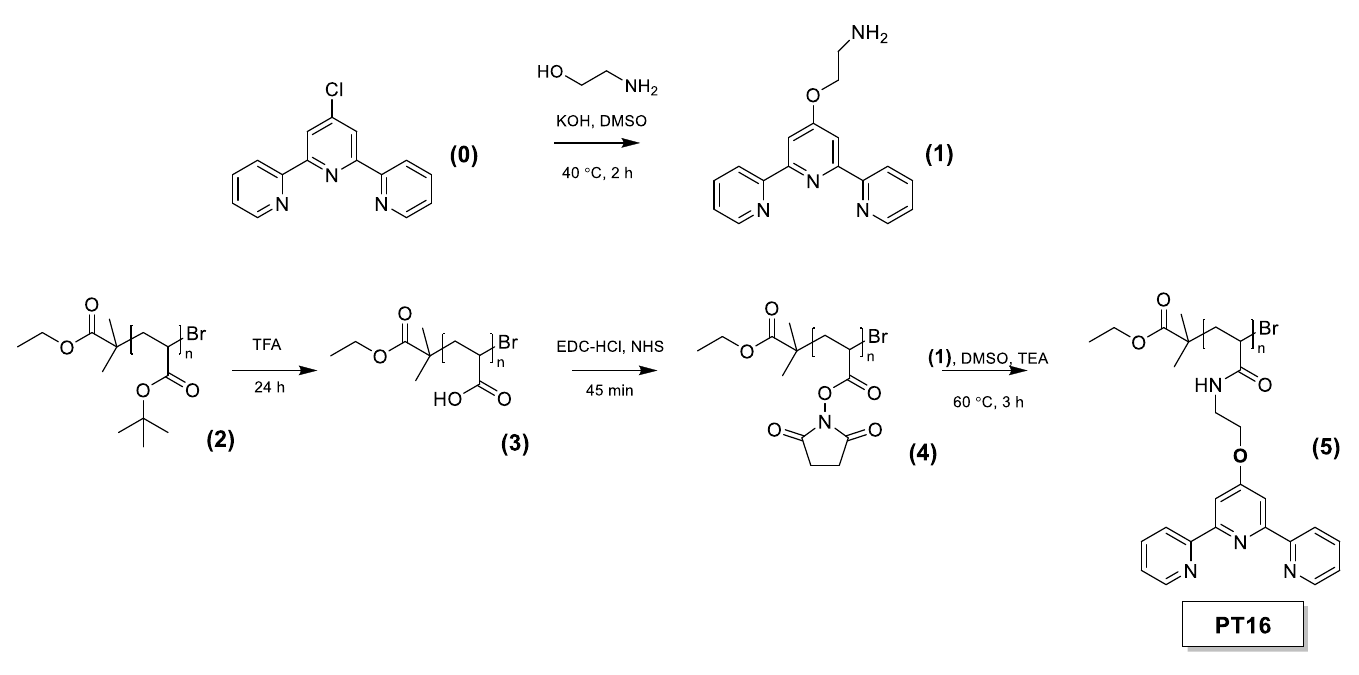}
    \caption{\textbf{Synthesis overview.} Steps of synthesizing the terpyridine-functionalized polymer with 16 repeating units (PT16).}
    \label{fig:synthesisoverview}
\end{figure}

The synthesis of the terpyridine-functionalized polymers was done in multiple steps. As an overview, illustrated in Figure \ref{fig:synthesisoverview}, the synthesis starts with synthesizing poly (\textit{tert} - butyl acrylate) and deprotecting it into poly(acrylic acid) (PAA). 

\noindent Subsequently, EDC/NHS activation was done on the carboxylic acid groups of PAA in order to synthesize poly (N-hydroxysuccinimide) (PNHS). To functionalize (PNHS) with terpyridine, Cl-T was synthesized with ethanolamine to yield an amine functionalized terpyridine, 2-(2,2':6',2"-terpyridine-4'-yloxy) ethylamine (ET), which was necessary in order to substitute the NHS groups on the polymer via nucleophilic substitution. Eventually, the terpyridine-functionalized polymers were synthesized and coded as PT16 with structure (5), shown in Figure \ref{fig:synthesisoverview}. Number 16 shows the number of monomer units. After each synthesis step, the product was characterized using the same instruments as described the Instrumentation Section for the HPE study.

\subsubsection{Synthesis of 2-(2,2':6',2"-terpyridine-4'-yloxy) ethylamine (ET)  (1)}

\begin{figure}[H]
    \centering
    \includegraphics[width=0.75\linewidth]{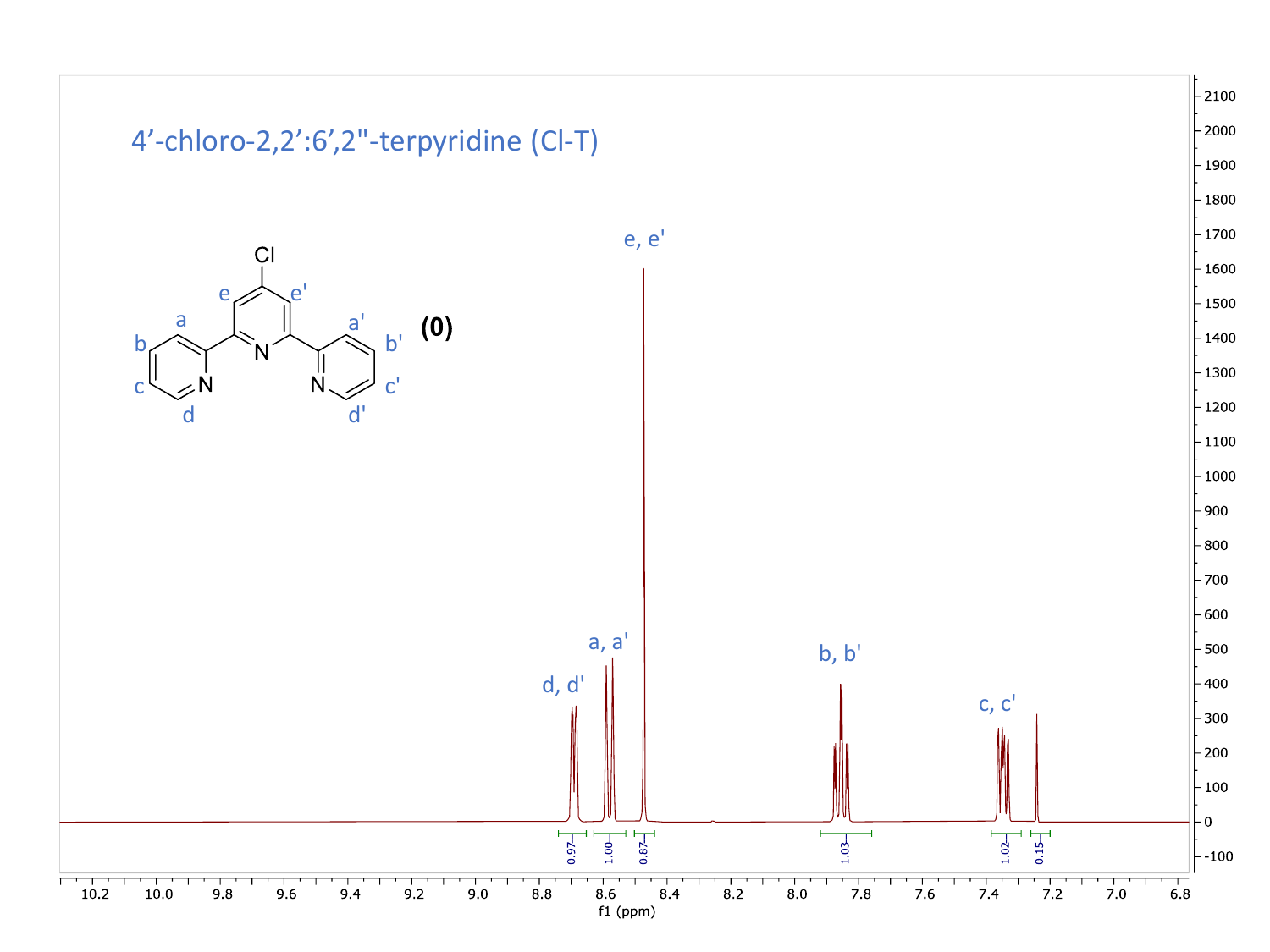}
    \caption{\textbf{\textsuperscript{1}H-NMR spectrum} of Cl-T (0).}
    \label{fig:NMR0}
\end{figure}

\begin{figure}[H]
    \centering
    \includegraphics[width=0.75\linewidth]{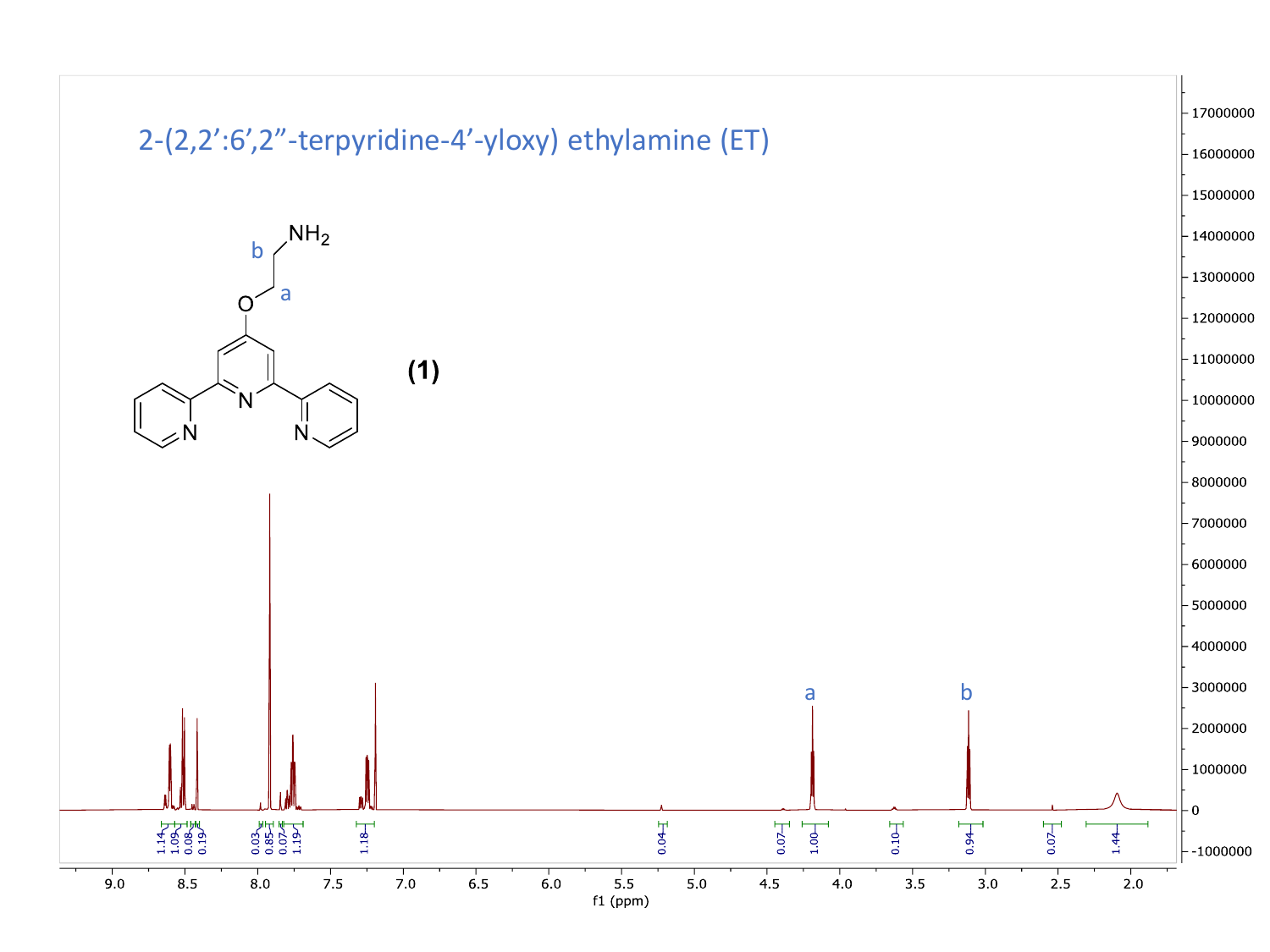}
    \caption{\textbf{\textsuperscript{1}H-NMR spectrum} of ET (1).}
    \label{fig:NMR1}
\end{figure}

Cl-T (267.71 mg, 1 mmol, \textsuperscript{1}H-NMR is shown for comparison in Figure\ref{fig:NMR0}) and ethanolamine (67.17 $\mu$l, 1.1 mmol) were added to a suspension of powdered KOH (280.55 mg, 5 mmol) in DMSO (5 ml) and stirred at 40 \degree C for 2 h. The reaction mixture was then added to 40 ml of DCM and washed with MQ water (3 $\times$ 40 ml) by liquid-liquid extraction. The DCM solution was dried over Na\textsubscript{2}SO\textsubscript{4} and the solvent was removed. 2-(2,2':6',2"- terpyridine-4'-yloxy) ethylamine (ET) was obtained as a light yellow solid and used subsequently without further purification (252.60 mg, 86.5\% yield).\textsuperscript{1}H-NMR (400 MHz in Chloroform-d\textsubscript{1}) is shown in Figure \ref{fig:NMR1}).

\subsubsection{SET-LRP Synthesis of poly(\textit{tert}-butyl acrylate) (P\textit{t}BA) (2)}

\begin{figure}[H]
    \centering
    \includegraphics[width=0.8\linewidth]{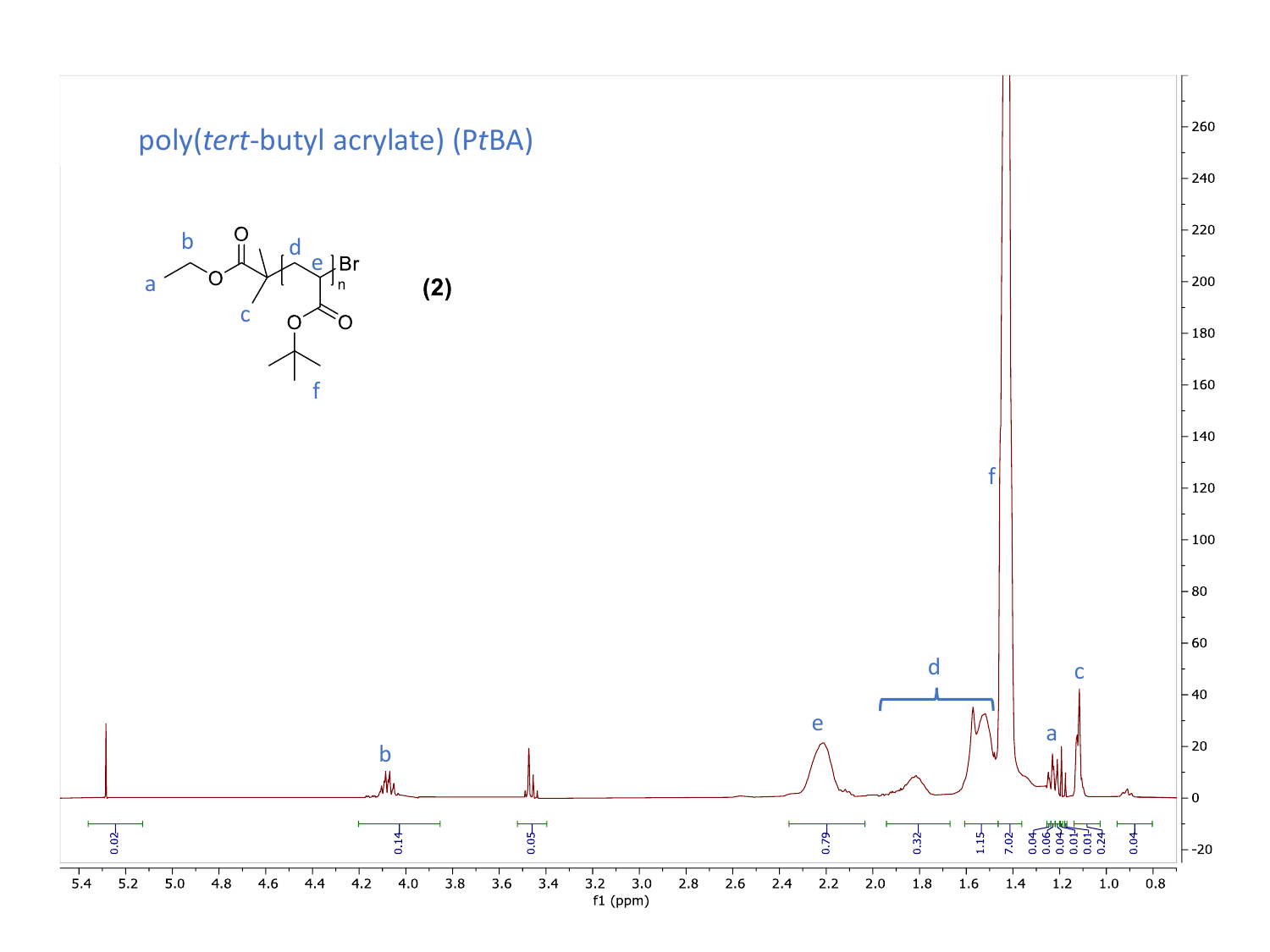}
    \caption{\textbf{\textsuperscript{1}H-NMR spectrum} of P(\textit{t}BA) (2).}
    \label{fig:NMR2}
\end{figure}

\begin{figure}[H]
    \centering
    \includegraphics[width=0.8\linewidth]{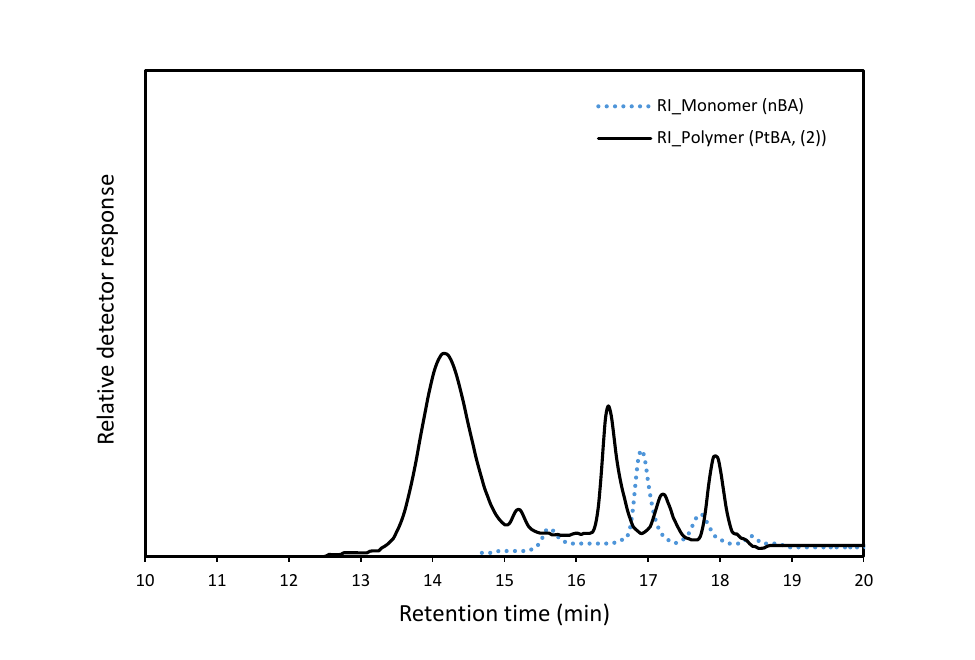}
    \caption{\textbf{SEC chromatograms} of P\textit{t}BA (2, solid black line), compared to BA (monomer, dotted blue line).}
    \label{fig:OMC_GPC}
\end{figure}

CuBr\textsubscript{2} (7.59 mg, 0.034 mmol, 0.05 equiv. per initiating sites), TFE (1 ml) and Me\textsubscript{6}TREN (33 $\mu$l, 0.123 mmol, 0.18 equiv. per initiating sites) were added in a 4-ml vial and the resulting solution was treated with ultrasonication for 30 minutes using a  CPX8800H ultrasonic cleaning bath (Branson Ultrasonics\textsuperscript{TM}, Brookfield, CT, USA). For Cu(0) source, 5cm of 0.25 mm thick Cu copper wire was cut and etched in HCl for 30-45 minutes. Afterwards, the wire was washed with acetone and dried with dust-free paper (three times). \textit{t}BA (2ml, 13.7mmol, 20 equiv.) was added, together with the initiator, EBiB (101 $\mu$l, 0.685 mmol, 1 equiv.) to the mixture and the reaction was activated by adding a magnetic stir bar wrapped by the copper wire. The reaction was degassed under N\textsubscript{2} flow for the first 20 min was left for 4 hours (in total). The product was then diluted with Et\textsubscript{2}O, and purified by passing through a 1:1 Al\textsubscript{2}O\textsubscript{3}:Silica column. Consequently, the product was dissolved in minimal amount of DCM and washed 4 times by a 4:1 MeOH:water solution by centrifugation at 3273 $\times$ g for 15 minutes using an Allegra X-12R Centrifuge (Beckman Coulter, Brea, CA, USA). Finally, the product was dried under N\textsubscript{2} flow and vacuum. The resulting material was a sticky viscous gel (80\% conversion, DPn $\approx$ 16).\textsuperscript{1}H-NMR (400 MHz in Chloroform-d\textsubscript{1}) and SEC chromatogram of the final product are shown in Figure \ref{fig:NMR2} and \ref{fig:OMC_GPC}, respectively.

\subsubsection{Deprotection of (2) into poly(acrylic acid) (PAA) (3)} \label{deprptectionof(2)to(3)}

\begin{figure}[H]
    \centering
    \includegraphics[width=0.8\linewidth]{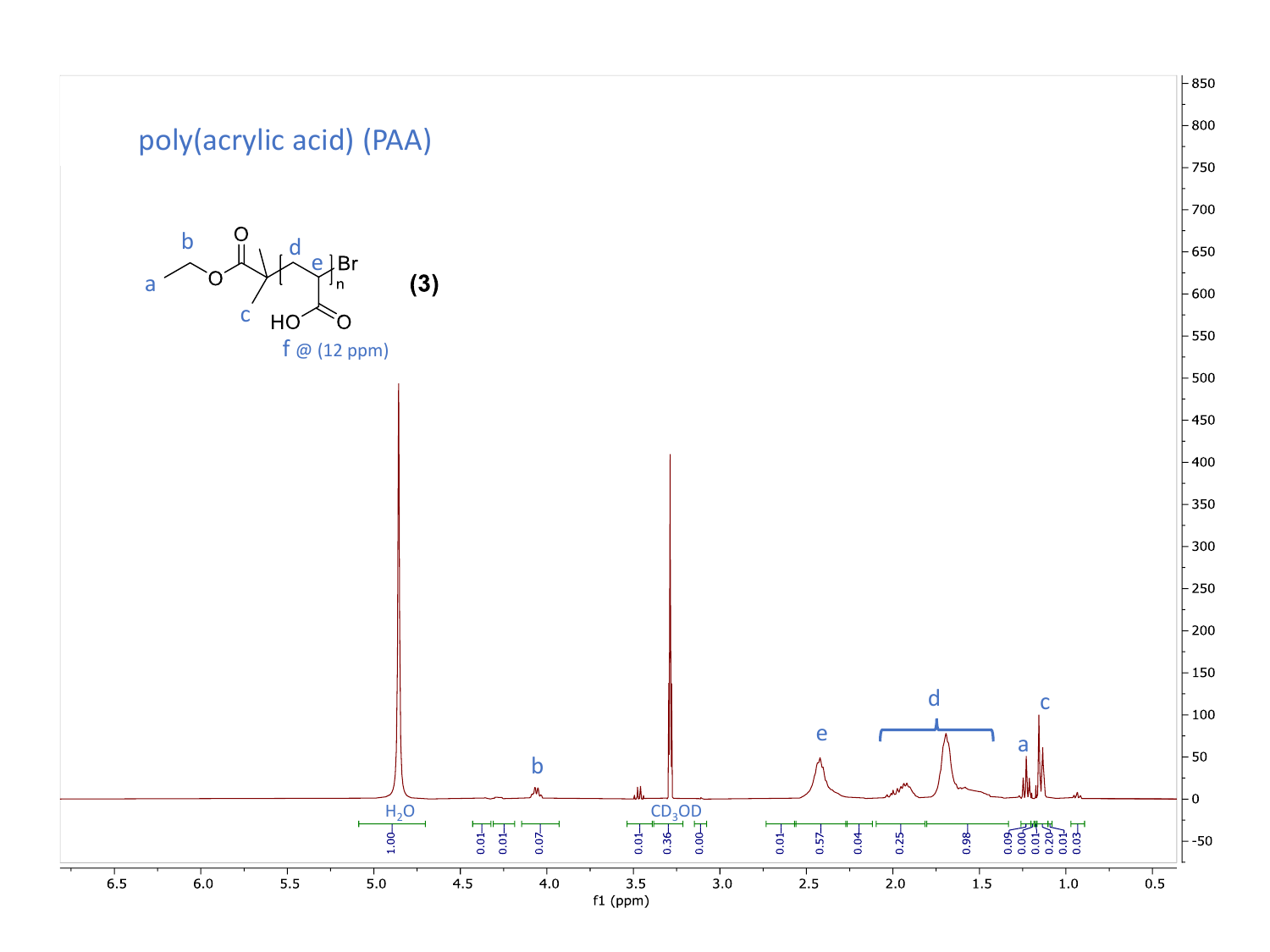}
    \caption{\textbf{\textsuperscript{1}H-NMR spectrum} of PAA (3).}
    \label{fig:NMR3}
\end{figure}

In a 4-ml vial, compound (2) (256.64 mg), TFA (2.6 ml) and a magnetic stir bar were added and left under stirring for 24 hours. Next, the mixture was diluted with minimal amount of DCM, added to an excess amount of Et\textsubscript{2}O and centrifuged at 3273 $\times$ g for 15 minutes. This was repeated three times. After centrifugation, the product was dried overnight and a white powder was obtained ($\approx$  80\% yield). \textsuperscript{1}H-NMR (400 MHz, methanol-d\textsubscript{4}) is shown in Figure \ref{fig:NMR3}.

\subsubsection{EDC/NHS Activation of (3) into PNHS (4)} \label{activationof(3)to(4)}

\begin{figure}[H]
    \centering
    \includegraphics[width=0.8\linewidth]{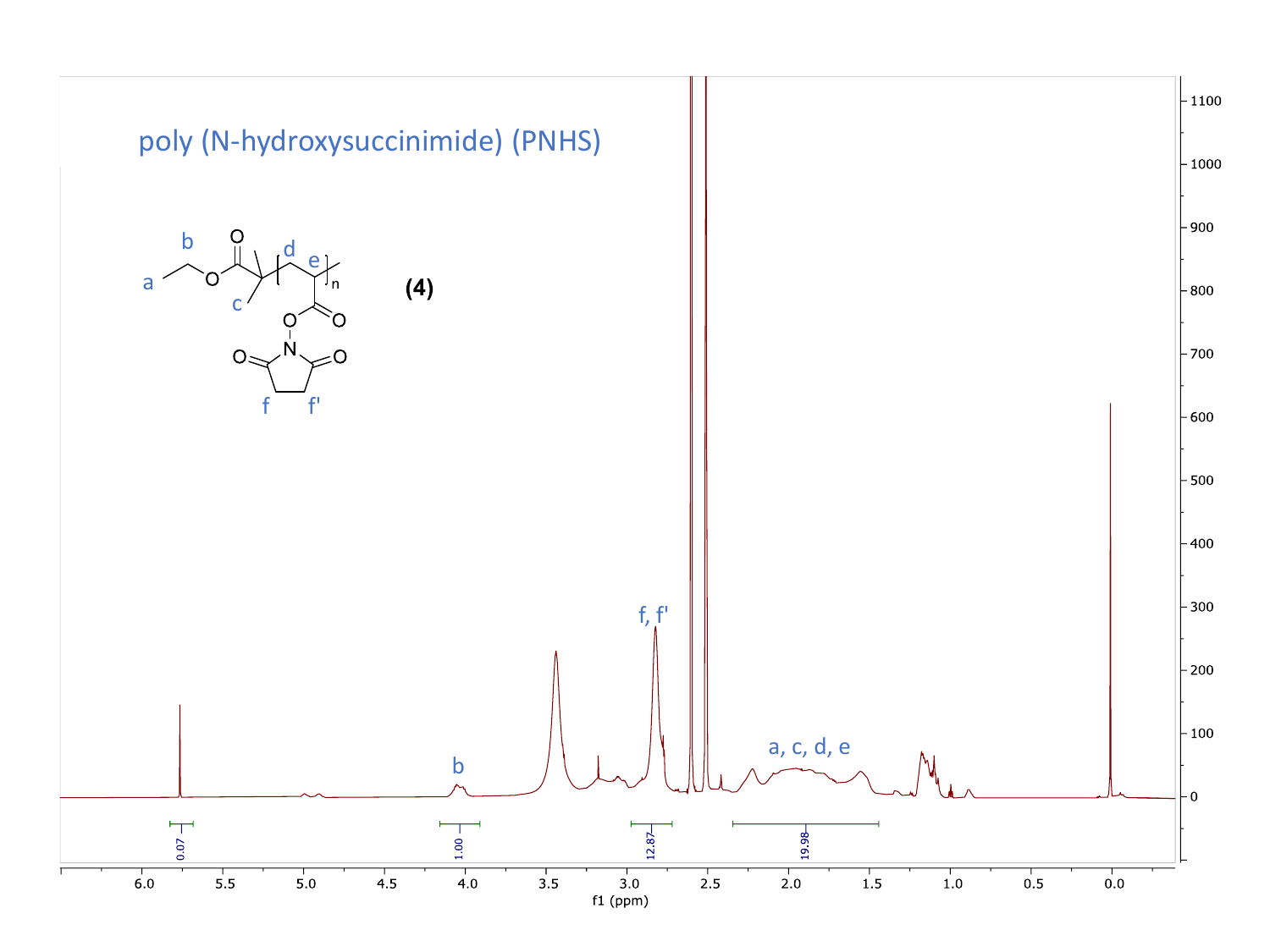}
    \caption{\textbf{\textsuperscript{1}H-NMR spectrum} of PNHS (4).}
    \label{fig:NMR4}
\end{figure}

Compound (3) (47.55 mg, 0.057 mmol) was added together with EDC-HCl (164.24 mg, 0.857 mmol, 1.5 equiv. per carboxylic acid groups) and NHS (98.60 mg, 0.857 mmol, 1.5 equiv. per carboxylic acid groups) in a 20-ml vial. Afterwards, 8.57 ml MQ water was added to have 0.1 M EDC-HCl / NHS in water. The mixture was stirred at room temperature for 45 minutes. Consequently, the mixture was added to 35ml MQ water, vortexed vigorously and centrifuged at 3273 $\times$ g for 20 minutes. This was repeated 3 times. The precipitate was dried overnight under a flow of N\textsubscript{2}. The outcome was a white powder (51.73 mg, 49.5\% yield). \textsuperscript{1}H-NMR (400 MHz, acetone-d\textsubscript{6}) is shown in Figure \ref{fig:NMR4}

\subsubsection{Functionalization of PNHS (4) with ET (1) to get PT16 (5)}

\begin{figure}[H]
    \centering
    \includegraphics[width=0.8\linewidth]{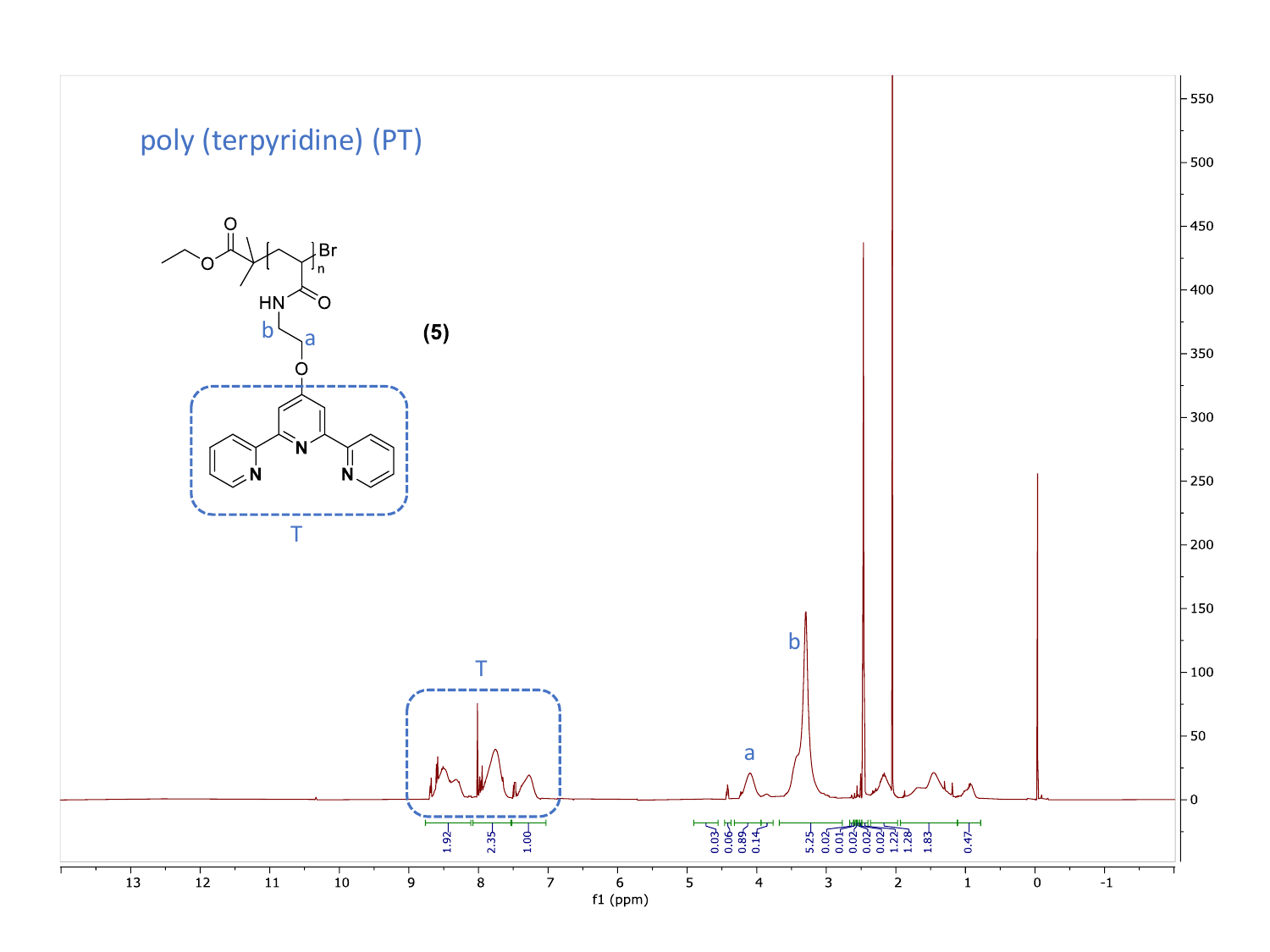}
    \caption{\textbf{\textsuperscript{1}H-NMR spectrum} of PT (5).}
    \label{fig:NMR5}
\end{figure}

In a 4-ml vial, compound (4) (30 mg, 0.010 mmol) and compound (1) (72.51 mg, 0.25 mmol, 1.5 equiv. per NHS group) were added, following by addition of a magnetic stir bar. The vial was sealed and degassed under N\textsubscript{2} flow. Next, anhydrous DMSO (0.3 ml) was added, and the mixture was stirred until it became homogeneous. Afterwards, 1 drop of TEA was added and the vial was kept in an oil bath for 3 hours at 60 \degree C. During the reaction the solution became red. Consequently, the product was precipitated in 20 ml ice cold acetone, and then centrifuged at 3273 $\times$ g for 30 minutes. The precipitate red gel was left overnight under N\textsubscript{2}. The final product was a red gel-like solid (17.3 mg, $\approx$ 30\%  \emph{\DJ}$\approx1.04$). \textsuperscript{1}H-NMR (400 MHz, acetone-d\textsubscript{6}) is shown in Figure \ref{fig:NMR5}.

\subsection{Monitoring the binding behavior of iron onto terpyridine in a two-phase water and oil set-up} \label{OMC_experimentalset-up_SI}

In order to monitor the binding behavior of the iron ions onto terpyridine monomer and the terpyridine-functionalized polymers, a two-phase water/oil set-up was designed, Figure \ref{fig:Intro}. The oil phase was chosen to be dichloromethane (DCM) due to its immiscibility with water. Solutions of terpyridine monomer or terpyridine-functionalized polymers in DCM with the same concentration of terpyridine groups (200 $\mu$M) were prepared. In addition, the aqueous phase contained various concentrations of FeCl\textsubscript{2} ranging between 10 and 200 $\mu$M. To ensure the reduction of the iron ions to Fe\textsuperscript{2+}, an excess amount of ascorbic acid (AA) was added to the water solutions. The described water and oil solutions were then exposed to each other in sealed vials and kept for 10 days after which the color of the aqueous phase in the vials turned pink. In the case of terpyridine-functionalized polymers, gel-like polymer networks were observed in the aqueous phase which is explained by the formation of bis (terpyridine) iron (II) complexes and therefore formation of crosslinked polymer chains with iron bridges. Finally, the water and oil phases were isolated and analyzed for iron and terpyridine concentrations, respectively (See Figure \ref{fig:OMCfoandhillplots} for summary of the results).

\subsubsection{Quantification of the free iron concentration in the water phase by inductively coupled plasma - atomic emission spectroscopy (ICP-AES) and ultraviolet-visible light spectroscopy (UV-Vis)} \label{quant_theta_SI}

\begin{figure}[H]
    \centering
    \includegraphics[width=\linewidth]{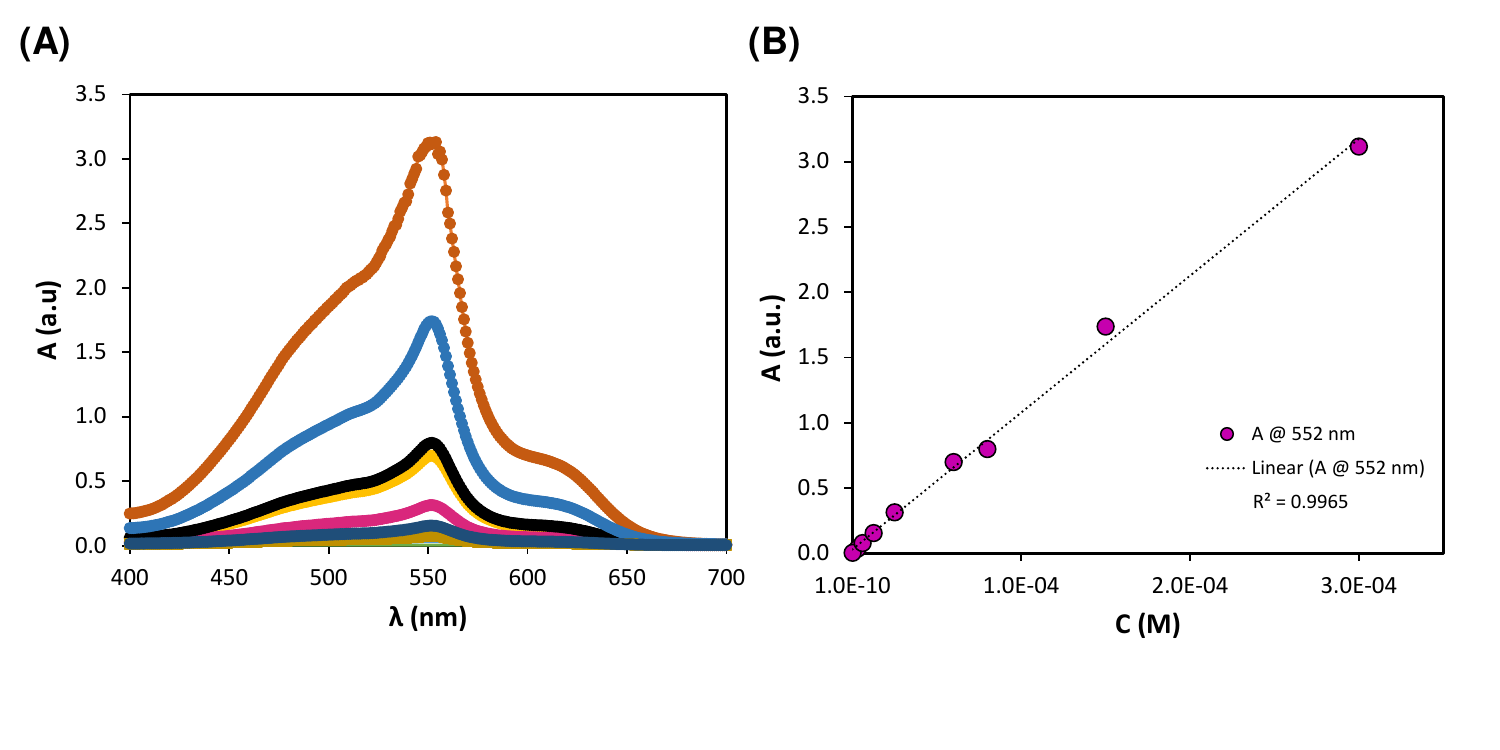}
    \caption{\textbf{Calibration used for quantification of the bis iron (II) terpyridine complexes in the aqueous state.} (A) Raw UV-Vis data of bis iron (II) terpyridine, (FeT\textsubscript{2})\textsuperscript{2+} complexes in the water phase. (B) Calibration curve of the (FeT\textsubscript{2})\textsuperscript{2+} complexes used for quantification of bound iron in the water phase (aqueous state).}
    \label{fig:UV_Cal_aq}
\end{figure}

The free iron concentration was obtained by subtraction of the iron in chelate complexes with terpyridine groups (i.e., bound iron ions) from the total iron in the water phase (i.e., bound + free iron ions). Therefore, free iron (i.e., not chelated by terpyridine) concentration in the water phase was measured by the difference between the measured iron concentrations with ICP-AES and UV-Vis spectroscopies. 
ICP-AES was performed on the solutions using an Optima 8300 instrument (PerkinElmer, Waltham, MA, USA) to obtain the total iron concentration in the water phase. Samples were dissolved in 10 ml of a 2\% HNO\textsubscript{3} solution to achieve optimal measurement concentration ranges. The measurements were performed in triplicate and the results were reported by an average and a standard deviation from duplicate measurements. 
UV-Vis measurements were done on the solutions at $\lambda_{max}$ = 552 nm at room temperature by a Lambda-35 spectrophotometer (PerkinElmer, Waltham, MA, USA), using quartz cuvettes. All measurements were performed in separate duplicates and quantification of the bound iron was performed with a calibration curve on a similar system (0.0125 – 0.3 mM, R\textsuperscript{2} $>$ 0.99), see Figure \ref{fig:UV_Cal_aq}.

\subsubsection{Quantification of terpyridine/terpyridine-functionalized polymers in the oil phase by ultraviolet-visible light spectroscopy (UV-Vis)} \label{quant_fH_SI}
\begin{figure}[H]
    \centering
    \includegraphics[width=\linewidth]{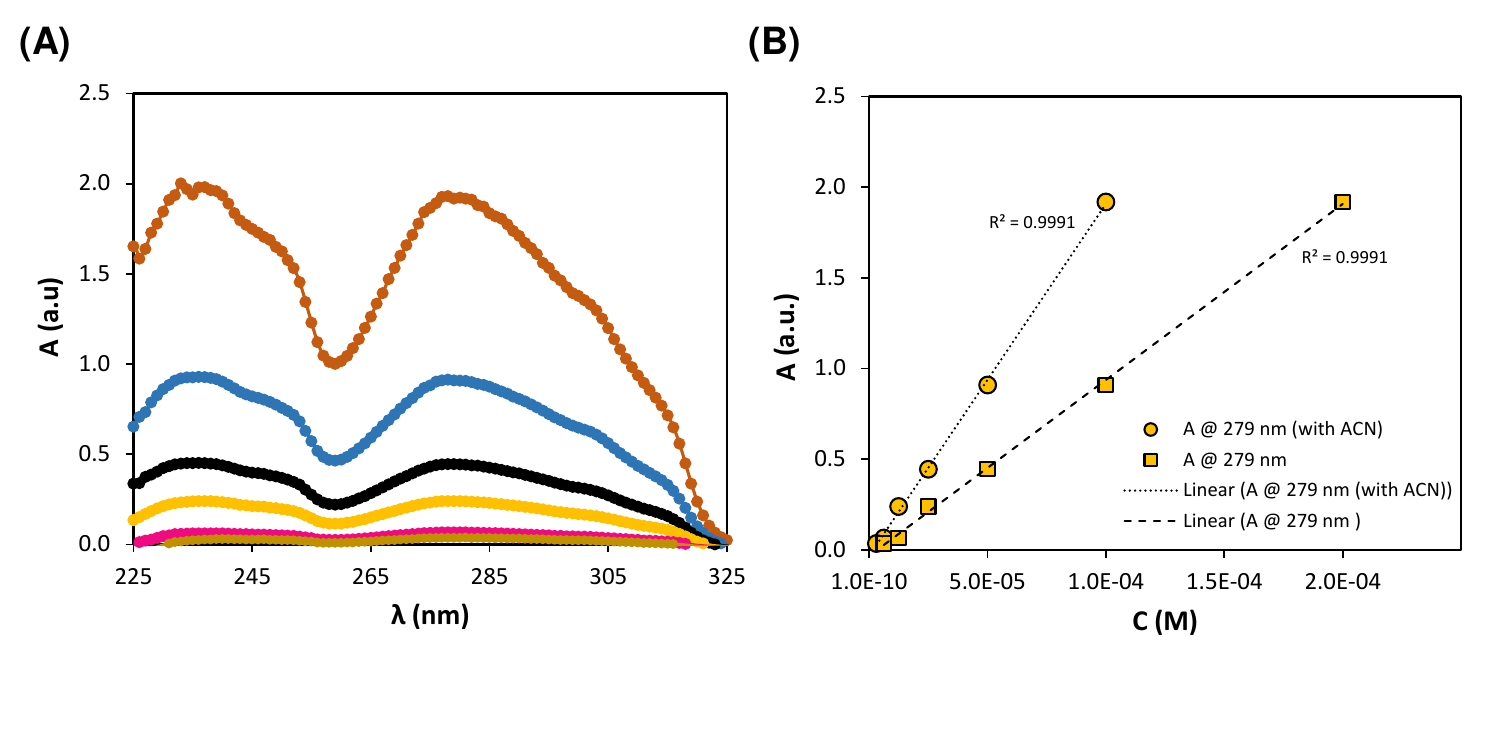}
    \caption{\textbf{Calibration used for quantification of the terpyridine groups in the hydrophobic state.} (A) Raw UV-Vis data of terpyridine groups in the oil (DCM/ACN with 1:1 volume ratio) phase. (B) Calibration curve of terpyridine in DCM/ACN (square markers) used for quantification of remained terpyridine in the oil phase (hydrophobic state). Addition of ACN helped with tracing lower concentration of terpyridine in the hydrophobic state, compared to pure DCM (circle markers).}
    \label{fig:UV_Cal_oil}
\end{figure}

The concentration of terpyridine groups (i.e., also indicative of the concentration of the polymers) in the oil phase was quantified by UV-Vis with $\lambda_{max}$ = 279 nm at room temperature. The oil solutions were mixed with acetonitrile (ACN) (1:1 volume ratio) prior to measurement, following a previously-reported procedure \cite{Hu2013_supp, Belhadj2013_supp}. UV-Vis spectra of the solutions were then recorded on a Lambda-35 spectrophotometer (PerkinElmer, Waltham, MA, USA), using quartz cuvettes. All the measurements were done in separate duplicates and quantification of the remained terpyridine fraction in the oil phase was performed with a calibration curve of terpyridine solutions in DCM/ACN with 1:1 volume ratio (0.00625 – 0.2 mM, R\textsuperscript{2} $>$ 0.99), see Figure \ref{fig:UV_Cal_oil}. 

\subsection{Determination of the hydrophobic contribution $g_H$ of terpyridine by its solubility in DCM and water} \label{TerpygH_SI}

The solubilities of monomeric terpyridine in water (with ascorbic acid, pH $\approx$ 3, saturated with DCM) and in DCM (saturated with water) are measured to be $c_T^{aq} = \text{3.00 $\pm$ 0.06 mM}$ and $c_T^{DCM} = \text{2.80 $\pm$ 0.03 M}$, respectively. Taking the chemical potential of terpyridine in the solid as $\mu^*$, we have $\mu_T^{0, aq} + k_BT\ln{(c_T^{aq}/c^0)} = \mu^*$. Likewise, in DCM, we have $\mu_T^{0, DCM} + k_BT\ln{(c_T^{DCM}/c^0)} = \mu^*$. Here the $\mu_T^{0, i}$ are the standard chemical potentials of terpyridine in $i = \text{aq, DCM}$ at concentration $c^0$. The reversible work to transfer a terpyridine molecule from DCM to the aqueous phase is given by $g_H = \mu_T^{0, aq} - \mu_T^{0, DCM} = k_B T \ln{(c_T^{DCM}/c_T^{aq})} \approx \text{6.8} k_BT$.

\vfill
\pagebreak

\subsection{Partitioning of terpyridine monomer in the two-phase system} \label{OMC_partition_T}

\begin{figure}[H]
    \centering
    \includegraphics[width=0.8\linewidth]{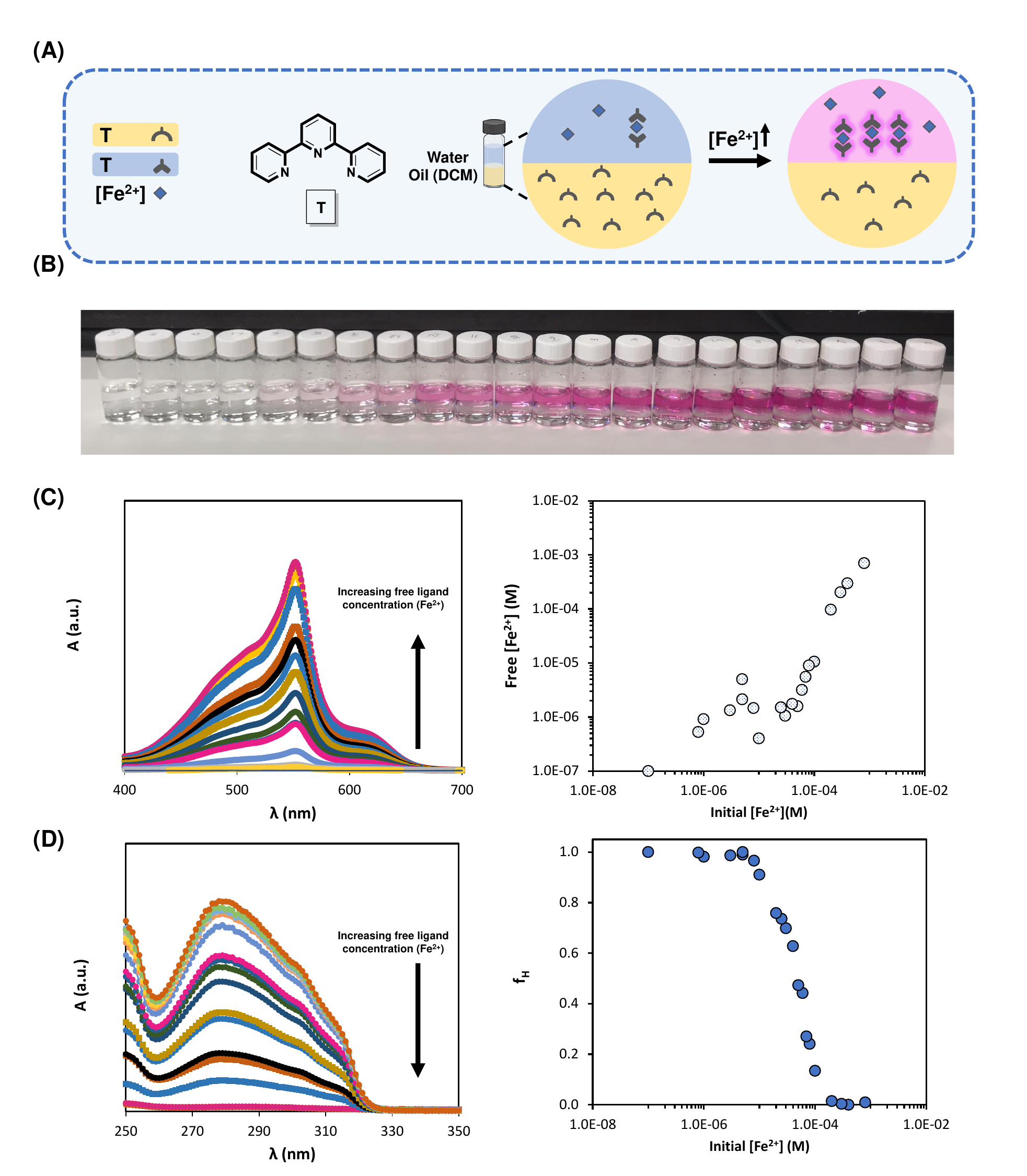}
    \caption{\textbf{Partitioning of terpyridine monomer in the two-phase water/DCM system.} (A) Schematic representation of the experimental setup. (B) Visualization of the experiment. Top and bottom phases are water and DCM, respectively. Complexation of terpyridine with iron ions in the water (top) phase upon increasing iron concentration in the water phase (from left to right) is evident by the pink color of the top phase. (C) UV-Vis spectroscopy of terpyridine iron (II) complexes in the water phase (left) and extraction of free iron ion concentration after subtracting the obtained Fe\textsuperscript{2+} concentrations (by UV-Vis) from the values obtained by ICP-AES (right). (D) UV-Vis spectroscopy of terpyridine in the hydrophobic phase (left) and extraction of $f_H$; fraction of terpyridine in DCM phase (right).}
    \label{fig:RandomTerpymonomer_SI}
\end{figure}

\vfill
\pagebreak

\subsection{Partitioning of the terpyridine-functionalized polymer in the two-phase system}\label{OMC_partition_PT}

\begin{figure}[H]
    \centering
    \includegraphics[width=0.8\linewidth]{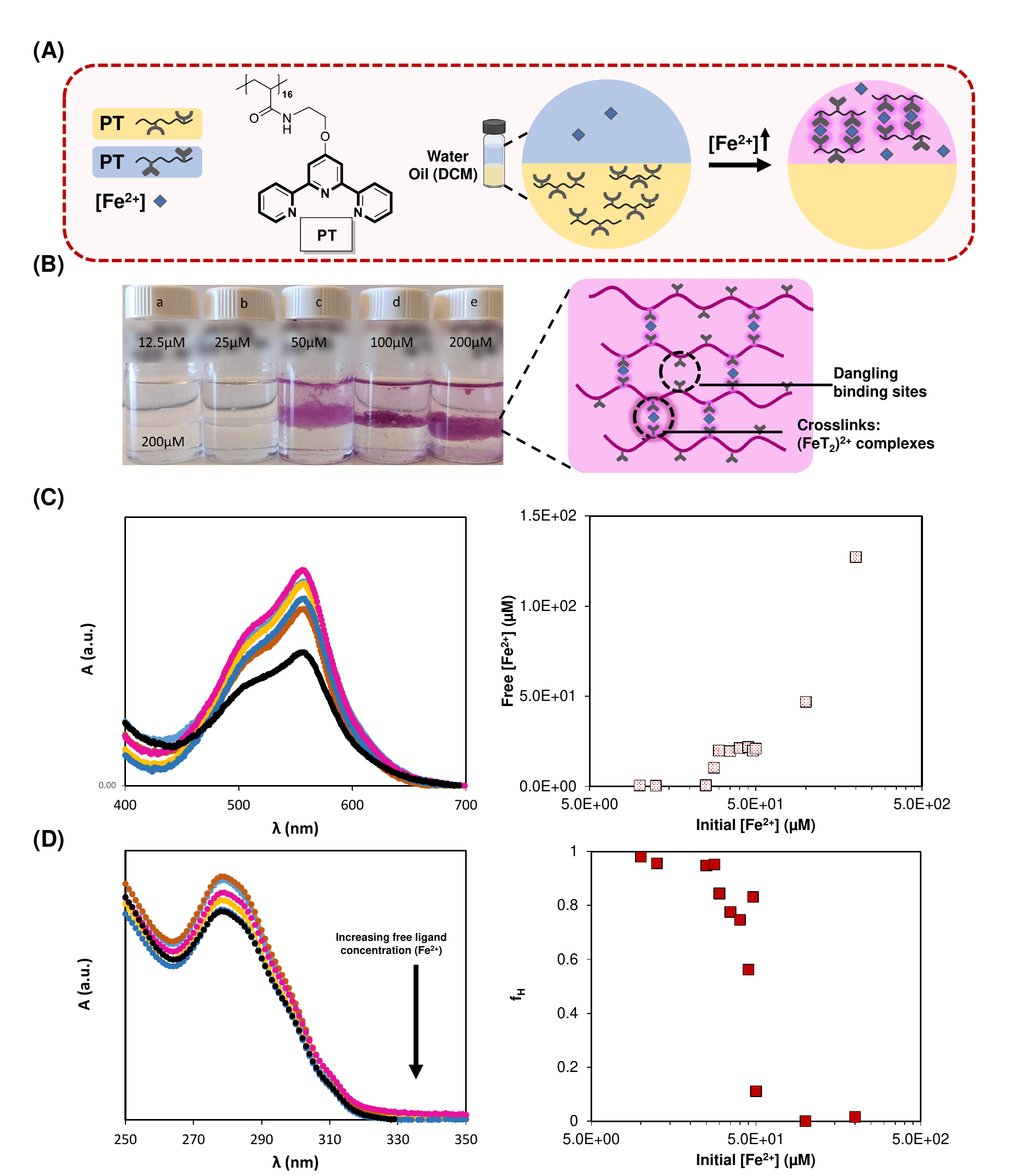}
    \caption{\textbf{Partitioning of terpyridine-functionalized polymer in the two-phase water/DCM system.} (A) Schematic representation of the experimental setup. (B) Visualization of the experiment. Top and bottom phases are water and DCM, respectively. Complexation of terpyridine functional groups with iron ions in the water (top) phase upon increasing iron concentration in the water phase (from left to right) is evident by the pink color of the top phase. Clusters of polymer chains bridged by iron ions in the bis terpyridine iron (II) complexes are visible to the eye. (C) UV-Vis spectroscopy of terpyridine iron (II) complexes in the water phase (left) and extraction of free iron ion concentration after subtracting the obtained Fe\textsuperscript{2+} concentrations (by UV-Vis) from the values obtained by ICP-AES (right). (D) UV-Vis spectroscopy of terpyridine in the hydrophobic phase (left) and extraction of $f_H$; fraction of terpyridine groups (equivalent to the fraction of the polymers) in DCM phase (right).}
    \label{fig:CooperativePolyTerpy_SI}
\end{figure}

\subsection{Terpyridine monomer partitioning between DCM and water} \label{T_monomervsdimer}

\begin{figure}[H]
    \centering
    \includegraphics[width=\linewidth]{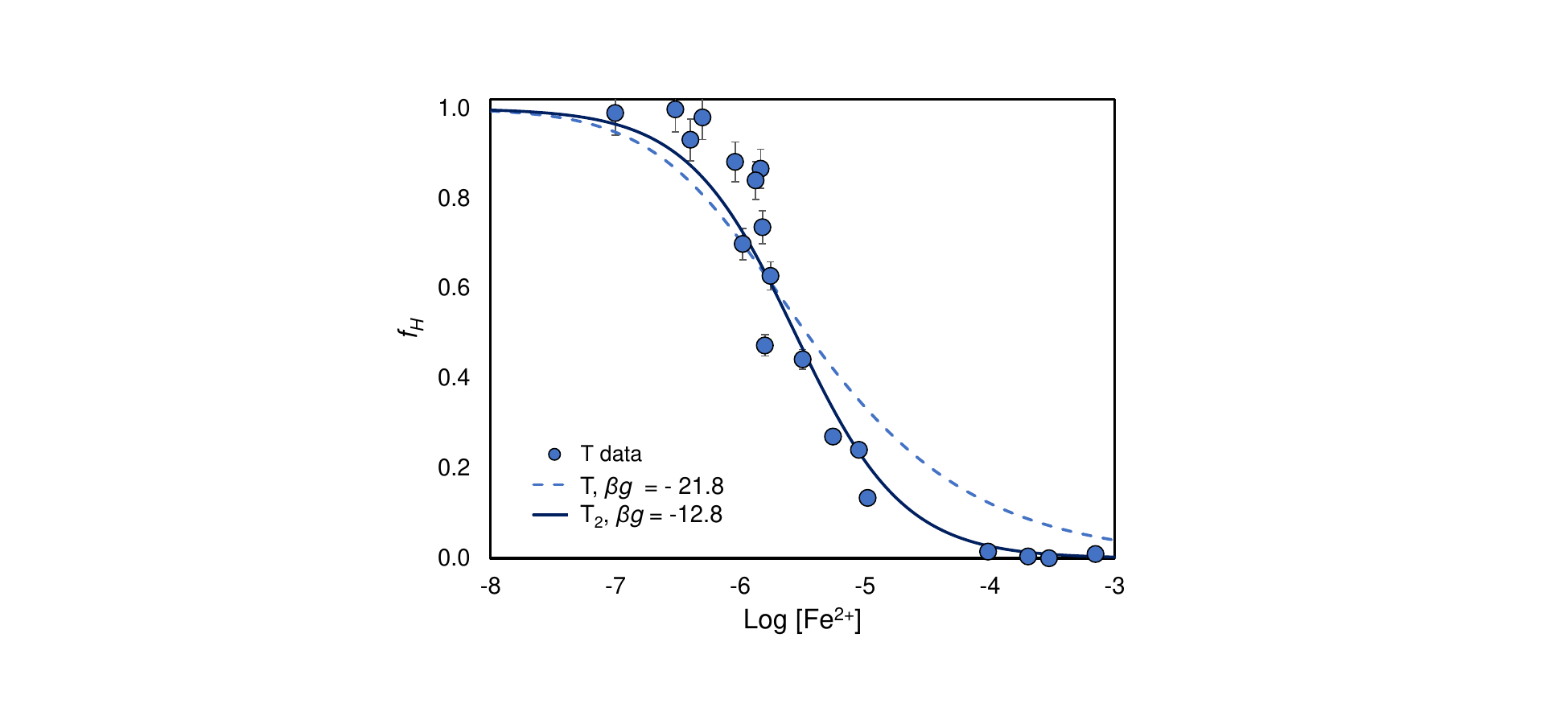}
    \caption{\textbf{Comparison between experiments and theory for partitioning of terpyridine monomer between oil (DCM) and water phases}. Fraction of terpyridine (blue markers, T data) in the oil phase ($f_H$) as a function of the free iron (II) concentration in the aqueous phase. Blue dashed line (T) is Equation \ref{foMono} which gives the best fit with $\beta g = - 21.8$. The blue solid line (T$_2$)which gives a better description of the data on terpyridine is Equation \ref{fodim} with $\beta g = - 12.8$.}
    \label{fig:T_monomervsdimer}
\end{figure}

The experimental results showed a gradual decrease in fraction of terpyridine in the hydrophobic state ($f_H$) from 1 to 0 upon increasing and over a broad range of free iron concentration in the aqueous phase, see the markers in Figure \ref{fig:T_monomervsdimer}.
As it was mentioned in the section \ref{terpyridine monomer}, the obtained fraction of terpyridine in the oil (DCM) phase experimentally showed clear deviation from Equation \ref{foMono}. The blue dashed line in Figure \ref{fig:T_monomervsdimer} is Equation \ref{foMono} with $[T_{tot}] = 2 \times 10^{-4} M$ (based on the experiments). The best fit based on Equation \ref{foMono} results in linear combination of binding and hydrophobic energies for each bis (terpyridine) iron (II) complex, $\beta g = - 21.8$.
However, our results of binding iron ions onto terpyridine pointed to a significantly steeper dependence of the terpyridine fraction in the oil phase on the free iron (II) concentration. To resolve the mentioned discrepancy, we assumed that terpyridine is not in monomeric form in the hydrophobic state, but in the form of pairs. 
The blue solid line in Figure \ref{fig:T_monomervsdimer} is the best fit of Equation \ref{fodim} to the experimental data of terpyridine (blue circles) with $\beta g = - 12.8$ and gives a better match for the experiment results, compared to Equation \ref{foMono}.

Overall, it is not clear to us why the experimental data are best described by taking the assumption of formation of terpyridine dimers in the solution in the presented model. Furthermore, it should be noted as well that the partitioning of terpyridine monomer in water in the presence of the oil (DCM) was not measurable due to the detection limit of the UV-Vis spectroscopy. 

\bibliography{references_supp} 